\pdfoutput=1
\documentclass[12pt,a4paper]{amsart} 
\usepackage{amsaddr}
\usepackage[T1]{fontenc}
\usepackage[margin=2cm]{geometry}
\usepackage[colorlinks,allcolors=blue,pdfusetitle]{hyperref}
\usepackage{microtype}  
 
\usepackage{amscd,amsmath,amssymb,amsfonts,xspace,mathrsfs,amsthm}
\usepackage{xcolor}

\usepackage{bbold}
\usepackage{latexsym}
\usepackage{graphicx}
\usepackage{dsfont}
\usepackage{longtable} 
\usepackage{tikz-cd}
\usetikzlibrary{bending}

\theoremstyle{definition}
\numberwithin{equation}{section}

\def\diag{{\rm diag}}

\def\({\left(}
\def\){\right)}
\def\<{\left\langle}
\def\>{\right\rangle}

\newcommand{\Omstar}{\Omega_\star}

\newcommand{\de}{\mathrm{d}}

\newcommand{\I}{\mathrm{i}}

\newcommand{\cA}{\mathcal{A}}

\newcommand{\cC}{\mathcal{C}}
\newcommand{\cD}{\mathcal{D}}
\newcommand{\cE}{\mathcal{E}}
\newcommand{\cF}{\mathcal{F}}
\newcommand{\cV}{\mathcal{V}}

\newcommand{\cI}{\mathcal{I}}

\newcommand{\cM}{\mathcal{M}}
\newcommand{\cW}{\mathcal{W}}

\newcommand{\cX}{\mathcal{X}}

\newcommand{\cO}{\mathcal{O}}
\newcommand{\cP}{\mathcal{P}}
\newcommand{\cR}{\mathcal{R}}

\newcommand{\cT}{\mathcal{T}}
\newcommand{\cU}{\mathcal{U}}

\DeclareSymbolFont{AMSa}{U}{msa}{m}{n}
\DeclareSymbolFont{AMSb}{U}{msb}{m}{n}
\DeclareMathSymbol{\fieldR}{\mathalpha}{AMSb}{"52}

\newcommand{\nn}{\nonumber}

\newcommand{\eps}{\epsilon}

\newcommand{\IR}{\mathds{R}}
\newcommand{\IC}{\mathds{C}}
\newcommand{\IZ}{\mathds{Z}}
\newcommand{\IQ}{\mathds{Q}}
\newcommand{\IN}{\mathds{N}}
\newcommand{\IH}{\mathds{H}}

\newcommand{\IP}{\mathds{P}}
\newcommand{\IF}{\mathds{F}}

\def\bea{\begin{eqnarray}}
\def\eea{\end{eqnarray}}
\def\be{\begin{equation}}
\def\ee{\end{equation}}
\def\ba{\begin{align}}
\def\ea{\end{align}}
\def\bse{\begin{subequations}}
\def\ese{\end{subequations}}

\def\({\left(}
\def\){\right)}
\def\[{\left[}
\def\]{\right]}

\DeclareMathOperator{\sign}{sign}

\DeclareMathOperator{\Li}{Li}
\DeclareMathOperator{\Hom}{Hom}

\DeclareMathOperator{\Rep}{Rep}
\DeclareMathOperator{\Ext}{Ext}
\DeclareMathOperator{\Stab}{Stab}

\DeclareMathOperator{\Coh}{Coh}

\DeclareMathOperator{\Fr}{Fr}
\newcommand{\GLt}{\widetilde{GL^+}(2,\IR)}

\makeatletter
\DeclareRobustCommand{\ch}{\@ifnextchar _{\textnormal{ch}}{\operatorname{ch}}}
\makeatother

\def\bOm{\bar\Omega}

\definecolor{varcolor}{rgb}{0.1,0.55,0.25}
\definecolor{functioncolor}{rgb}{0.1,0.35,0.75}
\definecolor{paper_blue}{rgb}{0.3,0.2,0.75}
\definecolor{paper_red}{rgb}{0.65,0.1,0.15}
\definecolor{paper_green}{rgb}{0.05,0.35,0.125}
\definecolor{paper_grey}{gray}{0.375}
\definecolor{perm}{rgb}{0.1,0.45,0.85}
\definecolor{deemph}{rgb}{0.7,0.7,0.7}
\usepackage{mathtools}
\newcommand{\mathtikz}[2][]{\begin{tikzpicture}[baseline=\the\dimexpr-\fontdimen22\textfont2\relax,#1]#2\end{tikzpicture}}

\newcommand{\Rgeo}{\cR^{\rm geo}}

\newcommand{\Ract}{\cR}
\newcommand{\Uact}{\cU}

\ifdefined\pdfsuppresswarningpagegroup\pdfsuppresswarningpagegroup=1 \fi

\title{BPS Dendroscopy on local \texorpdfstring{$\mathds{P}^1\times \mathds{P}^1$}{P1xP1}}

\author{Bruno Le Floch, Boris Pioline, Rishi Raj}

\address{
\vspace*{.8cm}
Laboratoire de Physique Th\'eorique et Hautes Energies (LPTHE, 
UMR 7589), \\
Sorbonne Universit\'e and CNRS, 
Campus Pierre et Marie Curie, \\
4 place Jussieu, F-75005, Paris, France}
\email{\href{mailto:blefloch@lpthe.jussieu.fr}{\textcolor{black}{blefloch}},\href{mailto:pioline@lpthe.jussieu.fr}{{\textcolor{black}{pioline}}},\href{mailto:raj@lpthe.jussieu.fr}{\textcolor{black}{raj@lpthe.jussieu.fr}}}

\begin{document}

\begin{abstract}

BPS states in type II string compactified on a Calabi-Yau threefold can typically be decomposed as moduli-dependent bound states of absolutely stable constituents, with a hierarchical structure labelled by attractor flow trees. This decomposition is best understood from the scattering diagram, an arrangement of real codimension-one loci (or rays) in the space of stability conditions where BPS states of given electromagnetic charge and fixed phase of the central charge exist. The consistency of the diagram when rays intersect determines all BPS indices in terms of the `attractor indices' carried by the initial rays. 
In this work we study the scattering diagram for a non-compact toric CY threefold known as 
local $\mathds{F}_0$, namely the total space of the canonical bundle over $\mathds{P}^1\times \mathds{P}^1$. We first construct the 
scattering diagram for the quiver, valid near the orbifold point, and for the large volume slice, valid
when both $\mathds{P}^1$'s have large (and nearly equal) area. We then combine the insights gained from these simple limits to construct the scattering diagram along the physical slice of $\Pi$-stability conditions, which carries an action of a $\mathds{Z}^4$ extension of the modular group $\Gamma_0(4)$. We sketch a proof of the Split Attractor Flow Tree Conjecture in this example, albeit for a restricted range of the central charge phase. 
Most arguments are similar to our early study 
of  local $\mathds{P}^2$ \cite{Bousseau:2022snm},  but complicated by the occurence of 
an  extra mass parameter and 
ramification points on the  $\Pi$-stability slice. 
\end{abstract}

\maketitle
\tableofcontents

\section{Introduction and summary}
The BPS spectrum in type II string theory compactified on a Calabi-Yau (CY) threefold $X$ has been a favorite playground for physicists and mathematicians alike. One reason is that the indices 
$\Omega_\sigma(\gamma)$ counting stable BPS states with fixed electromagnetic charge $\gamma$ 
have a rigorous definition, as the generalized Donaldson-Thomas invariants
associated to the derived category of coherent sheaves $\cC=D^b\Coh X$. 
They depend on a choice of Bridgeland stability condition\footnote{Here $Z$ is the usual central charge determining the mass of BPS states, and $\cA$ is the `heart', an Abelian subcategory of $\cC$
locally determined by $Z$ and 
and determining the allowed decays.}
 $\sigma=(Z,\cA)\in \Stab\cC$, which formalizes and 
generalizes the physical notion of stability \cite{Bridgeland:2006bzr}.  
For compact CY threefolds, direct computations
of DT invariants are generally out of reach, and explicit results typically rely on 
wall-crossing from an empty chamber in $\Stab\cC$ 
where $\Omega_\sigma(\gamma)$ vanishes (see e.g.~\cite{Toda:2011aa,Feyzbakhsh:2020wvm,Feyzbakhsh:2021rcv,Feyzbakhsh:2022ydn,Alexandrov:2023zjb} for recent progress using this strategy). Instead, for non-compact toric CY threefolds, the category $\cC$ is isomorphic to the derived category of representations of a quiver with potential $D^b\Rep(Q,W)$, and one may hope to describe the complete BPS spectrum at any point $\sigma\in\Stab\cC$,
or at least be able to compute the BPS index $\Omega_\sigma(\gamma)$ at any such point. This in turn determines the 
BPS spectrum of the five-dimensional superconformal field theory on $\mathbb{R}^{3, 1} \times S^1$ arising in the reduction of
M-theory on $X$, after compactifying further on a circle down to 3+1 dimensions.

\medskip

In \cite{Bousseau:2022snm}, we carried out this program for $X=K_{\IP^2}$, the total space of the canonical bundle over the projective plane, also known as local $\IP^2$; let us briefly recall
the salient results. In this case, since $b_2(\IP^2)=1$ the space of 
stability conditions has complex dimension 3, but the BPS spectrum is invariant under the action\footnote{More precisely, the space of stability conditions 
carries an action of the universal cover $\widetilde{GL(2,\IR)^+}$, but to avoid cluttering 
we omit the tilde.
} of $GL(2,\IR)^+$, effectively leaving a complex one-dimensional space of inequivalent 
stability conditions. We found that the BPS spectrum can be
efficiently encoded into the scattering diagram $\cD_\psi\subset \Stab \cC$, defined as the union
of the loci (also called rays) $\cR_\psi(\gamma)$ where there exist semi-stable objects of central charge of fixed phase $Z(\gamma)\in \I e^{\I\psi} \IR^+$, decorated at every point with a certain automorphism of the quantum torus algebra encoding the refined BPS index $\Omega_\sigma(\gamma,y)$.  This concept, introduced in the context of quivers in \cite{bridgeland2016scattering}, is the natural framework for the attractor flow tree representation of 
BPS bound states \cite{Denef:2001xn,Denef:2007vg,Alexandrov:2018iao,Arguz:2021zpx}, at least in the context of non-compact CY threefolds where the phase $\psi$ is constant along the 
flow \cite{Bousseau:2022snm}.  

Specifically, in {\it loc.~cit.}\ we constructed the scattering diagram $\cD_\psi$ along the large volume slice $\Lambda$, generalizing the analysis of  \cite{Bousseau:2019ift}
to arbitrary phase $\psi$, and along the physical slice $\Pi$ singled out by mirror symmetry. This slice
is isomorphic to the universal cover of the K\"ahler moduli space $\IH/\Gamma_1(3)$, which is equal to the Poincar\'e upper half-plane $\IH$ itself, and it admits an action of the modular group $\Gamma_1(3)$ by autoequivalences \cite{Bridgeland:2005my,Bayer:2009brq}. In particular,
the central charge has a contour integral representation in terms of a certain weight 3 Eisenstein series of 
$\Gamma_1(3)$, which allows for a straightforward analytic extension to the full Poincar\'e upper half-plane. 
The large volume slice $\Lambda$ is also parametrized by $\IH$, and is in fact related to $\Pi$ under $GL(2,\IR)^+$ above a certain curve which passes through the  conifold points $\tau\in \IZ$ but stays away from the orbifold points $\tau\in \tau_o +\IZ$ with $\tau_o=-\frac12+\frac{\I}{2\sqrt3}$. Along the slice $\Lambda$, the only initial rays
are those emanating from the conifold points $\tau=m\in \IZ$, associated to the fluxed D4 and anti-D4 branes $\cO_{\IP^2}(m)$ and
$\cO_{\IP^2}(m)[1]$, and the full BPS spectrum at large volume arises by iterated scattering of these rays. 
The scattering diagram along the slice $\Pi$ is much more complicated, since there are initial rays for all $\Gamma_1(3)$ images of $\cO_{\IP^2}$, emanating from every $\tau=\frac{p}{q}\in\IQ$ with $(p,q)=1, q\neq 0\mod 3$. This includes in particular all homological shifts $\cO_{\IP^2}(m)[k]$ at $\tau=m$. However, as long as $|\psi|$ is less than a certain critical phase $\psi_c$, the only
initial rays which can reach $\tau=\I\infty$ are the same as those in $\Lambda$. The scattering diagram of the orbifold quiver $(Q,W)$ is embedded in a neighbourhood of the orbifold point $\tau_o$ and
its $\Gamma_1(3)$ images, which shrinks as $|\psi|\to \frac{\pi}{2}$. When $|\psi|=\frac{\pi}{2}$,
the scattering diagram simplifies drastically as rays coincide with level sets of a certain slope function $s(\tau)$, which only intersect at $\tau_o$ and its images.

\medskip

\begin{figure}[t]
    \centering
\includegraphics[height=9cm]{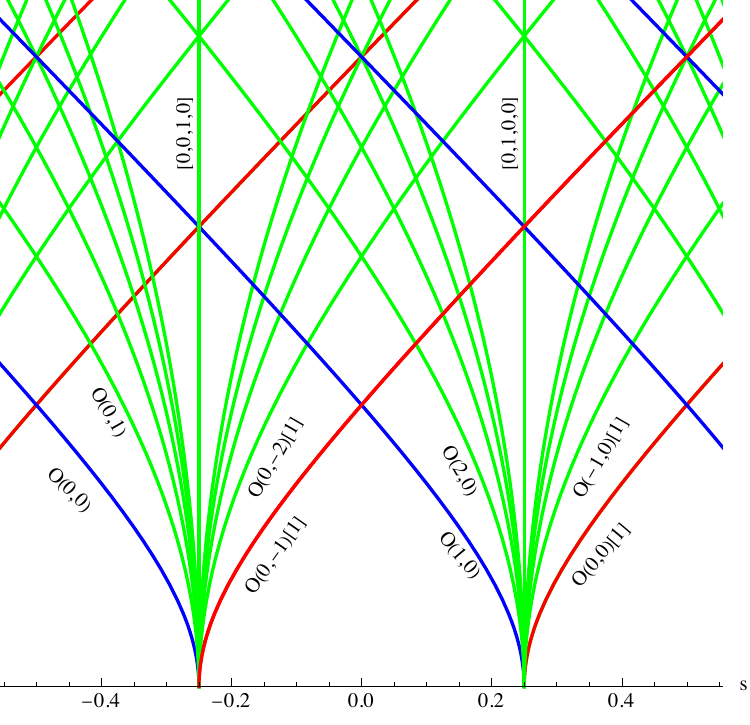}
    \caption{Initial rays for the large volume scattering diagram of local $\IF_0$ in the $(s,t)$ plane at $m=1/2$   \label{LVScatt_half}}
\end{figure}

Our aim in this paper is to extend this study to the next simplest toric CY threefold, namely $X=K_{\IF_0}$, the canonical bundle over the Hirzebruch surface $\IF_0=\IP^1\times \IP^1$, also known as local $\IF_0$. Unlike the local $\IP^2$ case, M-theory compactified on $\IF_0$ 
yields a weakly coupled five-dimensional gauge theory at low energies, with gauge group $SU(2)$ and vanishing discrete theta angle. 
Because $b_2(\IF_0)=2$, the space of stability conditions has complex dimension 4, reduced
to 2 after modding out by the action of $GL(2,\IR)^+$. 
Fixing only the $\IC^\times \subset GL(2,\IR)^+$ action, we can write the central charge 
for an object of Chern character $\gamma(E)=[r,d_1,d_2,\ch_2]$ as 
\be
Z(\gamma) = -2r T_D+ d_1 T_1 + d_2 T_2 -\ch_2
\ee
where $(T_1,T_2)\in \IH\times \IH$ can be viewed as the complexified K\"ahler parameters  measuring the area and $B$-field along the two $\IP^1$ factors.
The large volume slice $\Lambda\subset \Stab \cC$ is obtained by setting $T_D=\frac12 T_1 T_2$, and choosing a particular heart $\cA_{T_1,T_2}$ obtained from the Abelian category of coherent sheaves on $\IF_0$ by tilting (see \S\ref{sec_stabcond} for details). 
Since we shall mostly focus on the
canonical polarization $\Im T_1=\Im T_2$, we set $T_1=T, T_2=T+m$ and assume that $m$ is real, unless stated otherwise.
For fixed real $m$, the rays $\cR_\psi(\gamma)$ in the half-plane $(s\in \IR,t\in \IR^+)$ 
with $T=s-\frac{m}{2}+\I t$
are portions of hyperbolae centered on the real axis $t=0$,
while walls of marginal stability $\cW(\gamma,\gamma')$ are half-circles centered on the same axis, as in  \cite{Bousseau:2022snm}. However, unlike local $\IP^2$  where there are only a pair of rays
emanating from each integer point along the boundary $t=0$, we find that the
scattering diagram $\cD_{\psi,m}^{\Lambda}$ admits  an infinite set of initial rays 
emanating from each of the  points $s\in\IZ+\frac{m}{2}$ and $s=\IZ+\frac{m}{2}-\Fr(m)$ on the boundary, where the
following objects become massless (see Fig.~\ref{LVScatt_half} for $m=1/2$):
\be
\label{origLV}
\begin{split}
s=k+\tfrac{m}{2}:& \quad \cO(k+1+\lfloor m \rfloor+n,k), \quad \cO(k+\lfloor m \rfloor-n,k)[1], 
\quad GV_{-y-1/y}(1,0,k)  \\
s=k-\tfrac{m}{2}:& \quad   \cO(k,k-\lfloor m \rfloor +n), \quad \cO(k, k-\lfloor m \rfloor-n-1)[1], \quad 
GV_{-y-1/y}(0,1,k)
 \end{split}
\ee
for all $n\geq 0$. Here $\Fr(m):= m - \lfloor m \rfloor\in[0,1)$ denotes the fractional part of $m$, 
$\cO(k_1,k_2)$ is the line bundle with first Chern class $(k_1,k_2)$ on $\IP^1\times \IP^1$, corresponding to a pure D4-brane with $(k_1,k_2)$ units of flux, 
while $GV_\Omega(d_1,d_2,k)$ denotes a D2-D0 brane with Chern vector $[0,d_1,d_2,k]$ and 
BPS index $\Omega$. Fortunately, upon extending the scattering diagram slightly below the $t=0$ axis\footnote{More precisely, below the
parabola $y=-\frac12x(x+\mu)$  in affine coordinates introduced in \eqref{defxymu}, defined such that rays become straight lines in the $(x,y)$ plane.}, 
one finds that each of these infinite sets of rays originate from the scattering of just two rays with Dirac product 
$\langle\gamma_1,\gamma_2\rangle=\pm 2$, each of them carrying $\Omega(\gamma_i)=1$,
corresponding to the objects with $n=0$ in \eqref{origLV}. \footnote{In particular, we observe that i) the initial rays $O(k,k),O(k+1,k),O(k+1,k+1),\cO(k+2,k+1)$ in an interval of length 2 (corresponding to tensoring with the canonical class $K_{\IF_0}=\cO(2,2)$) form the foundation of a 
geometric helix~\cite{bridgeland2010helices}, and ii)
the large volume scattering diagram coincides with the scattering diagram
controlling  the Gromov-Witten invariants of the log Calabi-Yau surface $(\IF_0,D)$, where $D$ is a smooth very ample anticanonical divisor,
obtained by 
unfolding the spanning polytope of the toric fan of $\IF_0$~\cite{graefnitz2020tropical}.\label{fooGW}} 
With the knowledge of these initial rays, one can determine the BPS spectrum at any point $(s,t)$ along the slice $\Lambda$ by enumerating all possible scattering sequences in the past `light-cone', as in \cite{Bousseau:2022snm}. As $m$ varies,  the location of the initial rays also varies and some sequences may (dis)appear, leading to further wall-crossing phenomena beyond those occurring
at fixed $m$. 
For $m$ integer (the values relevant for duality with Chern-Simons on the lens space $L(1,2)$ \cite{Aganagic:2002qg}), the sets of initial rays coalesce in pairs and the structure of the initial rays becomes
more complicated, see Fig.~\ref{LVScatt0}.

\medskip

\begin{figure}
\begin{tikzpicture}[inner sep=2mm,scale=2]
  \node (a) at ( -1,1) [circle,draw] {$1$};
  \node (b) at ( 1,1) [circle,draw] {$2$};
  \node (c)  at ( 1,-1) [circle,draw] {$3$};
  \node (d)  at ( -1,-1) [circle,draw] {$4$};
 \draw [->>] (a) to node[auto] {$ $} (b);
 \draw [->>] (b) to node[auto] {$ $} (c);
 \draw [->>] (c) to node[auto] {$ $} (d);
 \draw [->>] (d) to node[auto] {$ $} (a);
\end{tikzpicture}
\caption{Quiver associated to the Ext-exceptional collection \eqref{extcoll}, sometimes known as phase II or orbifold quiver. This is accompanied
by the quartic superpotential $W$ in \eqref{Worb}.
\label{fig_orbquiver}}
\end{figure}
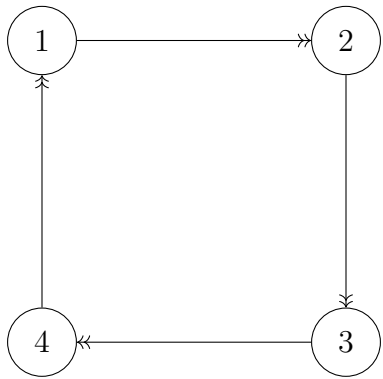

\begin{figure}[t]
    \centering
\includegraphics[height=7cm]{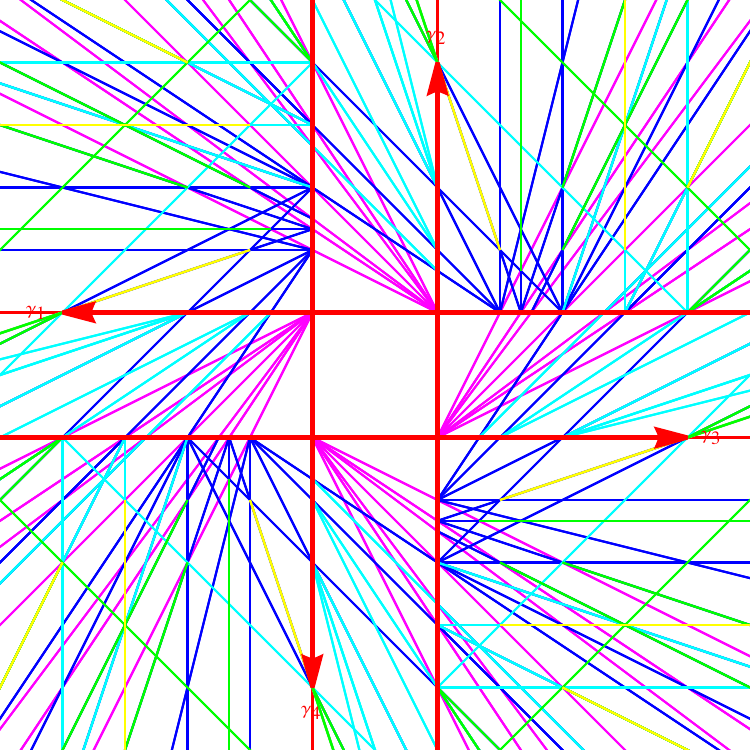}
    \caption{Scattering diagram of the orbifold quiver for $m=1/2$, keeping rays with height $\sum_i N_i\leq 8$.
      \label{QuiverScatt}}
\end{figure}

\begin{figure}[t]
    \centering
\includegraphics[height=10cm]{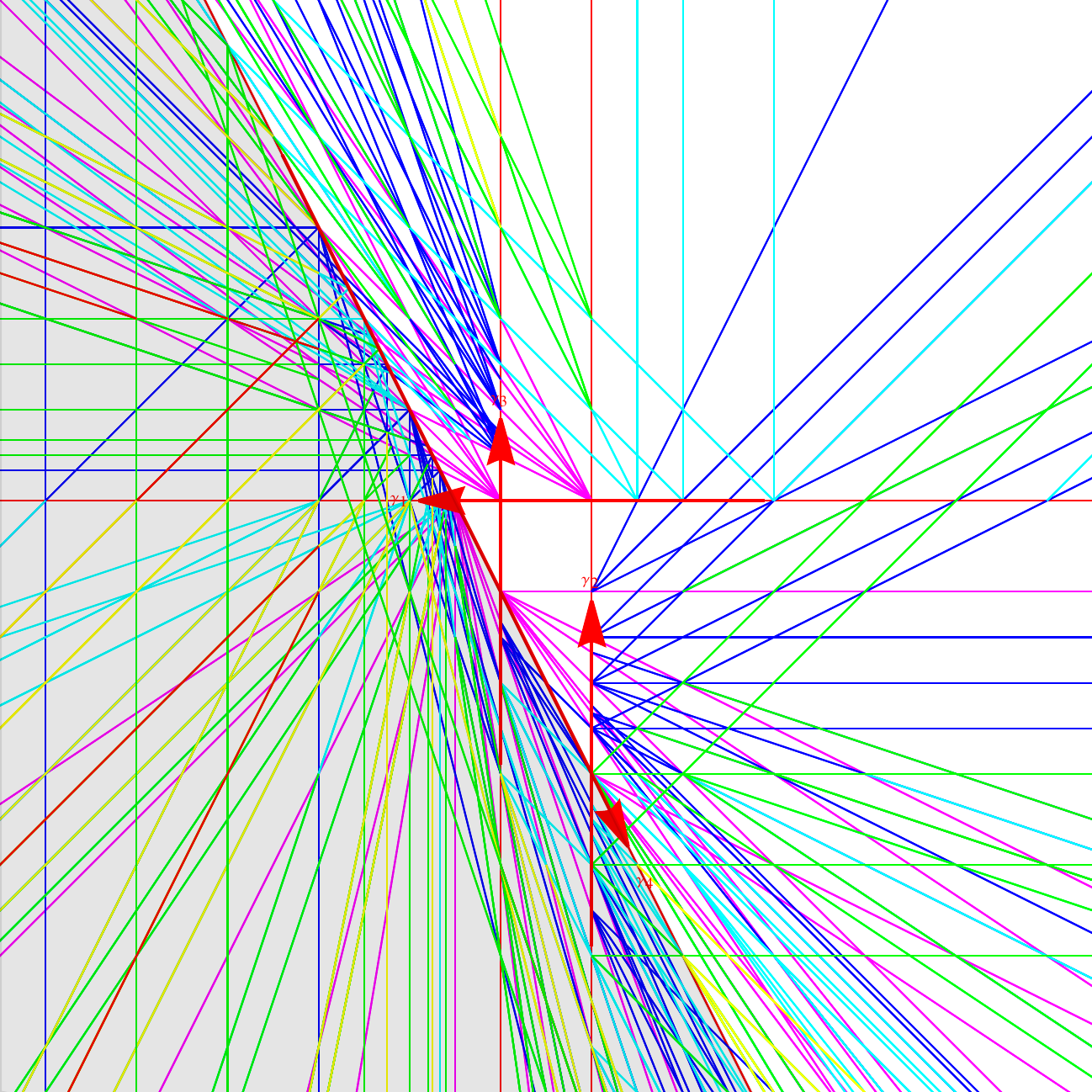}
    \caption{Scattering diagram  $\cD^I_\mu$  for the phase I quiver at $\mu=1/2$, keeping rays with height $\sum_i N'_i\leq 8$, with $\max(n_1,n_2)\leq 4$ at each intersection.  Up to truncation effects, it 
    coincides with the 
    scattering diagram for the orbifold quiver in Fig.~\ref{QuiverScatt} when $\zeta'_4<0$, i.e.\ outside the shaded region.
      \label{QuiverScattI}}
\end{figure}

In order to rule out initial rays entering the large volume slice $\Lambda$ through the intervals 
in between the points 
$s=k\pm \frac{m}{2}$ on the boundary, we follow the same strategy as in \cite{Bousseau:2019ift} and use the isomorphism $D^b\Coh(X)\sim D^b \Rep(Q,W)$ with the derived category of representations of the `orbifold' quiver (see Fig.~\ref{fig_orbquiver}) associated to the Ext-exceptional collection generated by the objects 
\be
\label{extcoll}
E_1 = \cO(0,0), \quad E_2=\cO(-1,0)[1], \quad 
E_3=\cO(1,-1)[1], \quad E_4=\cO(0,-1)[2]
\ee
Physically, the quiver with potential $(Q,W)$ describes the fractional branes on the $\IZ_2$ orbifold of the resolved conifold, which arises in a particular degeneration of local $\IF_0$. 
In fact, the region of validity\footnote{Namely, the region in $\Stab \cC$ where the objects $E_i$ are stable
simultaneously and have central charges in a common half-plane, such that the heart $\sigma$ is related by 
a tilt to the quiver heart. \label{fooqvalid}}
 of the orbifold quiver itself does not overlap with the large volume slice $\Lambda$, but suitable mutations of $(Q,W)$, including the phase I quiver shown in  Fig.~\ref{fig_quiverI}, do. 
By the same reasoning as in \cite[\S 5.2]{Bousseau:2022snm}, one can show that the
orbifold quiver satisfies the Attractor Conjecture \cite{Beaujard:2020sgs,Mozgovoy:2020has},
namely that the only non-vanishing attractor invariants are those associated to the simple dimension vectors $\gamma_i=\ch E_i$ and to D0-branes\footnote{Since the dimension vector $\delta=(1,1,1,1)$ for D0-branes (or skyscraper sheaves) is in the kernel of the Dirac pairing, the associated rays decouple
 from the scattering diagram and the value of $\Omstar(k\delta)$ does not affect other indices.
 We thank P. Descombes for confirming that the proof of the Attractor Conjecture in
 \cite[\S 5.2]{Bousseau:2022snm} carries over to this model. }
\be
\label{Omstar}
\Omstar(\gamma_i)=1, \quad \Omstar(k\delta) = -y^3-2y-1/y
\ee
with $\Omstar(\gamma)=0$ otherwise. Using these initial rays, it is easy to
construct the scattering diagram $\cD^{o}_m$ for the orbifold quiver
(see Fig.~\ref{QuiverScatt}), as well as for the related type I quiver
(see Fig.~\ref{QuiverScattI}). 
 We note that the BPS spectrum of 
local $\IF_0$ in special chambers of the quiver region has been studied previously in \cite{Longhi:2021qvz,DelMonte:2021ytz,DelMonte:2022kxh} and proven rigorously in \cite{Xiong:2025isclocal,Bridgeland:2024iscCY3}.

\begin{figure}[t]
    \centering
\includegraphics[height=7cm]{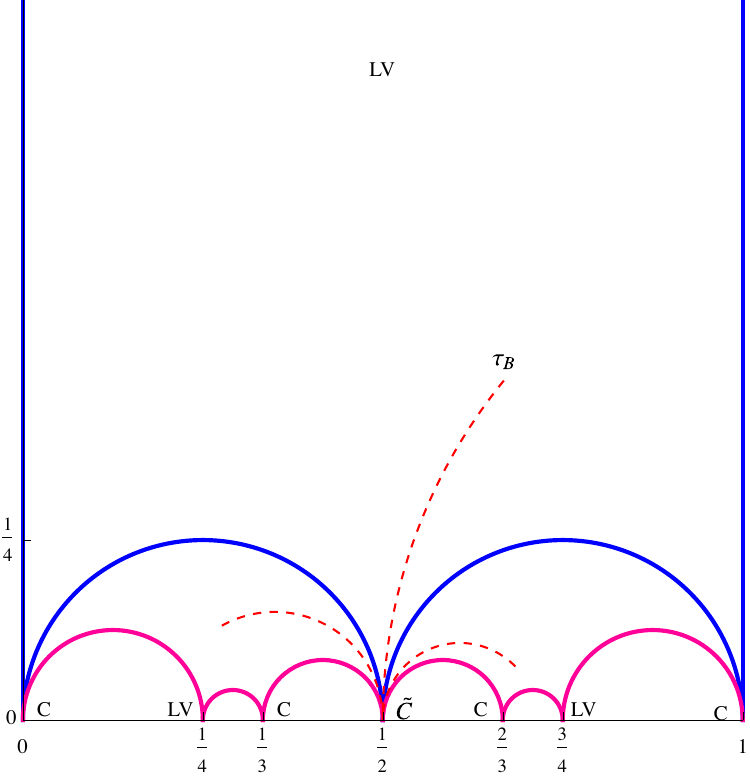}
    \caption{Fundamental domain $\cF$ (in blue) for the action of $\Gamma_0(4)$ on the upper half-plane, and its images under $\tau\mapsto \frac{\tau-1}{4\tau-3}$ and $\tau\mapsto \frac{3\tau-1}{4\tau-1}$. 
    The cusps at $\tau=(\I\infty,0,\frac12)$ correspond 
    to the large volume limit $z=0$, 
    conifold point $z=\tfrac{1}{4(1+\sqrt{\lambda})^2}$ 
    and  dual conifold point $z=\tfrac{1}{4(1-\sqrt{\lambda})^2}$, 
    while the branch point $\tau_B$ corresponds to $z=\infty$. The dotted lines show the cuts joining $\tau_B$ and its images to $\tau=\frac12$ along geodesic circles. 
        \label{figG04}}
\end{figure}

\medskip

In order to study the BPS spectrum along the $\Pi$-stability slice, it is convenient to trade the coordinate $T$ for the modulus $\tau$
of the genus one mirror curve. From the point of view of the five-dimensional 
$SU(2)$ gauge theory compactified on a circle, $\tau$ parametrizes the Coulomb branch 
while $m$ keeps track of the 5D gauge coupling, in units of the circle radius. 
The study of the mirror curve in Appendix \ref{sec_periods}
shows that the Coulomb branch is a double cover of $\IH/\Gamma_0(4)$, ramified over a point
$\tau_B$ determined in terms of $m$ by the 
condition\footnote{An equivalent modular parametrization was proposed in \cite{Kim:2025pidkdi},
see \eqref{quarticcurve} for the dictionary.}
\be
J_4(\tau_B)=-8\cos\pi m, \quad J_4(\tau) := 8 + \left(\frac{\eta(\tau)}{\eta(4\tau)}\right)^8
\ee
where 
$J_4(\tau)$ is the Hauptmodul 
mapping $(\I\infty,0,\frac12)$ to $(\infty,8,-8)$ (we choose $\tau_B$ such that it lies in the fundamental domain $\cF$ shown in Fig.~\ref{figG04}). The point $\tau=\tau_B$ corresponds to the limit where
$K_{\IF_0}$ degenerates into a $\IZ_2$ orbifold of the resolved conifold.  For $m$ real, both $\tau_B$ and its $\Gamma_0(4)$ image $\tilde\tau_B=\frac{\tau_B-1}{4\tau_B-3}=1-\overline{\tau_B}$ lie on the half-circles at the boundary of the fundamental domain. At the special points  $m\in \IZ$ (or $\lambda:=e^{2\pi\I m}=1$, which much of the literature on local $\IF_0$ focuses on), the ramification point $\tau_B$ coalesces with the conifold point $\tau=\frac12$.

\medskip

\begin{figure}[t]
    \centering
\includegraphics[height=7cm]{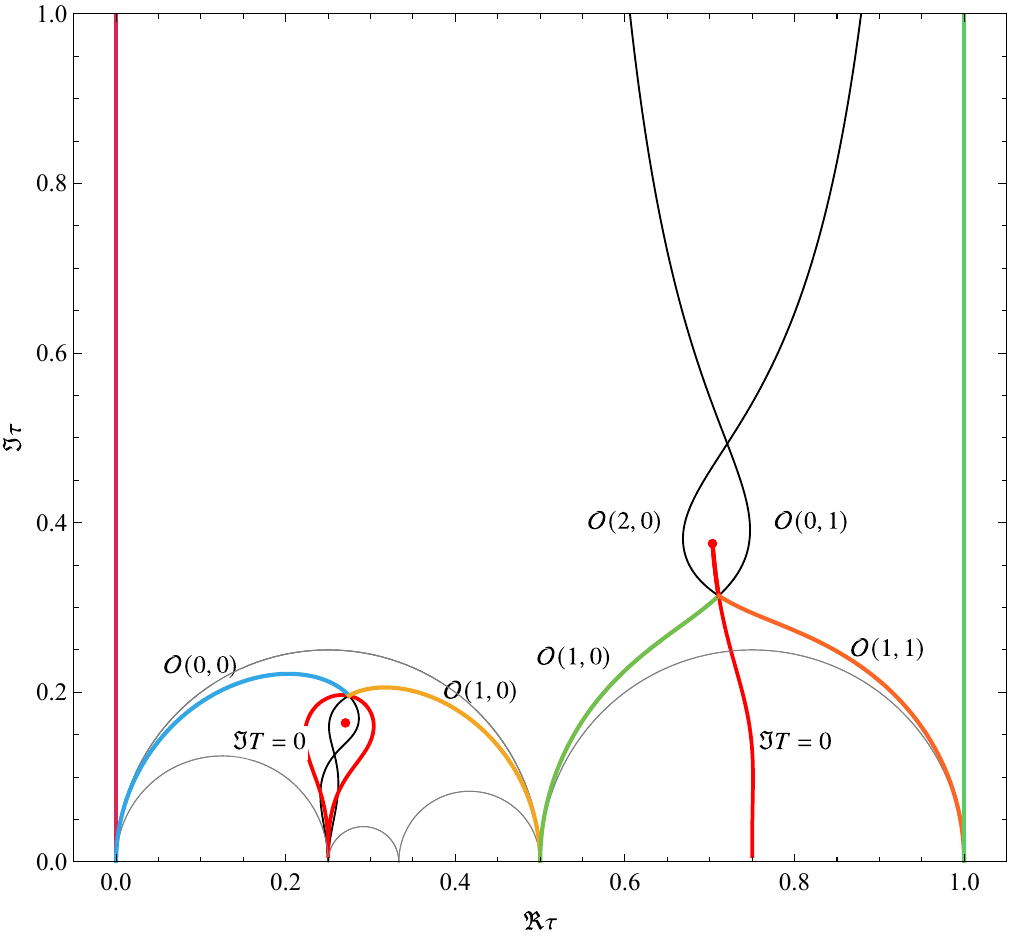}
    \caption{(Putative) boundary of the space of geometric stability conditions, for $m=0.4+0.3\I$. 
    The red solid line is the locus 
    $\Im T=0$. Along the blue, yellow, green and orange lines, the central charge associated to 
    $\cO(0,0)$, $\cO(1,0), \cO(0,1)$ (or rather  its image $\cO(1,0)$ through the cut) and $\cO(1,1)$ become real and negative, destabilizing skyscraper sheaves. The green and orange lines
 intersect  at a point $\tau_b$ where $\Im T=\Im T_D=0$, and similarly  
the blue and yellow lines intersect  at $\tilde\tau_b$ where  where $\Im T=\Im T_D=0$. 
Along the segment of the locus $\Im T=0$ joining $\tau_b$ to $\tau_B$,  skyscraper sheaves are instead destabilized by D2-branes. The fundamental domain $\widehat{\cF}$ is defined as the 
region above the blue, yellow, green, orange lines with the red segment from $\tau_b$ to $\tau_B$ excised.  \label{fig_geobound}}
\end{figure}

The space of $\Pi_m$-stability conditions $(Z,\cA)$ is defined as the universal cover of K\"ahler moduli space
at fixed $m$, equipped with a central charge function $Z:\IZ^4\to \IC$ determined by periods of a certain one-form with logarithmic singularities on the mirror, and a compatible heart $\cA$ which coincides with the large volume heart $\cA_{T,T+m}$ for $\Im\tau$ large enough.  For $m\in\IZ$, the universal
cover of K\"ahler moduli space is simply the Poincar\'e upper half-plane $\IH$, and the central charge
is given by a contour integral of a weight 3 modular
form under\footnote{More precisely, $C(\tau,0)$ is modular under $\Gamma_1(4)$, as 
$\Gamma_0(4)$ admits no modular forms of odd weight. Note that the quotient 
$\IZ_2=\Gamma_1(4)/\Gamma_0(4)$ acts trivially on $\tau$. }
$\Gamma_0(4)$, similar to \cite{Bousseau:2022snm}, 
\be
\label{Eichlercst0}
\begin{pmatrix} T \\ T_D  \end{pmatrix}(\tau)
= \begin{pmatrix} T(\tau_0)  \\ T_D(\tau_0) \end{pmatrix} 
+  \int_{\tau_0}^{\tau} \begin{pmatrix} 1 \\u \end{pmatrix} \, 
C(u, m)  
\de u
\ee
with $m=0$ and 
\be
C(\tau,0) = \frac{\eta(\tau)^4 \eta(2\tau)^6}{\eta(4\tau)^4} = 1-4\sum_{n\geq 1} \frac{n^2 \chi_4(n) q^n}{1-q^n}  
\ee
where $\chi_4(n)$ is the Dirichlet character modulo 4, equal to $\pm 1$ for $n=\pm 1\mod 4$ and 0 otherwise. 
In the fundamental domain and its translates  $\cF+\IZ$, the heart $\cA_\tau$ is constructed by 
the usual tilting construction, and extended to $\IH$ by the action of auto-equivalences.
For $m\notin \IZ$, the fundamental domain of $\Gamma_0(4)$ can still be unfolded to $\IH$, but at the cost of introducing $\IZ_2$ branch cuts between the $\Gamma_0(4)$ images 
of $\tau_B$ and the corresponding $\Gamma_0(4)$ images
 of $\tau=\frac12$. Since the monodromies around different ramification points do not commute, the physical  slice $\Pi_m$
is now an infinite cover of $\IH$, with sheets labeled by a rank 4 lattice\footnote{More precisely,
the monodromy group $\Gamma$ preserving a generic $m$ is an extension $1\to \IZ^4 \to \Gamma \to \Gamma_0(4) \to 1$. \label{fooGamma}}
 $\IZ^4$, ramified at all $\Gamma_0(4)$ images of $\tau_B$. The  central charge along $\Pi_m$ 
 is still given by contour integral of the form \eqref{Eichlercst0}, but now
\be
\label{defC3}
C(\tau,m):=  \frac{\eta(\tau)^4 \eta(2\tau)^6}{\eta(4\tau)^4} 
\sqrt{\frac{J_4(\tau)+8}{J_4(\tau)+ 8\cos\pi m}  }
\ee
has square root singularities at $\tau_B$ and images thereof.
The integral now depends on the choice of path avoiding the $\IZ_2$-ramification points. 
We refer to the sheet reached by analytic continuation along a path from $\tau_0$ near $\I\infty$ to $\tau$ which does not cross any of the branch cuts\footnote{In order to evaluate 
the periods $T,T_D$, we find it convenient to choose branch cuts along hyperbolic geodesics from $\tau_B$ to $\tau=\frac12$ and $\Gamma_0(4)$ images thereof, which differ slightly from the locus where the argument of the square root in \eqref{defC3}
crosses the negative real axis. In view of the discussion of geometric stability conditions in \S\ref{sec_stabgeo},
another natural choice of branch cut would be the locus $\Im T=0$ in $\cF$ and  $\Gamma_0(4)$ images thereof, but it is less convenient for numerics.}
 as the principal sheet.  In \S\ref{sec_stabgeo} 
 we explain how to construct the heart $\cA_\tau$ by tilting 
 in a fundamental domain $\widehat \cF$ which differs slightly from $\cF$ when $m$ is complex (see Fig.~\ref{fig_geobound}),
 and then extend it to the full slice of $\Pi$-stability conditions by auto-equivalences. \footnote{For the mathematically minded reader, we find evidence that the space of geometric stability conditions (modulo tensoring with line bundles) provides a fundamental domain in a connected component of the space of Bridgeland stability conditions modulo the action of auto-equivalences, similar to 
 the local $\IP^2$ case \cite{Bayer:2009brq}.}  

\begin{figure}[t]
    \centering
\includegraphics[height=7cm,trim=0 20 0 0,clip]{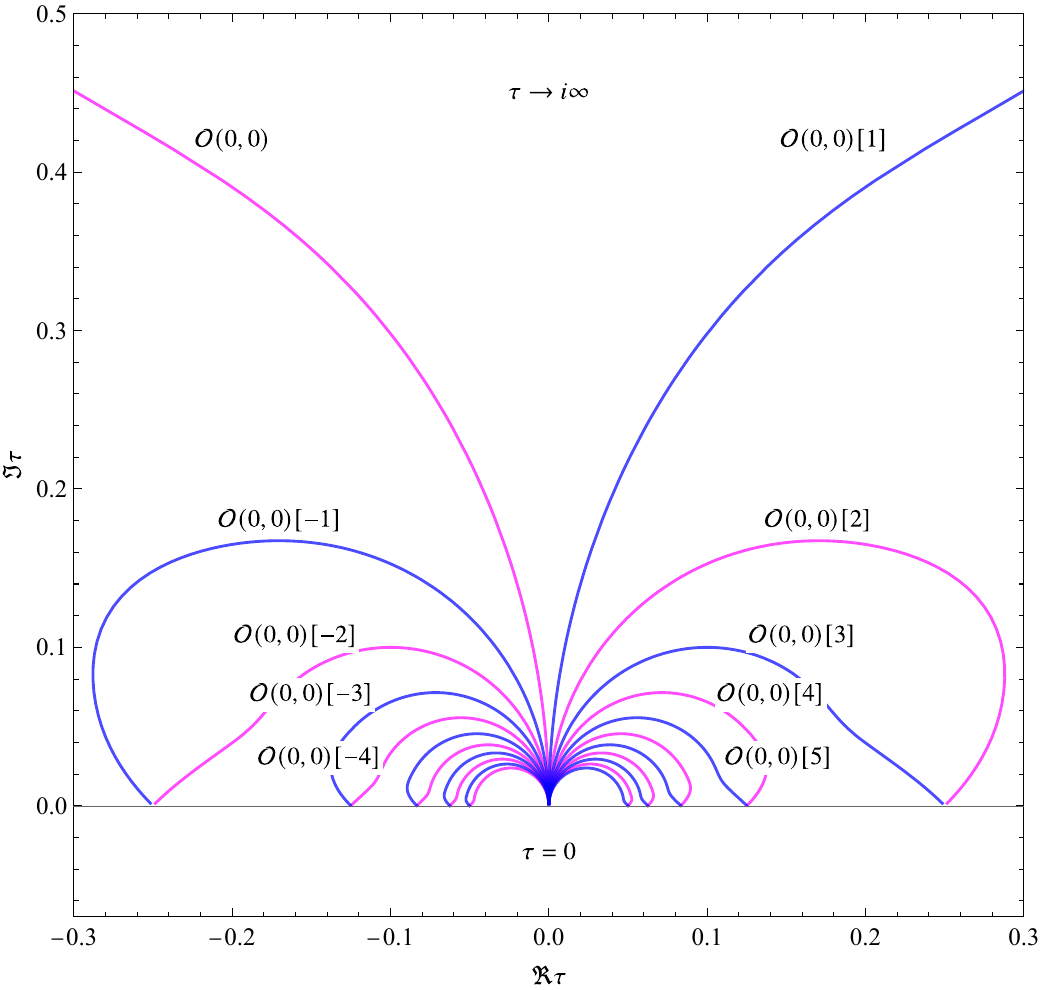}
\includegraphics[height=7cm,trim=0 20 0 0,clip]{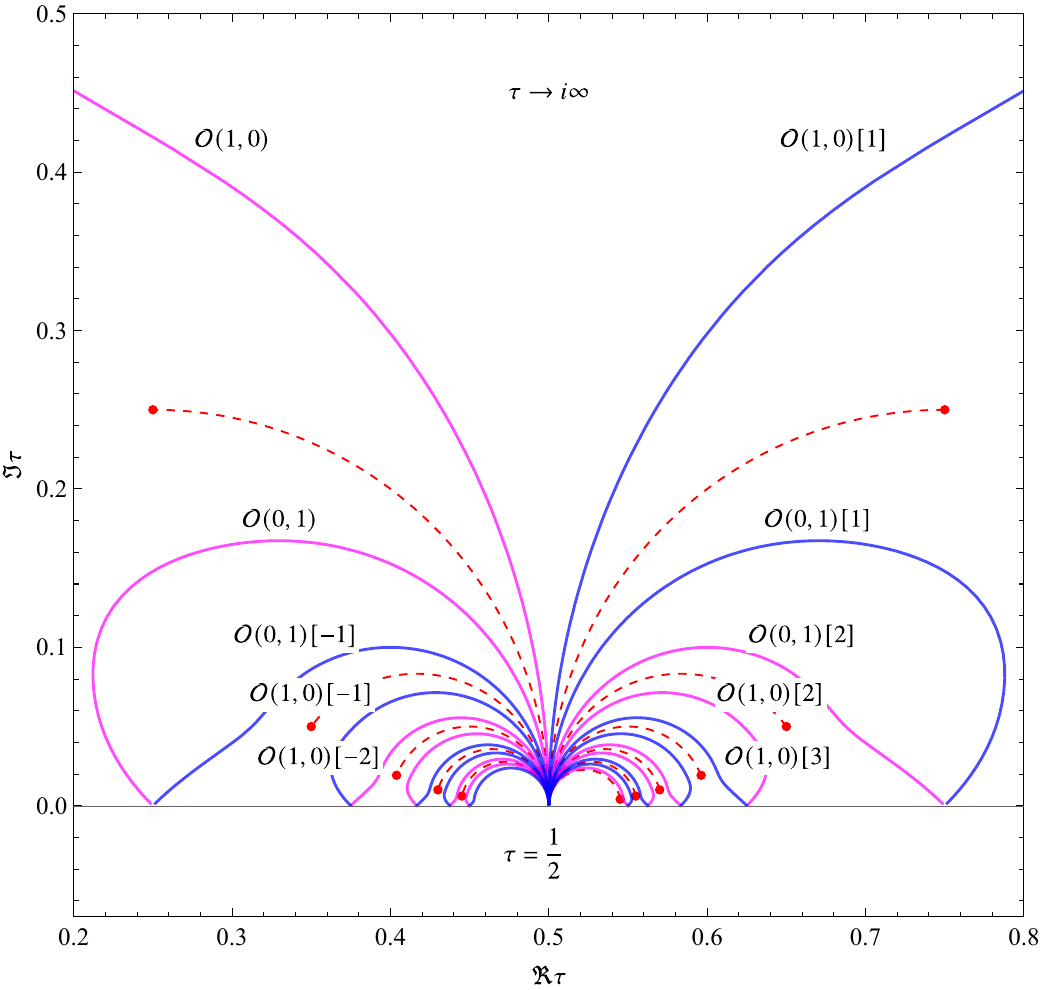}
    \label{HomRays}
    \caption{Initial rays emitted from $\tau=0$ (left) and $\tau=\frac12$ (right), for $m=1/2$ and $\psi=0$. Rays with positive and negative rank $r$ are shown in magenta and blue, respectively. 
    \label{fig_iniPi}}
\end{figure}

\medskip

The scattering diagram $\cD_{\psi,m}^{\Pi}$ along $\Pi$, for fixed values of the mass parameter 
$m$ and phase $\psi$, is then uniquely determined by requiring that it reduces to the large volume scattering diagram $\cD_{\psi,m}^\Lambda$ 
as $\tau\to\I\infty$, to the orbifold scattering diagram $\cD^{o}_m$ as $\tau\to \tau_B$,
and by requiring invariance under the monodromy group $\Gamma$. In particular, it includes
an infinite set of rays emanating from every conifold point $\tau=\frac{p}{q}$ with $(p,q)=1, 
q\neq 0 \mod 4$. The structure of the initial rays near $\tau=0$ and $\tau=\frac12$ on the particular 
sheet reached by analytic continuation from $\tau=\I\infty$ is shown in Fig.~\ref{fig_iniPi}.

\begin{figure}[t]
    \centering
\includegraphics[height=8cm]{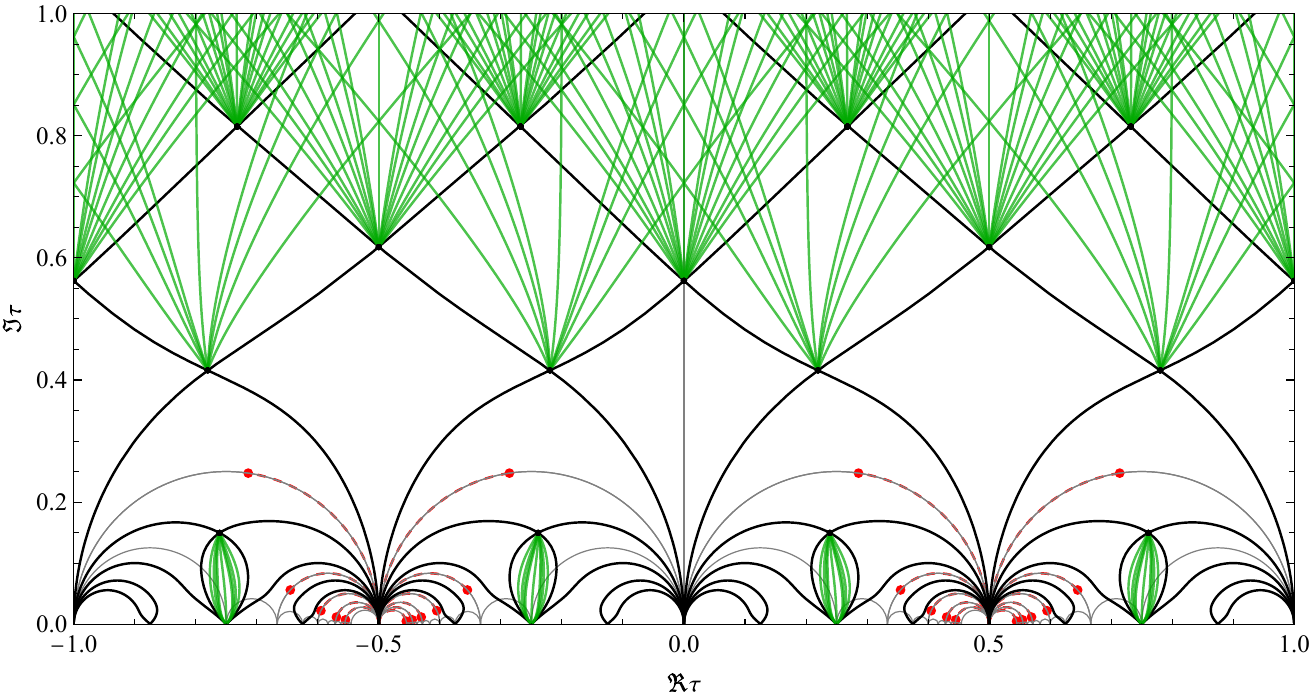}
    \caption{Scattering diagram $\cD^\Pi_{m,\psi}$ along the $\Pi$-stability slice for $m=0.4$, $\psi=0$.
    For clarity, only primary scatterings are shown.  \label{fig_raypsi0} }
\end{figure}

\begin{figure}[t]
    \centering
\includegraphics[height=8cm]{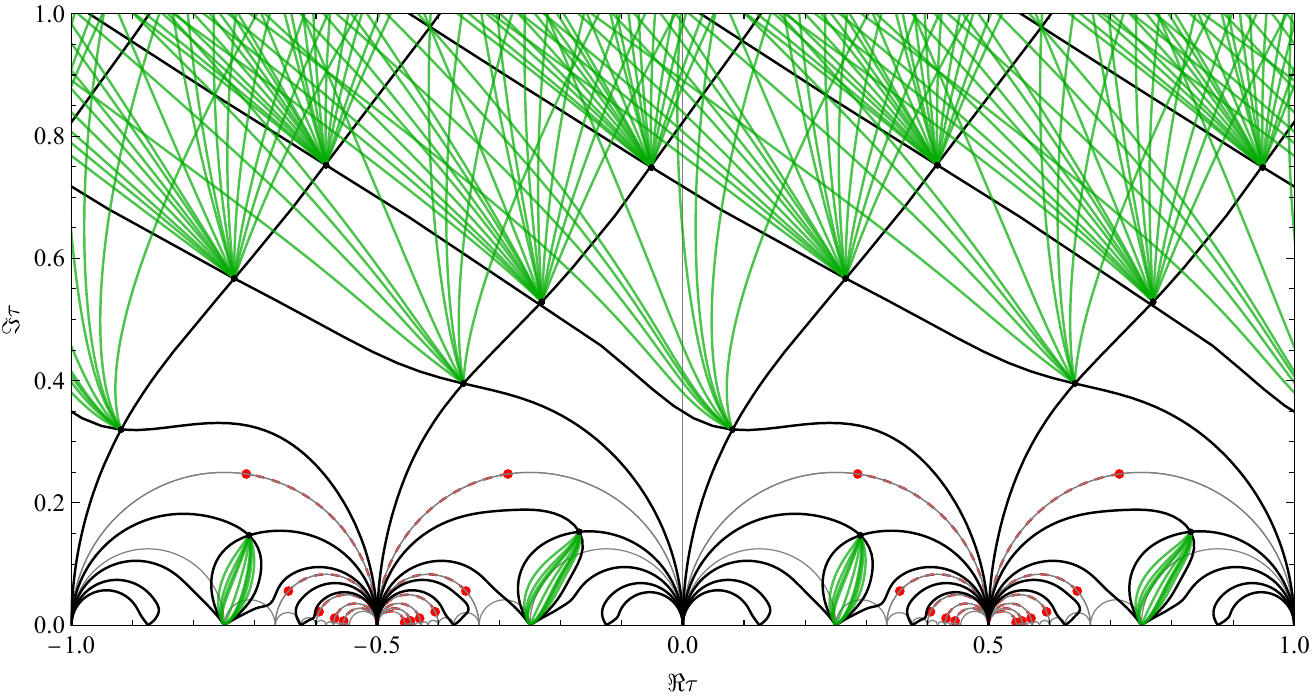}
    \caption{Scattering diagram  $\cD^\Pi_{m,\psi}$ for $m=0.4$, $\psi=0.4$ below
    the first critical value $\psi_{cr}\simeq 0.545$.
\label{fig_raypsi1} }
\end{figure}

\begin{figure}[t]
    \centering
\includegraphics[height=8cm]{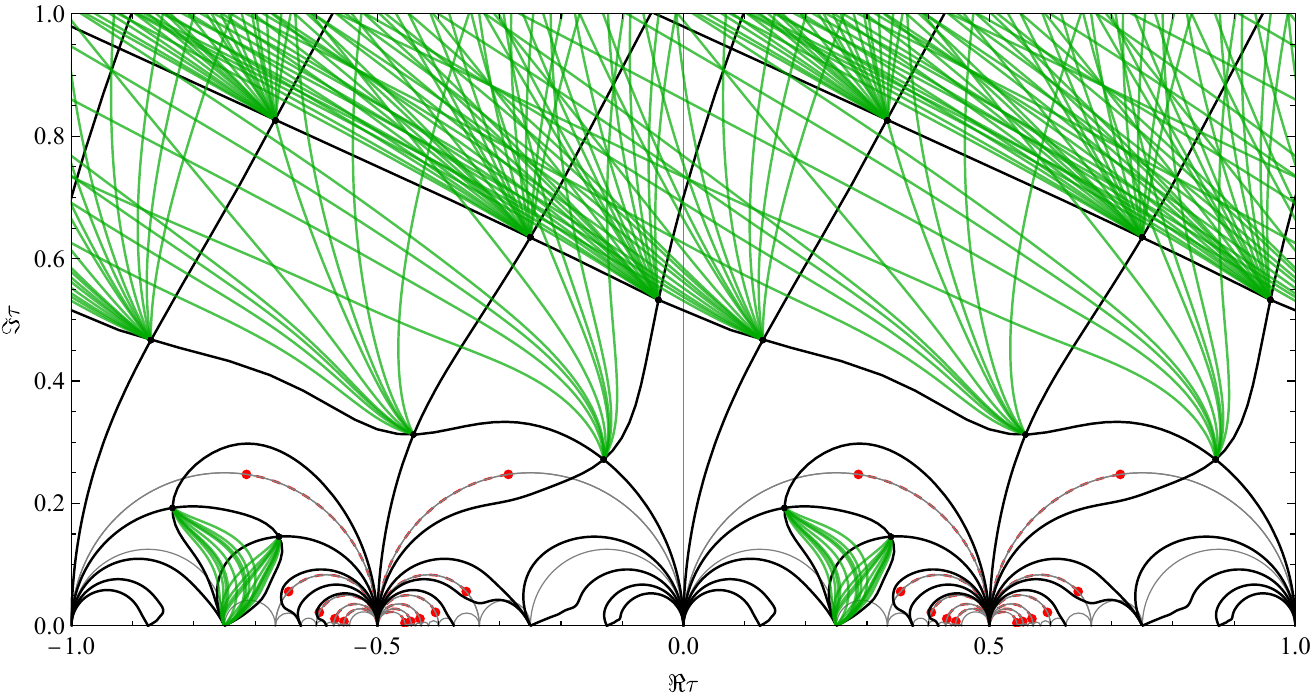}
    \caption{Scattering diagram  $\cD^\Pi_{m,\psi}$for $m=0.4$, $\psi=0.68$ 
    between the first two critical values $\psi_{cr}\simeq 0.545$ and $\tilde\psi_{cr}\simeq 0.832$.
 \label{fig_raypsi2} }
\end{figure}

\begin{figure}[t]
    \centering
\includegraphics[height=8cm]{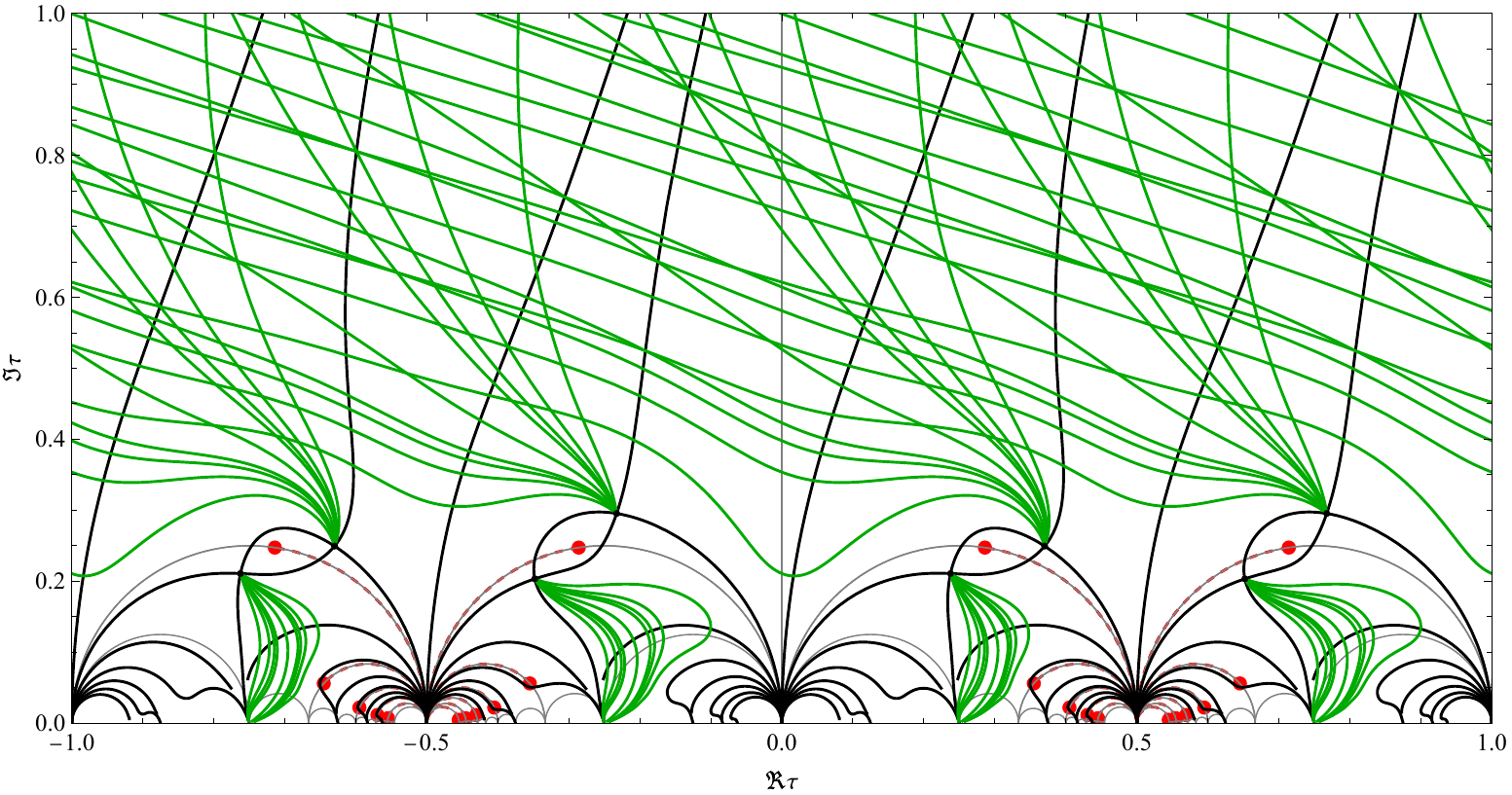}
    \caption{Scattering diagram  $\cD^\Pi_{m,\psi}$ for $m=0.4$, $\psi=0.98$ 
    above $\tilde\psi_{cr}\simeq 0.832$. 
   \label{fig_raypsi3} }
\end{figure}

\begin{figure}[t]
    \centering
\includegraphics[height=8cm]{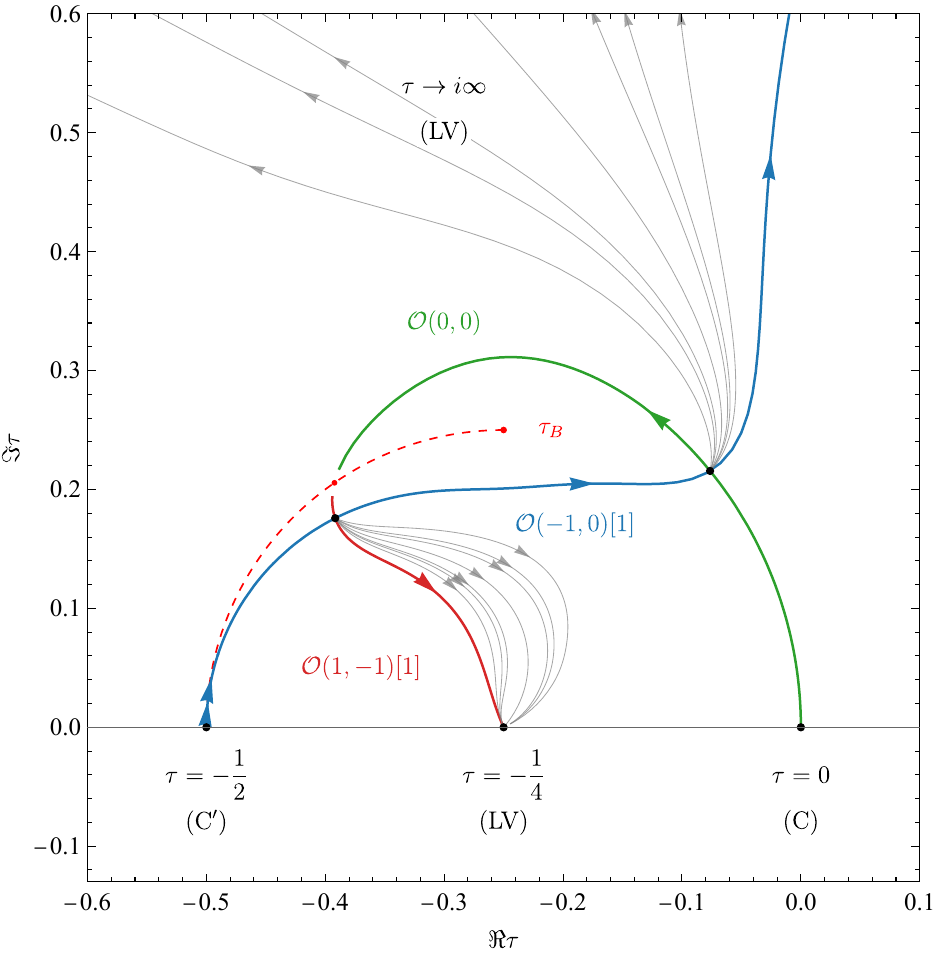}
\includegraphics[height=8cm]{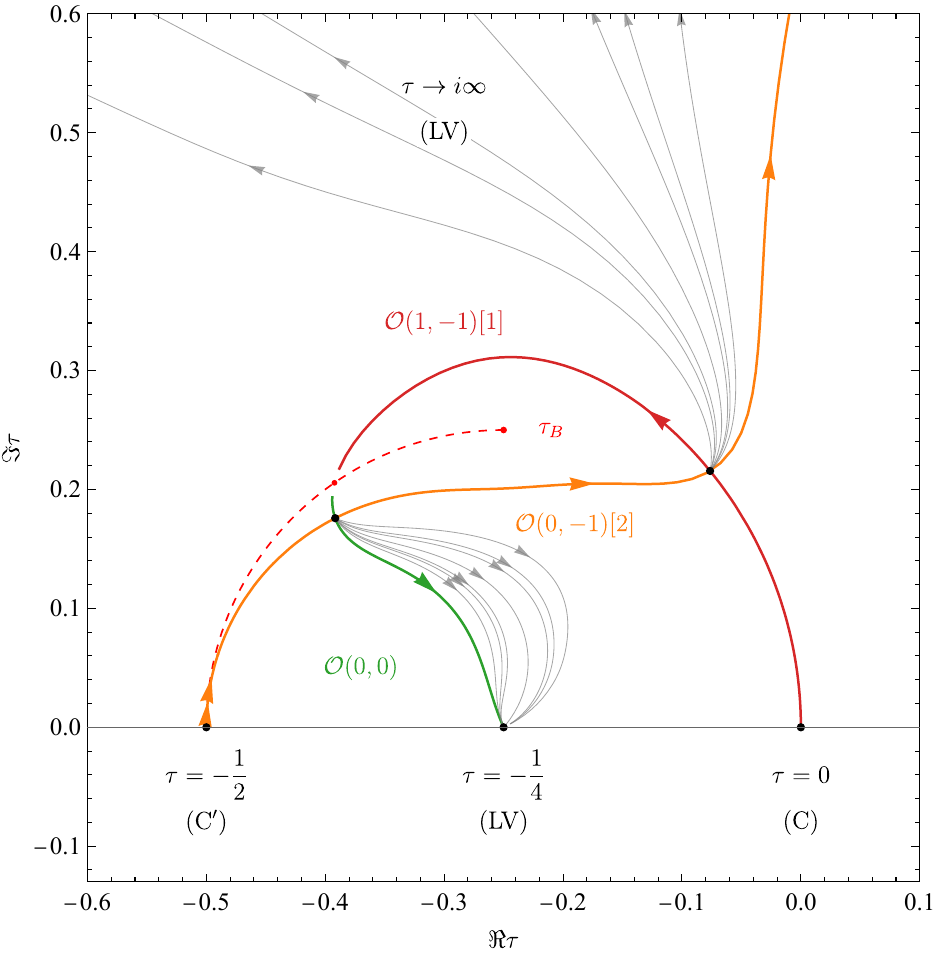}
    \label{OrbSheets}
    \caption{Embedding of the orbifold scattering diagram into the $\Pi$-stability scattering diagram
     around $\tau_B-1=\frac{\I-1}{4}$, for $m=1/2$ and $\psi=0.72$. The initial rays $\cR(\gamma_i)$ for the exceptional collection \eqref{orbcoll1} are shown in green, blue, red and orange.  The
scatterings of $\{\gamma_1,\gamma_2\}$ and $\{\gamma_2,\gamma_3\}$ take
place on one sheet, and those of $\{\gamma_3,\gamma_4\}$ and
$\{\gamma_4,\gamma_1\}$ on another sheet,  glued along the branch cut (shown in red dashed line). \label{fig_orb2sheets}}
\end{figure}

\begin{figure}[t]
    \centering
\includegraphics[height=7cm]{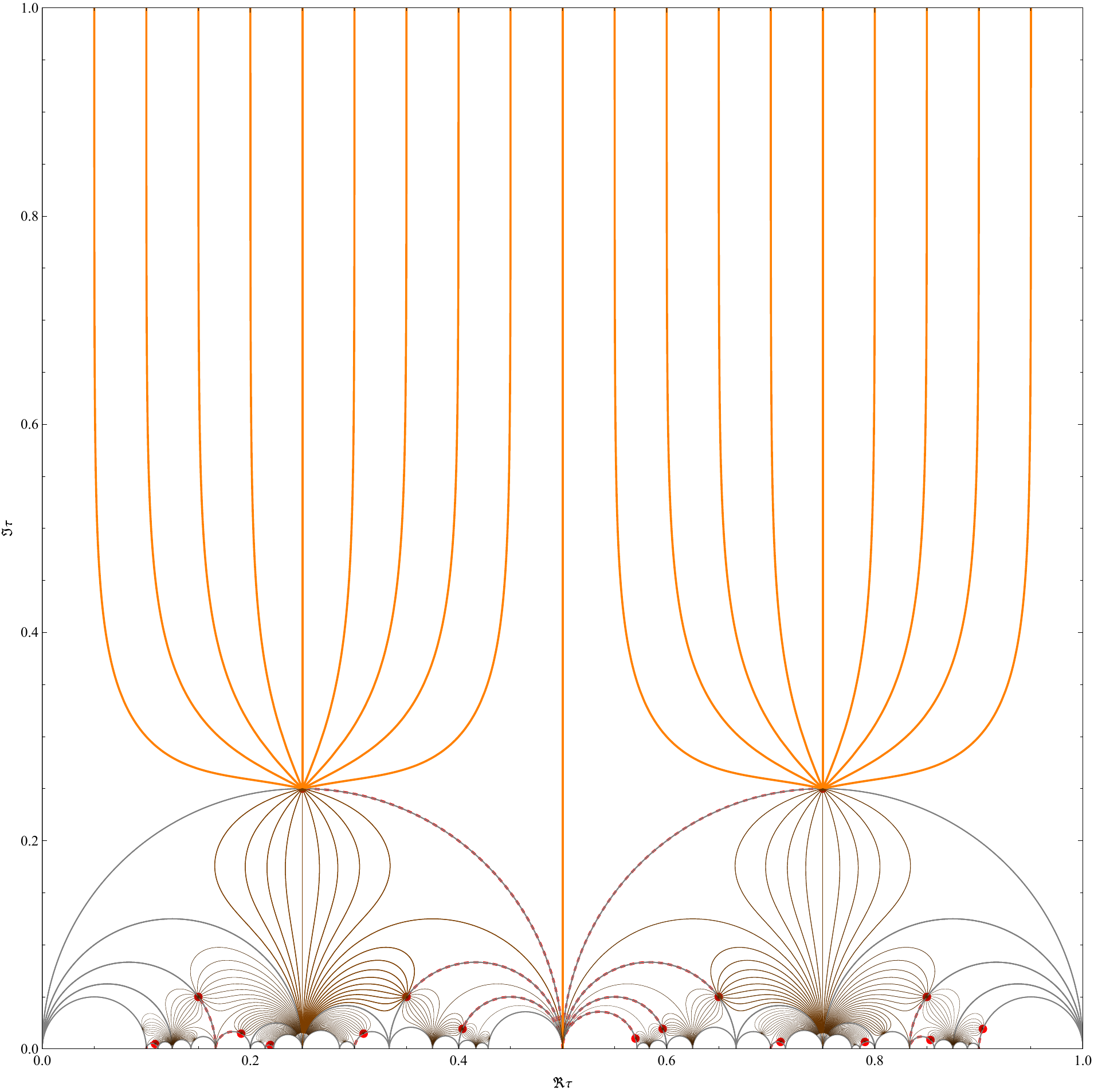}
    \caption{Contour lines of $s=\frac{\Im T_D}{\Im T}$ for $m=1/2$.  \label{fig_scont} }
\end{figure}

For small  phase $|\psi|$,
the  structure  of  $\cD_{\psi,m}^{\Pi}$  along the fundamental domain $\cF$ 
and its translates turns out to be
similar to that along the extended large volume slice, with initial rays $\cO(k,k), \cO(k,k)[1]$ emitted
at $\tau=k$ and $\cO(k+1,k), \cO(k+1,k)[1]$ emitted at $\tau=k+\frac12$ (see Fig.~\ref{fig_raypsi0}). These rays do not cross the
cuts and asymptote to $\tau=\I\infty$ as in the large volume slice. As $\psi$ increases, some of these initial rays curl back
and end up at $\Gamma_0(4)$ images of the large volume point, potentially after crossing cuts, while suitable homological shifts of the original rays manage to escape to $\I\infty$ (see Figs.~\ref{fig_raypsi1}-\ref{fig_raypsi3}).
 In particular, for $\psi$ bigger than both\footnote{
Here we assume $\psi>0$, the critical values for $\psi<0$ can be found in \eqref{psicritical}.
 For $m=1/2$, the two critical  values coincide, $\psi_{\rm cr}=\tilde\psi_{\rm cr} \simeq 0.68641$.}
\be
\label{psicritical0}
\psi_{\rm cr} = \arctan\left( \frac{\Im \cV(m)}{\Re\cV(m)}\right),  \quad 
\tilde\psi_{\rm cr} = \arctan\left( \frac{\Im[ \tilde\cV(m)+ \I(1-m)]}{\Re[\tilde\cV(m)+\I(1-m)]}\right),
\ee
where $\cV(m)$ and $\tilde\cV(m)$  are the `quantum volumes' defined in \eqref{defVm} and \eqref{defVtm}, 
the rays $\cR(\gamma_i)$ 
associated to the simple objects in the Ext-exceptional collection \eqref{extcoll} (tensored by $\cO(1,1)$)
circle around the ramification point $\tau_B$, intersecting exactly as predicted by the orbifold 
scattering diagram, before exiting to the large volume points $\tau=\I\infty$ or $\tau=1/4$.
It is worth noting that the rays $\gamma_i$ and $\gamma_{i+2}$ live on different sheets, although they project to the same ray on the $\tau$-plane (see Fig.~\ref{fig_orb2sheets}). For $\psi=\frac{\pi}{2}$ and real $m$, the scattering diagram simplifies drastically since geometric rays coincide with contours of $s=\Im T_D/\Im T$,
and intersect only  at $\tau_B$ and its $\Gamma_0(4)$ images, see Fig.~\ref{fig_scont}.
As in \cite{Bousseau:2022snm}, this explains why the quiver index agrees with the Gieseker index for
normalized torsion-free sheaves in canonical polarization, as observed in \cite{Beaujard:2020sgs}.\footnote{ For $m$ complex, this simple argument no longer works since  the polarization 
$\eta=1+\frac{\Im m}{\Im T}$ is now $\tau$-dependent, and objects with the same slope $s=\frac{d_1+\eta d_2}{2r}$ may have non-trivial Dirac product. } 
When $m$ takes integer values, the structure of the diagram on the one hand simplifies due to the absence of branch cuts, but on the other hand is complicated by coalescence of various intersection points, due to initial rays related by fiber/base duality, see Fig. \ref{fig_raypsim0}.

\medskip

In general, we conjecture 
 that the scattering diagram in $\cF+\IZ$ (translates of the fundamental domain) arises from the scattering of rays associated to orbifold quivers with lone rays emanating from $\tau\in \IZ$ and $\tau\in \IZ+\frac12$.  In other words,
 attractor flow trees (see \S\ref{ssec_safc}) can always be decomposed into a finite set of `shrubs' (i.e.\ trees rooted at elements in the exceptional collection \eqref{extcoll} or one of its translates or images under fiber/base duality, and supported in the corresponding quiver region) and a finite set of `lone branches' emanating from  $\tau\in \IZ$ and $\tau\in \IZ+\frac12$. We sketch a possible strategy for proving this claim, and hence the validity of the Split Attractor Flow Conjecture, at least in some range  $\psi\in\cI_{\rm cvx}$ around $\psi=0$ where we can use convexity arguments to rule out rays entering quiver regions.

\medskip

We conclude this introduction and summary by listing a few open questions for future 
research \cite{lpr-in-progress}. 
First, we have mostly focused on the scattering diagram at fixed mass parameter $m$, which at large $\Im T$ determines the counting of Gieseker semi-stable sheaves on $\IF_0$ with canonical polarization $H\propto c_1(S)$. It would be of interest to understand the scattering diagram at $m=M_1+M_2 T$ with fixed $(M_1,M_2)$, as it should determine the counting of $\eta$-Gieseker semi-stable sheaves with generic polarization. Second, it would be interesting to extend our study 
to higher del Pezzo surfaces, at least along the large volume slice, and see whether the
coincidence between the large volume scattering diagram and the unfolding of the spanning polytope
of the log CY surface, noted  in footnote \ref{fooGW}, 
continues to hold. Third, it would be useful to clarify the relation between attractor flow trees and exponential spectral networks, introduced in \cite{Eager:2016yxd} and applied to the case of local $\IF_0$ in \cite{Banerjee:2020moh}. Finally, it would be of great interest to determine the spectrum of framed BPS states in these geometries. Indeed, as the phase $\psi$ of the framing defect is varied,
framed BPS indices jump precisely along the rays of the  scattering diagram $\cD_{\psi,m}$
\cite{Gaiotto:2010be,Andriyash:2010qv}, and one expects that the generating series of 
framed BPS indices should interpolate between the generating series of molten crystals near the orbifold point \cite{Okounkov:2003sp,Mozgovoy:2008fd,Mozgovoy:2020has} 
and the topological string amplitude in the large volume limit, generalizing the picture
of \cite{Jafferis:2008uf} to the case of CY threefolds with compact divisors.


\medskip

The rest of this article is organized as follows. In \S\ref{sec_gen} we recall relevant properties of bundles and coherent sheaves on $\IF_0$, collect some known results for
the Gieseker index of rank 0 sheaves, and their relation to Gopakumar-Vafa invariants,
and outline the construction of Bridgeland stability conditions along the slice of $\Pi$-stability conditions. 
In \S\ref{sec_quiv}, we introduce two useful exceptional collections and their respective quivers, related by mutation, and construct a two-dimensional slice in their respective scattering diagrams. 
In \S\ref{sec_LV}, we construct the scattering diagram in the large volume slice with fixed 
 parameter $m$, identifying the initial rays, and use it to reproduce some known Gieseker indices
for coherent sheaves with low rank. 
In \S\ref{sec_Pi}, we combine the orbifold and large volume scattering diagrams with invariance under 
$\Gamma_0(4)$, and construct the scattering diagram in the slice of $\Pi$ stability conditions, for fixed real mass parameter $m$, and discuss phase transitions as the phase $\psi$ is varied.
We sketch the proof of the SAFC, and illustrate the complexity of the scattering diagram by 
 identifing the scattering sequences contributing to the index of the structure sheaves $\cO_S:=\cO(0,0)$ and $\cO_C$ for varying values of $\psi$. In Appendix \S\ref{sec_periods} we establish the contour integral representation  \eqref{Eichlercst0} 
for the central charge, and use it to obtain expansions near cusp and 
ramification points in the fundamental domain. A Mathematica package {\tt F0Scattering.m} 
and accompanying worksheets allowing to reproduce our computations,  in particular evaluating the central charge on the principal sheet over  any point in the Poincar\'e upper half-plane, is provided in a publicly available repository.
\footnote{\url{https://github.com/bpioline/F0Scattering} \label{foogit}}

\medskip

{\noindent \bf Acknowledgements:} We are grateful to Pierrick Bousseau, Andrea Brini, Cyril Closset, Pierre Descombes, Horia Magureanu, Jack Huizenga, Heeyeon Kim, Jan Manschot and Thorsten Schimannek for useful discussions or correspondence related to this project. This research is supported by the Agence Nationale de la Recherche under contract number ANR-21-CE31-0021. {\it For the purpose of Open Access, a CC-BY public copyright licence has been applied by the authors to the present document and will be applied to all subsequent versions up to the Author Accepted Manuscript arising from this submission.}

\section{Generalities \label{sec_gen}}

\subsection{Coherent sheaves on \texorpdfstring{$K_{\IF_0}$}{K(F0)}}

The Hirzebruch surface $S=\IF_0$ is biholomorphic to $\IP^1\times \IP^1$. We denote by $(C,F)$ the
 classes in $H_2(S,\IZ)$ corresponding to the two $\IP^1$ factors, such that $C^2=F^2=0$ and $C.F=1$. 
 The effective cone and nef cone are both spanned by $C$ and $F$. 
Every ample divisor is an integer multiple of $H = C+ \eta F$ for some rational $\eta>0$, which we also call the polarization.
 For any coherent sheaf $E$ on $S$, we denote  the Chern character by $\gamma(E)=[r(E),d_1(E),d_2(E),\ch_2(E)]\in \IZ^4$, with $d_1(E)=C.c_1(E)$, $d_2(E)=F.c_1(E)$, and $\ch_2(E) := \int_S \ch(E)$. 
When the rank $r(E)$ is non-zero, we define the slope and  Bogomolov discriminant by
\be
\label{slopedel}
\mu_\eta(\gamma)= \frac{d_1+ \eta d_2}{2r}, 
\quad 
\Delta(\gamma)=\frac{d_1 d_2}{r^2} -\frac{\ch_2}{r}
\ee
and set $\mu_\eta(\gamma)=+\infty$ if $r=0$. 
Since the canonical class is
$K_S=-c_1(S)=-2C-2F$, the canonical slope is obtained by setting $\eta=1$. 
The Euler form is given by the Grothendieck-Riemann-Roch theorem, 
\be
\begin{split}
\chi(E,E') & = \dim \Hom(E,E') - \dim \Ext^1(E,E') + \dim \Ext^2(E,E') \\
& =
 r r'+ r (d_1'+d_2') - r' (d_1+d_2)
+ r \ch_2' + r'\ch_2 - (d_1 d_2'+d_1' d_2)
\end{split}
\ee
We define the antisymmetrized Euler form (or Dirac product) as
\be
\label{antiEuler}
\langle \gamma,\gamma'\rangle = 2r (d_1'+d_2') - 2r' (d_1+d_2)
\ee

\medskip
When the moduli space $\cM_\eta(\gamma)$ 
of Gieseker-semistable sheaves\footnote{Recall that a torsion-free coherent sheaf $E$ is Gieseker-semistable
if $\mu_\eta(F)\leq \mu_\eta(E)$ for every subsheaf $F\subset E$, and if $\frac{\ch_2(F)}{r(F)} +\frac{c_1(S).c_1(F)}{2 r(F)} \leq \frac{\ch_2(E)}{r(E)}+\frac{c_1(S).c_1(E)}{2 r(E)} $ whenever 
$\mu_\eta(F)= \mu_\eta(E)$, see e.g.~\cite[\S 2]{Manschot:2011ym}. \label{fooGieseker}
} with 
Chern character $\gamma=[r,d_1,d_2,\ch_2]$ is non empty, then it 
is an irreducible projective variety of complex dimension
\be
\label{dCgam}
d_\IC(\gamma) = 1-\chi(E,E) = 2 ( d_1 d_2 - r \ch_2) - r^2 + 1  
\ee
which is smooth whenever $\gamma$ is primitive, i.e.\ $\gcd(r,d_1,d_2,\ch_2)=1$.
For $r>0$, the classical Bogomolov identity implies that $\cM_\eta(\gamma)$ is empty unless
$\Delta(\gamma)\geq 0$. A sharp lower bound analogous to the Dr\'ezet-Le Potier condition 
in the projective plane case \cite{drezet1985fibres} was established
in \cite{coskun2021existence} using exceptional bundles, which we summarize below. 

\medskip

\begin{figure}[t]
    \centering
\includegraphics[height=8cm]{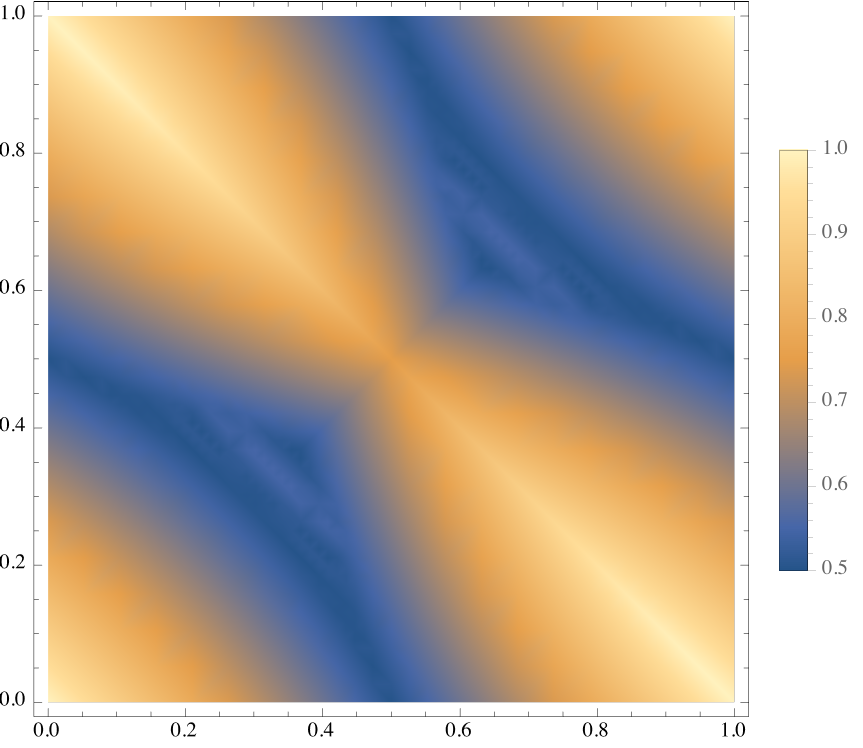}
    \caption{Lower bound $\delta_{\rm DLP}(\eta,a,b)$ on the discriminant of non-exceptional H-stable sheaves for $\eta=1$, obtained by retaining exceptional sheaves with rank up to 19. Figure
    adapted from  \cite[Fig. 5]{coskun2021existence}.
     \label{fig_DLPF0}}
\end{figure}

Exceptional bundles are defined by the conditions $\Hom(E,E)=\IC$
and $\Ext^{i>0}(E,E)=0$. 
Since $d_\IC(\gamma)=0$, their discriminant is
$\Delta(E)=\frac12(1-\frac{1}{r^2(E)})$ by \eqref{dCgam} hence $r(E)$ is
necessarily odd \cite[Lemma 6.7]{coskun2021existence}. 
The discriminant vanishes in the rank 1 case $E=\cO(d_1,d_2)$, and is otherwise 
at least $\frac49$. 
The first few exceptional bundles with rank $\leq 19$ have (up to fiber/base duality and tensoring
with line bundles) Chern characters $[2k+1,1,k,-k]$, along with sporadic cases
 $[11,4,4,-4]$, $[17,5,5,-7]$ and $[19,4,7,-8]$  (see \cite[Table 1]{coskun2021existence}). 
All exceptional bundles are stable for the canonical polarization $\eta=1$, but they 
get destabilized when $|\eta-1|$ is large enough, except in the rank 1 case.
The list $\cE_\eta$ of Gieseker-stable exceptional bundles then determines a lower bound
$\Delta(E)\geq \delta_{\rm DLP}(\eta,a,b)$ 
on the discriminant of non-exceptional Gieseker-stable torsion-free sheaves with 
slope $(a,b)=(d_1,d_2)/r$, given by  \cite[Proposition 6.15]{coskun2021existence}\footnote{The parameter $m$ in  \cite{coskun2021existence} corresponds to $\eta$ in our notations.}
\be
\label{deltaDLP}
\begin{split}
\delta_{\rm DLP}(\eta,a,b) :=
{\rm sup}_{F\in \cE_\eta, |2\mu_\eta(F) - a-\eta b|\leq  \eta+1}
\Bigl[ P_\epsilon \Bigl(a-\tfrac{d_1(F)}{r(F)},b-\tfrac{d_2(F)}{r(F)}\Bigr) - \Delta(F) \Bigr]
\end{split}
\ee
where $P_\epsilon(a,b)=(\epsilon a+1)(\epsilon b+1)$ and $\epsilon(F)=\sign[2\mu_\eta(F) - a-\eta b]$.
Note that $\delta_{\rm DLP}(\eta,a,b)$ is invariant under $(a,b)\to(a+1,b),(a,b+1),(-a,-b),(b,a)$ and is at least $1/2$ by \cite[Corollary 8.15]{coskun2021existence}. 
Conversely, $\cM_\eta(\gamma)$ is non-empty if $\gamma$ is the Chern character of an exceptional stable bundle (or an integer multiple thereof) or if $\Delta(\gamma)\geq \delta_{\rm DLP}(\eta,d_1/r,d_2/r)$ \cite[Corollary 7.6]{coskun2021existence}. 
The function $\delta_{\rm DLP}(M,a,b)$ for the canonical polarization $\eta=1$ is plotted in 
 Fig.~\ref{fig_DLPF0}.\footnote{We are grateful to Jack Huizenga for sharing the code used for producing this picture.}

\medskip

For $\gamma$ primitive, we define the refined Gieseker index $\Omega_\eta(\gamma)$ as the Poincar\'e-Laurent polynomial
for the moduli space $\cM_\eta(\gamma)$,
\be
\label{eqn:defGieseker}
\Omega_\eta(\gamma,y) = \sum_{p= 0}^{2d_\IC(\gamma)}  b_p(\cM_\eta(\gamma))\, (-y)^{p-d_\IC(\gamma)}
\ee
For $r\neq 0$ and arbitrary polarization $\eta>0$, the Gieseker indices were determined in \cite{Manschot:2011ym}, using wall-crossing from the degenerate polarization $\eta=0$, where the indices are computable using the Hall algebra of $\IP^1$ \cite{Mozgovoy:2013zqx}. For $r=1$, the 
indices are independent of $\eta$, and determined by the modular generating series 
\be
\begin{split} 
\sum_{n\geq 0} 
\Omega_\eta([1,0,0,-n]) \, q^n = & \frac{\I q^{1/6} (y-1/y)}{\eta(q)\, \theta_1(q,y^2)} 
 =  
1+ ( y^2+2+1/y^2) q + (y^4+3y^2+6 + \dots ) q^2 \\ & + (y^6+3y^4+9y^2+14+\dots ) q^3+\cO(q^4) 
\end{split}
\ee
where $\eta(q)=q^{1/24}\prod_{n\geq 1}(1-q^n), \theta_1(q,y)=\I\sum_{r\in\IZ+\frac12} (-1)^{r-1/2} q^{r^2/2} y^r$ are the usual Dedekind and Jacobi theta series, and  
the dots are determined by invariance under Poincar\'e duality, $y\to 1/y$. Explicit results
for rank $r\leq 4$ can be found in \cite{Manschot:2011ym,Manschot:2016gsx,Beaujard:2020sgs}. 

\medskip

For rank 0, the Gieseker invariants $\Omega_\eta(0,d_1,d_2,\ch_2)$ 
are independent of $\ch_2$~\cite{maulik2023cohomological}, presumably also of the polarization $\eta$, and are related to the refined BPS invariants $N_{d_1,d_2}^{j_L,j_R}$ by
\be
\Omega_{\eta}(0,d_1,d_2,\cdot,y) = \sum_{j_L,j_R\in\frac12 \IZ_{\geq 0}}
(-1)^{2j_L+2j_R}  \chi_{j_L} (y) \chi_{j_R} (y) \, N_{d_1,d_2}^{j_L,j_R}
\ee
where $\chi_j(y)=(y^{2j+1}-y^{-2j-1})/(y-1/y)$. In particular, for $y=1$ they reduce to the 
genus 0 GV invariant $GV_{d_1,d_2}^{(0)}$. 
The  GV invariants for local $\IF_0$ were first computed at genus 0 in \cite{Chiang:1999tz},
and higher genus in \cite{Aganagic:2002qg}. 
The refined BPS invariants $N_{d_1,d_2}^{j_L,j_R}$ were obtained in \cite[p34]{Iqbal:2007ii}, \cite[Table 4]{Huang:2013yta}, and are summarized in the following table of 
multiplets $(j_L,j_R)$ for low degrees $(d_1,d_2)$:
\be\def\tmpstrut{\rule[-6pt]{0pt}{18pt}}
\begin{array}{|c|c|c|c|c|}
\hline 
d_1 \backslash d_2 &  0 & 1 &2  &3 \\ \hline 
0 & 0 & (0,\tfrac12) & 0 & 0 \tmpstrut\\ \hline 
1 & (0,\tfrac12) & (0,\tfrac32)  & (0,\tfrac52)  &  (0,\tfrac72) \tmpstrut \\ \hline 
2 & 0 & (0,\tfrac52)  &  (0,\tfrac72\oplus\tfrac52)\oplus (\tfrac12,4) & 
  \begin{array}{@{}c@{}} (1,\tfrac{11}{2})\oplus (\tfrac12,5\oplus4) \tmpstrut \\ 
\oplus (0,\tfrac92^{\oplus 2}\oplus\tfrac72\oplus\tfrac52) \tmpstrut \end{array}
\\ \hline 
3 & 0 & (0,\tfrac72)  & \begin{array}{@{}c@{}} (1,\tfrac{11}{2})\oplus (\tfrac12,5\oplus4) \tmpstrut \\ 
\oplus (0,\tfrac92^{\oplus 2}\oplus\tfrac72\oplus\tfrac52) \tmpstrut \end{array} & 
\begin{array}{@{}c@{}} 
(2,\tfrac{15}{2})\oplus(\tfrac32,7\oplus6) \tmpstrut \\
\oplus(1,\tfrac{13}{2}^{\oplus 3}\oplus \tfrac{11}{2}^{\oplus 2}\oplus\tfrac92 ) \tmpstrut \\
\oplus (\tfrac12, 7\oplus 6^{\oplus 3}\oplus5^{\oplus 3}\oplus 4^{\oplus 2}\oplus3) \tmpstrut \\
\oplus (0,\tfrac{11}{2}^{\oplus 4}\oplus\tfrac{9}{2}^{\oplus 3}\oplus\tfrac{7}{3}^{\oplus 4}\oplus\tfrac52\oplus\tfrac32) \tmpstrut
\end{array}
\\ \hline 
\end{array}
\ee
leading to the refined D2-D0 indices 
\be
\label{D2D0F0}
\begin{aligned}
\Omega_\eta(0,1,d,\cdot) &=-(y^{2d+2}-y^{-2d-2})/(y-1/y) \stackrel{y\to 1}{\rightarrow}-(2d-2) \\
\Omega_\eta(0,2,2,\cdot) &= -y^9-3 y^7-4 y^5-4 y^3-4   y- \dots
 \stackrel{y\to 1}{\rightarrow}-32 \\
\Omega_\eta(0,2,3,\cdot) &= -y^{13}-3 y^{11}-8 y^9-10 y^7-11 y^5-11 y^3-11 y+\dots
 \stackrel{y\to 1}{\rightarrow}-110 \\
\Omega_\eta(0,3,3,\cdot) &= -y^{19}-3 y^{17}-10 y^{15}-22 y^{13} \\
   &\quad -41 y^{11}-53 y^9-60 y^7-62 y^5-63 y^3-63 y+\dots\ \stackrel{y\to 1}{\rightarrow}-756 
\end{aligned}
\ee
The leading power of $y$ matches the expected dimension $d_\IC(\gamma)=2d_1 d_2+1$. Moreover,
$\Omega_\eta(0,d_1,d_2,\ch_2)$ vanishes if $d_1 d_2<0$, i.e.\ if the curve class lies outside the Mori cone.

\subsection{Derived category of coherent sheaves on \texorpdfstring{$K_{\IF_0}$}{K(F0)}}
The (compactly supported, bounded) 
derived category $\cC=D^b_c {\rm Coh} K_{\IF_0}$ is the category 
of finite length complexes $\cE= \dots \to E_k \to E_{k+1} \dots$ (up to quasi-isomorphisms)
where the component $E_k$
in cohomological degree $k$ is a coherent sheaf on $K_{\IF_0}$ 
supported set-theoretically  on $\IF_0$. The
category $\cC$ is  graded by the lattice $\Gamma:=\IZ^4$
corresponding to the Chern character 
$\gamma(\cE):=[r(\cE),d_1(\cE),d_2(\cE),\ch_2(\cE)]:=
\sum_k (-1)^k \gamma(E_k)$. 
The Euler form on $\cC$ coincides with the antisymmetrized Euler form $\langle -,- \rangle$ in \eqref{antiEuler}.
 
 \medskip
 
The category $\cC$ admits several natural auto-equivalences: 
\begin{itemize}
\item Homological shift $\cE\mapsto \cE[1]$, mapping $\gamma(\cE)\mapsto -\gamma(\cE)$
 \item Tensoring by the line bundle $\cO(k_1,k_2)$, also known as spectral flow, mapping 
\be
\label{tensorch}
[r,d_1,d_2,\ch_2] \mapsto [r, d_1+rk_1,d_2+r k_2, \ch_2+d_2 k_1+d_1 k_2+r k_1 k_2]
\ee
\item `Fiber-base duality' exchanging the two $\IP^1$ factors, and acting on the charge vector as 
\be
\label{FB}
FB: \quad [r,d_1,d_2,\ch_2] \mapsto [r, d_2,d_1,\ch_2]
\ee
\item Seidel-Thomas twist $ST_{S}$ by any spherical object $S$, acting as\footnote{Although the Chern vector $\ch S$ is invariant under \eqref{STtwist}, the object $S$ itself is not invariant, rather it is mapped to its homological shift $S[2]$.}
\be
\label{STtwist}
\ch E\mapsto \ch E - \langle \ch S, \ch E \rangle \ch S 
\ee 
\end{itemize}
In particular, the composition $FB \circ ST_{\cO(0,0)} \circ  (\cO(0,1) \otimes - )$ 
gives an order 4 automorphism, acting as 
\be
\label{Z4aut}
\IZ_4: \quad [r,d_1,d_2,\ch_2] \mapsto  [-2 d_1 - 2 d_2 - r, d_2 + r, d_1, d_1+\ch_2]
\ee
In addition, under derived duality $\cE \mapsto \cE^\vee$, 
the charge vector $\gamma(\cE)$ transforms as 
\be
[r,d_1,d_2,\ch_2] \mapsto [-r,d_1,d_2,-\ch_2] 
\ee

\subsection{Bridgeland stability conditions\label{sec_stabcond}}

Donaldson-Thomas invariants $\Omega_\sigma(\gamma)$ depend on a charge vector $\gamma\in\Gamma=K(\cC)$ and a choice of Bridgeland stability condition $\sigma=(Z,\cA)$ on the derived category $\cC=D^b_c {\rm Coh} X$. The latter consists of a group homomorphism $Z: \Gamma \to \IC$, called the central charge, and an Abelian  subcategory $\cA \subset \cC$, often called the heart, satisfying certain axioms \cite{Bridgeland:2006bzr} (see e.g.~\cite{macrì2019lecturesbridgelandstability} for an introduction of stability conditions,
or the lightning review in \cite[\S 2.4]{Bousseau:2022snm}). Key among those is the fact that for any nonzero element $E\in\cA$, the central charge is either in the strict upper half-plane, $\Im Z(E)>0$, or on the negative real semi-axis, $Z(E)<0$. The space of such stability conditions $\Stab\cC$ forms a complex manifold of dimension equal to the rank of the lattice $\Gamma$. It admits a left-action by auto-equivalences of $\cC$, and a right action by (the universal cover of) $GL(2,\IR)^+$, acting on the central charge $Z$ viewed as a vector in $\IR^2$. Given a stability condition $\sigma$ and a charge vector $\gamma\in\Gamma$, one defines the moduli space (or rather, moduli stack) $\cM_\sigma(\gamma)$ of $\sigma$-semi-stable objects $E\in \cA$ with class $\gamma(E)=\gamma$ (i.e.\ objects $E\in\cA$ such that $\arg Z(F)\leq \arg Z(E)$ for any subobject $F\subset E$). The refined (or motivic) DT invariant $\Omega_\sigma(\gamma,y)$ is an invariant of $\cM_\sigma(\gamma)$, which reduces to the (shifted) Poincar\'e polynomial
\be
\label{defOm}
\Omega_\sigma(\gamma,y) = \sum_{p= 0}^{2d_\IC(\gamma)} \dim H^p(\cM_\sigma(\gamma), \mathbb{Q}) \, (-y)^{p-d_\IC(\gamma)}
\ee
whenever $\cM_\sigma(\gamma)$ is a smooth projective variety of complex dimension $d_\IC(\gamma)$ \cite{ks,joyce2012theory}. The invariant $\Omega_\sigma(\gamma,y)$ is a Laurent polynomial with integer coefficients whenever $\sigma$ is generic (which means that the only subobjects which can saturate the condition $\arg Z(F)\leq \arg Z(E)$ are those such that $\gamma(F)$ divides $\gamma(E)$). The invariant $\Omega_\sigma(\gamma,y)$ is the mathematical counterpart of the BPS index in physics, which counts stable BPS states in type II string theory on $\IR^{3,1}\times X$, with a fugacity $y$ conjugate to angular momentum $J_3$ in $\IR^3$.
Being integer valued, it is a locally constant function of $\sigma$, but may jump across walls of marginal stability $\cW(\gamma,\gamma')$, where the phase of a semistable object $E$ of charge $\gamma$ becomes aligned with the phase of  a subject $F\subset E$ of charge $\gamma'$. It is convenient
to introduce the \textit{rational} invariant $\bOm_\sigma(\gamma, y)$
\be
\label{defOmb}
\bOm_\sigma(\gamma, y) = \sum_{k|\gamma} \frac{y - y^{-1}}{k (y^k - y^{-k})} \Omega_\sigma(\gamma/k, y^k)\ ,
\ee
which has simpler transformation properties under wall-crossing~\cite{ks,joyce2012theory,Manschot:2010qz}. 
The Gieseker indices $\Omega_\eta(\gamma, y)$ defined in \eqref{eqn:defGieseker}, counting $H$-semi-stable coherent sheaves on a surface $S$ for the polarization $H=C+\eta F$, coincide with the DT invariants $\Omega_\sigma(\gamma, y)$ for the local CY threefold $X=K_S$ with K\"ahler class $H$ in a suitable large volume limit. 

\medskip
In the present case $X=K_{\IF_0}$, the  space of Bridgeland stability conditions $\Stab\cC$ has complex dimension 4. Upon fixing 
the $\IC^\times\subset GL(2,\IR)^+$ action, we get a three-dimensional slice with central charge 
\be
\label{Zgen}
 Z(\gamma) = - 2 r T_D + d_1 T_1 + d_2 T_2 -\ch_2  
\ee
parametrized by $T_1,T_2$ and $T_D$ in some suitable domain  $U\subset \IC^3$. 
The parameters $T_1$ and $T_2$ can be viewed as the complexified K\"ahler parameters of
the two factors in $S=\IP^1\times \IP^1$, while $T_D$ measures the size of $S$. In terms of the 
rank one 5D gauge theory engineered by 
M-theory compactified on $K_{S}$, $T_1+T_2$ parametrizes the Coulomb branch while
$T_1-T_2$ corresponds to the 5D gauge coupling (or mass deformation of the superconformal point).
Thus, it is natural to define 
\be
T_1=T, \quad T_2=T+m
\ee 
and introduce the `period vector' (so called because it is determined by periods of the mirror curve along the $\Pi$-stability slice)
$\Pi=(1,m,T,T_D)$ such that 
\be
 Z(\gamma) =  \gamma \Sigma \Pi^t \ ,\quad 
\Sigma=\begin{pmatrix}
 0 & 0 & 0 & -2 \\
 0 & 0 & 1 & 0 \\
 0 & 1 & 1 & 0 \\
 -1 & 0 & 0 & 0 \\
\end{pmatrix} 
\ee
In order for the central charge $Z(\gamma)$ to stay invariant under an autoequivalence acting
as $\gamma\mapsto \gamma g$ on the Chern vector, the period vector should transform as 
$\Pi^t \mapsto M \Pi^t$ with $M=\Sigma^{-1}g^{-1}\Sigma$. For the autoequivalences listed in the previous subsection, the matrices
$M$ are given by 
\be
\label{Monm1m2} 
M_{k_1,k_2}=\begin{pmatrix}
 1 & 0 & 0 & 0 \\
 k_1-k_2 & 1 & 0 & 0 \\
 k_2 & 0 & 1 & 0 \\
 \frac{k_1 k_2}{2} & \frac{k_2}{2} & \frac{k_1+k_2}{2} & 1
\end{pmatrix},
\quad
M_{FB} =
\begin{pmatrix}
 1 & 0 & 0 & 0 \\
 0 & -1 & 0 & 0 \\
 0 & 1 & 1 & 0 \\
 0 & 0 & 0 & 1 \\
\end{pmatrix}
\ee
\be
\label{MConZ4}
M_C=\begin{pmatrix}
 1 & 0 & 0 & 0 \\
 0 & 1 & 0 & 0 \\
 0 & 0 & 1 & -4 \\
 0 & 0 & 0 & 1 \\
\end{pmatrix}
,\quad
M_{\tilde C}=\begin{pmatrix}
 1 & 0 & 0 & 0 \\
 0 & 1 & 0 & 0 \\
 0 & 0 & 3 & -4 \\
 0 & 0 & 1 & -1 \\
\end{pmatrix}
\ ,\quad
M_{\IZ_4} = \begin{pmatrix}
 1 & 0 & 0 & 0 \\
 1 & -1 & 0 & 0 \\
 0 & -1 & -1 & -4 \\
 0 & \frac{1}{2} & \frac{1}{2} & 1 \\
\end{pmatrix}
\ee
where $M_C$ and $M_{\tilde C}$ implement the Seidel-Thomas twist with respect to $\cO(0,0)$
and $\cO(1,0)$, respectively.
For later reference, we note that the only invariant period vector under 
$M_{\IZ_4}$ is $(1,m,T,T_D)=(1,\frac12,-\frac12,\frac18)$, while invariance
under the order 2 autoequivalence 
\be
\label{MBm1}
M_B:=(M_{\IZ_4})^2 =  M_C \cdot M_{\tilde C} \cdot M_{1,1} =
\begin{pmatrix}
 1 & 0 & 0 & 0 \\
 0 & 1 & 0 & 0 \\
 -1 & 0 & -1 & 0 \\
 \frac{1}{2} & -\frac{1}{2} & 0 & -1 \\
\end{pmatrix}
\ee
only requires $(T,T_D)=(-\frac12, \frac{1-m}{4})$, irrespective of the value of $m$.
The matrices which preserve a generic $m$ are all of the form
\be
M= \begin{pmatrix}
 1 & 0 & 0 & 0 \\
 0 & 1 & 0 & 0 \\
 n_1 & n_2  & 1 & 0 \\
 \frac{\ell_1}{2} & \frac{\ell_2}{2} & 0 & 1 \\
\end{pmatrix}
\cdot 
\begin{pmatrix}
 1 & 0 & 0 & 0 \\
 0 & 1 & 0 & 0 \\
0& 0  & d & c \\
0 & 0 & b & a \\
\end{pmatrix}
\ee
where $a,b,c,d,n_i,\ell_i$ integers,  $ad-bc=1, c=0 \mod 4$. In particular, 
the monodromy group $\Gamma$ for fixed $m$ is an extension of $\Gamma_0(4)$ by a rank 4 lattice,
as anticipated in footnote \ref{fooGamma}. When $m$ is integer or half-integer, $\Gamma$ gets
extended by the generator $M_{k_1,k_2} M_{FB}$ with $2m=k_1-k_2$.

\medskip

There is a simple two-dimensional slice inside $U$ which is preserved under tensoring
with $\cO(k_1,k_2)$,
namely the locus $T_D=\frac12 T(T+m)$. This is recognized as (minus half)
 the central charge of a D4-brane 
wrapped on $\IF_0$ in the large volume limit. We thus define the large volume central charge 
by
\be
\label{ZLV}
Z^{\rm LV}_{T,m}(\gamma) = -r T(T+m) + d_1 T + d_2 (T+m) -\ch_2 
\ee
In addition to being invariant
under tensoring with $\cO(k_1,k_2)$, the central charge \eqref{ZLV} is also
invariant under fiber-base duality,
\be
\label{fbduality}
Z^{\rm LV}_{T,m}(r,d_1,d_2,\ch_2) = Z^{\rm LV}_{T+m,-m}(r,d_2,d_1,\ch_2)  
\ee
and transforms into (minus) its complex conjugate under derived duality,
\be
\label{derduality}
Z^{\rm LV}_{-\overline{T},-\overline{m}}(-r,d_1,d_2,-\ch_2) = - \overline{ Z^{\rm LV}_{T,m}(r,d_1,d_2,\ch_2)  }
\ee
We note that this slice does not contain any invariant point under $M_{\IZ_4}$ or $M_B$.

\medskip

Another way to reduce the dimension of the space of Bridgeland stability conditions is to consider
orbits under the action of the (universal cover of the) group $GL(2,\IR)^+$,
acting  by positive-determinant linear transformations of $(\Re Z, \Im Z)$, and corresponding tilting of the heart. 
Using this action,
one may transform the general central charge
\eqref{Zgen} into $Z^{\rm LV}_{T^{\rm LV}_1,T^{\rm LV}_2-T^{\rm LV}_1}$
provided
\be
\frac{2\Im( T_D \overline{T_1})}{\Im T_2} = |T^{\rm LV}_1|^2\ ,\quad
\frac{\Im( T_D + \tfrac12 \overline{T_1} T_2)}{\Im T_2} = \Re T^{\rm LV}_1
\ee
for some $T^{\rm LV}_1$ with $\Im T^{\rm LV}_1>0$,
and the same equation with indices $1$ and $2$ exchanged.
This is possible provided 
\be
\label{condLV}
\Im T_1>0, \quad \Im T_2>0,  \quad 
2 \Im( T_D \overline{T_1}) \, \Im T_2  - \left( \Im( T_D +  \tfrac12 \overline{T_1} T_2)\right)^2 >0 
\ee
Note that this is invariant under the exchange of $T_1$ and $T_2$, thanks to the identity
\be
\Im T_1\, \Im( T_D \overline{T_2}) - \Im T_2\, \Im( T_D \overline{T_1}) = 
\Im T_D\, \Im( T_1 \overline{T_2})
\ee
It will be useful to define  the $GL(2,\IR)^+$-invariant coordinates
\be
\label{defsw}
s:=\frac{\Im T_D}{\Im T}, \quad 
w:=  \frac{\Im (T_D \overline{T})}{\Im T}, \quad
M_1:=-\frac{\Im (m \overline{T})}{\Im T}, \quad
M_2:=\frac{\Im m}{\Im T}
\ee
such that $T_D=-w+s T, m=M_1+M_2 T$,  or alternatively
\be
\label{defsw1}
\eta:=M_2+1, \quad 
s_1:=\frac{s-\frac{M_1}{2}}{\eta}, \quad 
w_1:=\frac{w}{\eta} -\frac12 s_1^2 
\ee
In terms of these coordinates, one has
\be
2 \Im( T_D \overline{T_1}) \, \Im T_2  - \left( \Im( T_D +  \tfrac12 \overline{T_1} T_2)\right)^2  = 
2 \eta^2  (\Im T)^2 w_1 
\ee
so
the inequalities  \eqref{condLV}  can be written as 
\be
\label{lvregion}
\Im T>0, \quad \eta>0, \quad 
w_1>0
\ee
Under spectral flow \eqref{tensorch}, $\Im T$ and $\eta$ stay invariant while 
\be
\label{specflowsw}
(s,w,M_1) \mapsto \left( s+ \tfrac{k_1+\eta k_2}{2}, 
w + k_2 (  s-\tfrac{M_1}{2} ) + \tfrac{\eta k_2^2}{2} , 
 M_1 + k_1 - \eta k_2  \right)
\ee
leaving the inequalities \eqref{lvregion} invariant. 

\medskip

Now, let us discuss the construction of a heart $\cA$ for the general central charge 
\eqref{Zgen}, such that  $(Z,\cA)$ provides an actual Bridgeland stability condition.
As recalled in the beginning of this section, a necessary condition on the heart
$\cA$ is that, for all non-zero objects $E\in \cA$, $\Im Z(E)\geq 0$ 
and $\Re Z(E)<0$ whenever $\Im Z(E)=0$. 
Assuming that $r(E)\neq 0$, the imaginary part of the central charge is expressed 
in terms of the slope
defined in \eqref{slopedel} via
\be
\label{ImZ}
\Im Z(\gamma)=  \left( -2 r s + d_1 + \eta d_2 \right) \Im T = 
-2 r  \left(s - \mu_{\eta}(\gamma) \right) \Im T
\ee
Moreover, when $\Im Z(\gamma)$ vanishes, the real part evaluates to 
\bea
\label{eqReZ2}
\Re Z(\gamma) &=& 2 r w + d_2 M_1 - \ch_2 \nn\\
&=& r \left[ 2w -\frac1{\eta} (s-\tfrac{M_1}{2})^2 + \Delta + \frac{(d_1-\eta d_2-M_1 r)^2}{4\eta r^2} \right]
\eea
Although this is guaranteed by construction, one may check using \eqref{specflowsw} that \eqref{eqReZ2} is invariant under
spectral flow, as long as the condition $s= \mu_{\eta}(\gamma)$ is imposed. 
If $r=0$, then $\Im Z(\gamma)=(d_1+d_2 \eta)\Im T $, which is always positive as long
as $\Im T>0, \eta>0$ and $d_1, d_2\geq 0$.

\begin{figure}[t]
    \centering
\includegraphics[height=8cm]{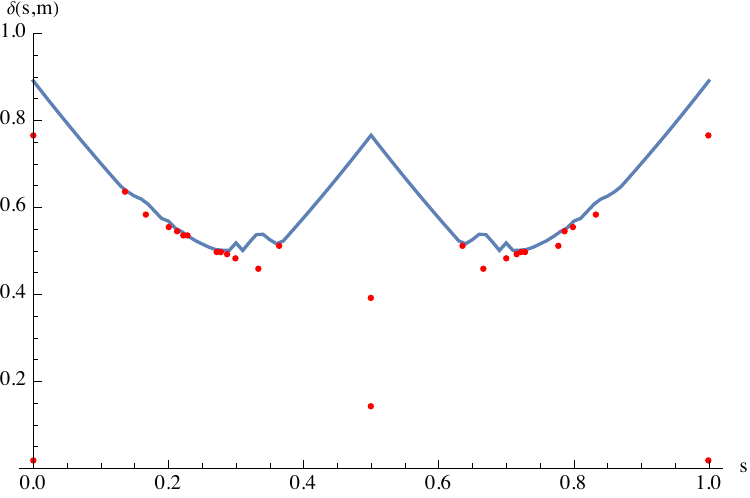}
    \caption{Approximate shape of $\delta(s,\eta,M_1)$ for $\eta=1, M_1=1/4$. The red dots indicate the corresponding lowest values arising from exceptional stable bundles.
     \label{fig_deltamuF0}}
\end{figure}

Let us now define
\be
\label{defdelmu}
\delta(s,\eta,M_1) := {\rm inf}_E \left[ \Delta(E) + \frac{(d_1(E)-\eta d_2(E)-M_1 r(E))^2}{4\eta r^2(E)} \right] 
\ee
where the infimum runs over all slope-semistable sheaves $E\in\cE_\eta$ of positive rank and slope $\mu_\eta(E)=s$. The function \eqref{defdelmu} is related to \eqref{deltaDLP}
through\footnote{Observe that if $ \delta_{\rm DLP}(\eta,a,b)$ was constant, the infimum would be obtained for $(a,b)=(s+\frac{M_1}{2},\frac{1}{\eta}(s-\frac{M_1}{2}))$. }
\be
\label{deltafromDLP}
\delta(s,\eta,M_1)= \inf_{a+\eta b=2s} \left[ \hat\delta_{\rm DLP}(\eta,a,b) + \frac1{4\eta} (a-\eta b-M_1)^2 \right]
\ee
where $\hat\delta_{\rm DLP}(\eta,a,b)=\min( \delta_{\rm DLP}(\eta,a,b), \Delta(F) )$
where $F\in \cE_{\eta}$ runs over all Gieseker-stable exceptional bundles with slope $(a,b)$, 
see Fig.~\ref{fig_deltamuF0} for an illustrative case $M_1=1/4,\eta=1$. 
Note that the term in bracket is a piecewise linear function (in general discontinuous for $\eta\neq 1$) of $(a,b)$ along the line $a+\eta b=2s$, thanks to a cancellation between
the  quadratic terms in \eqref{deltafromDLP} and \eqref{deltaDLP}. Moreover, 
$\delta(s,\eta,M_1)\geq \frac49$ except at points where there exists
integer values $(a,b)$ such that $a+\eta b=2s$ and $|a-\eta b-M_1|<(4/3)\eta^{1/2}$.

Then, inside the domain (which contains the large volume region \eqref{lvregion}) 
\be
\label{Ugeo}
U_{\rm geo} = \Bigl\{ \Im T>0, \quad \eta>0, \quad w_1 > - \frac{\delta(s,\eta,M_1)}{2\eta}  \Bigr\} \subset U
\ee 
we can define the heart $\cA$ by the standard tilt construction, i.e.\  generated by length-two complexes
$\cE \rightarrow \cF$ where $\cE$ (in degree $-1$) is a slope-semistable torsion-free sheaf of slope 
$\mu_\eta(\cE)\leq s$ and $\cF$ (in degree $0$) is a slope-semistable torsion-free sheaf of slope 
$\mu_\eta(\cF)> s$ or a torsion sheaf. This produces a stability condition such that skyscraper sheaves are stable, also known as a {\it geometric} stability condition. Along the loci
 $\Im T=0$ or $\eta=0$, the central charge of D2-branes wrapping $C$ or $F$ becomes
 aligned with the central charge of D0-branes, corresponding to  a boundary of
the space of geometric stability conditions. The loci where the last inequality in \eqref{Ugeo}
becomes saturated correspond to boundaries where skyscraper sheaves are destabilized by 
torsion-free sheaves.

%
%

\subsection{Stability conditions on the \texorpdfstring{$\Pi$}{Pi}-stability slice\label{sec_stabgeo}}

As explained in Appendix \ref{sec_periods}, mirror symmetry singles out a particular complex two-dimensional slice in $\Stab \cC$ where the coefficients $T, T_D$ are expressed in terms of 
the modular parameter $\tau\in \IH$ of the mirror curve and the mass parameter $m$ via the
contour integral representation \eqref{Eichlercst0}. The general strategy to construct the corresponding heart is to check that the conditions \eqref{Ugeo} are satisfied in a fundamental domain of $\Gamma_0(4)$, use
the tilt construction in that domain, and extend it to the full ($\IZ^4$ cover over the) Poincar\'e upper half-plane using autoequivalences of $\cC$. The details however differ 
depending whether $m$ is integer, real or complex valued.

\subsubsection{$m\in \IZ$}
Upon tensoring with $\cO(0,m)$ we can assume that $m=0$. In this case, 
the modular form \eqref{defC3} is an ordinary weight 3 Eisenstein series for $\Gamma_1(4)$,
with no square root singularities. The universal cover of the K\"ahler moduli space (with integer $m$) is therefore the Poincar\'e upper half plane, and it suffices to construct stability conditions
for $\tau$ in the fundamental domain $\cF$ bounded by the
vertical lines $\tau\in\I\IR, \tau\in1+\I\IR$ and the half-circles 
$\cC:=\cC(\frac34,\frac14), \tilde\cC:=\cC(\frac14,\frac14)$, and extend them elsewhere by acting
with the autoequivalences $ST_{\cO(0,0)}$ and $\cO(1,1) \otimes -$, which generate the group $\Gamma_0(4)$. 

As for the inequality $\Im T>0$, the $q$-series expansion \eqref{TTDLV} with $m=0$ shows that it holds at the boundaries
$\tau\in\I\IR^+, \tau\in1+\I\IR^+$, and numerical analysis shows that it also holds along the half-circles
$\cC$ and $\tilde\cC$, although it becomes saturated at the dual conifold point $\tau=\frac12$. Since $T$ is holomorphic inside $\cF$, the inequality holds everywhere in $\cF$ (and its translates $\cF(k)$ for $k\in\IZ$).

\begin{figure}[t]
    \centering
\includegraphics[height=8cm]{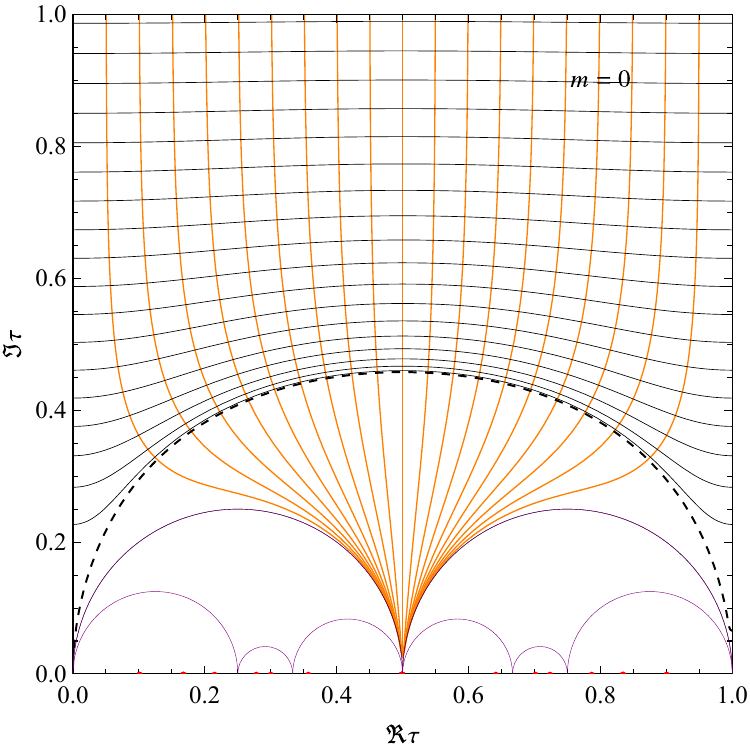}
    \caption{For $m=0$ (hence $\lambda=1$), the contours with fixed $s=\tfrac{\Im T_D}{\Im T}$ in the strip $\tau_1\in (0,1)$, shown in orange, all converge to $\tau=\frac12$. The contours with fixed (and real) $t=\sqrt{2w-s^2}$
approach $\tau=0$ and $\tau=1$ when $t\to 0$ (shown in dashed line), but stay away from $\tau=\frac12$. In the region above the dashed line, $\Pi$ stability is related to large volume stability 
by $GL(2,\IR)^+$.      \label{Contours1} }
\end{figure}

\begin{figure}[t]
    \centering
\includegraphics[height=8cm]{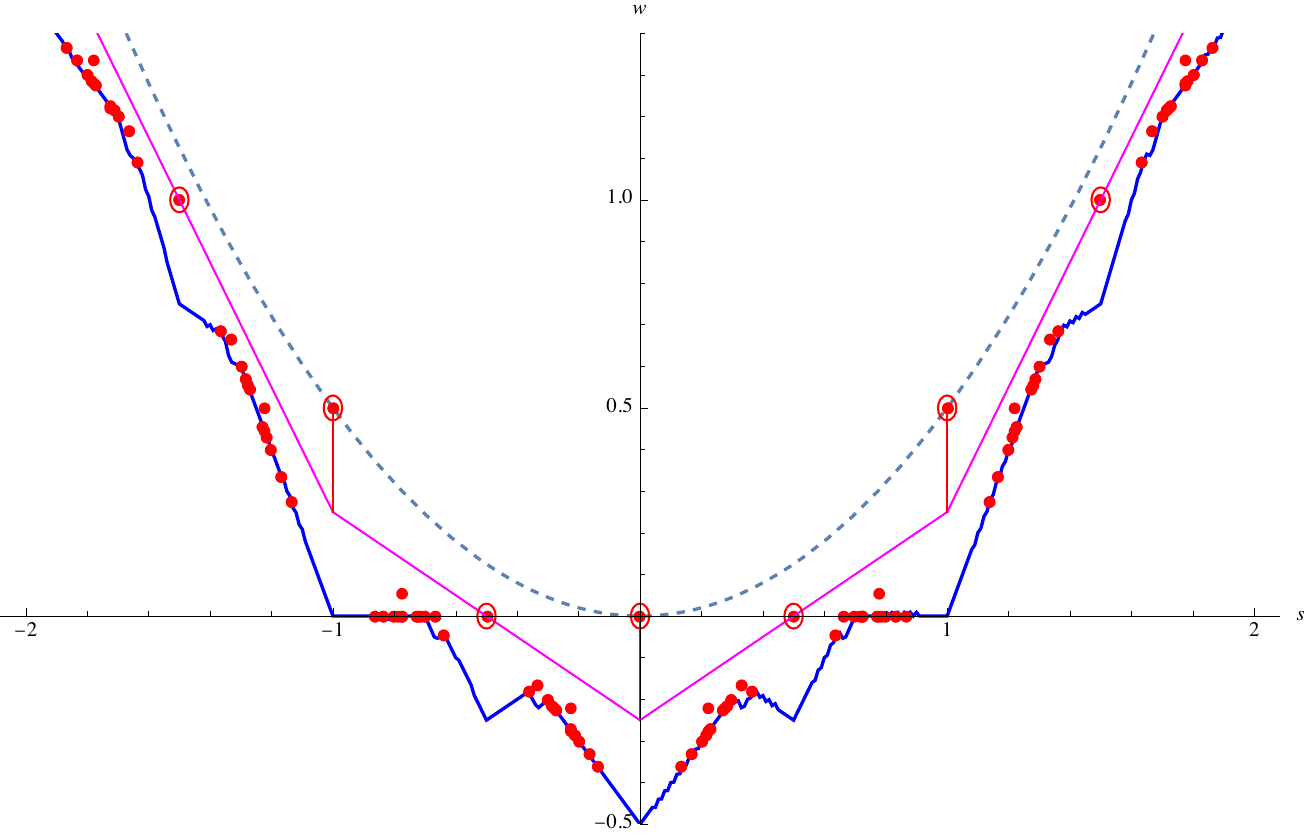}
    \caption{Image of the fundamental domain $\cF$ and its translates into the $(s,w)$ plane for $m=0$. The dashed line is the parabola $w=\frac12 s^2$, the red vertical segments are the images of the half-circles on the boundary of the fundamental domain and its translates
and the segments in magenta correspond to the values of $(s,w)$ around the dual conifolds 
$\tau\in\frac12+\IZ$.  The blue curve denotes the  (continuous part of the) lower bound 
$\frac12 s^2-\frac12\delta(s,1,0)$ of $w$ in the geometric domain \eqref{Ugeo},
and the red dots
are the discontinuous values occurring at slopes of exceptional stable bundles.  The circled ones are those which are massless at the conifold points $\tau=k$ and $\tau=k+\frac12$.  
 \label{fig_sw0}}
\end{figure}

Using the reality conditions  \eqref{TTDreal}, we see that $s=\Re\tau$ along the vertical lines
$\Re\tau\in\IZ$ and $\Re\tau\in\IZ+\frac12$. Moreover, similar to \cite[(A.57)]{Bousseau:2022snm}, we can use the fact that $M_{C}$ restricts to 
$\tau\mapsto -\bar\tau$ along $\tilde\cC$, to conclude that $\Im T_D=0$ along $\cC$, hence $s=0$.
Similarly, since  $M_{\tilde C}$ restricts to $\tau\mapsto 1-\bar\tau$ along $\cC$, one finds $\Im T_D=\Im T$ hence $s=1$. In the vicinity of the dual conifold point, $s$ varies from 0 on $\tilde\cC$ to $\frac12$
on the vertical axis $\frac12+\I\IR^+$ to 1 on $\cC$ (see Fig.~\ref{Contours1}). Using \eqref{TTDcond} and the equality $w=s \Re T-\Re T_D$, it follows that $w=\frac{s}{2}-\frac14$ grows linearly with $s$ 
around the cusp $\tau=\frac12$. Tensoring with $\cO(k,k)$, we find that the union of the fundamental domains $\cF(k)$  maps to the region in the $(s,w)$ plane above the piecewise linear curve joining the points $(k, \tfrac12 k^2 -\frac14)$
with $k\in\IZ$, see Fig.~\ref{fig_sw0}. 

In order to apply the tilting construction explained below \eqref{defsw}, we need
to check that the condition \eqref{Ugeo} (with $\eta=1, M_1=0$)
\be
\label{condws0}
w -\frac12 s^2 > - \frac12 \delta(s,1,0)
\ee
is also satisfied. The lower bound is the blue jagged curve shown in Figure
\ref{fig_sw0} along the red dots corresponding to exceptional bundles.
Since the left-hand side of \eqref{condws0} is greater than $-\frac14$, attained at the 
points $(k, \tfrac12 k^2 -\frac14)$, while 
$\delta(s,1,0)\geq \frac12$ except when $s$ is the slope of exceptional bundles of 
rank 1 or 3; in both cases, inspection of Fig. \eqref{fig_sw0} shows that the bound still holds.  
Thus, every point in the interior of the fundamental domain and its translates is  endowed with a geometric stability condition, while the skyscraper sheaves become destabilized 
by $\cO(k,k)$ and $\cO(k+1,k+1)$ along the half-circles $\tilde\cC(k)$ and $\cC(k)$, respectively. 

\subsubsection{$m\in \IR\backslash \IZ$}

When $m$ is real but not integer, we can tensor with $\cO(0,\lfloor m \rfloor)$ and assume $0<m<1$. 
The modular form \eqref{defC3} now has square root singularities
on the boundary of the fundamental domain at $\tau_B\in \cC$
and\footnote{For $m$ real, $\frac{\tau_B-1}{4\tau_B-3}$ actually coincides with $1-\bar\tau_B$.}
 $\tilde\tau_B:=\frac{\tau_B-1}{4\tau_B-3} \in \tilde\cC$. As indicated before, we choose 
branch cuts  along semi-circles  from $\tau_B$ to $\tau=\frac12$, and similarly
from $\tilde\tau_B$ to $\tau=\frac12$. Using the reality conditions  \eqref{TTDreal} implies that $s=\Re\tau$ along the vertical lines
$\Re\tau\in\IZ$ and $\Re\tau\in\IZ+\frac12$. Moreover, one may check that  $\Im T>0$
on the  fundamental domain $\cF(k)$ for any $k\in\IZ$ (except at the branch points
on $\tilde\cC(k)$ and $\cC(k)$).  When $m=0$, $\tau_B$ collides with the dual conifold singularity at $\tau=\frac12$ and the branch cuts disappear. 

\begin{figure}[t]
    \centering
\includegraphics[height=8cm]{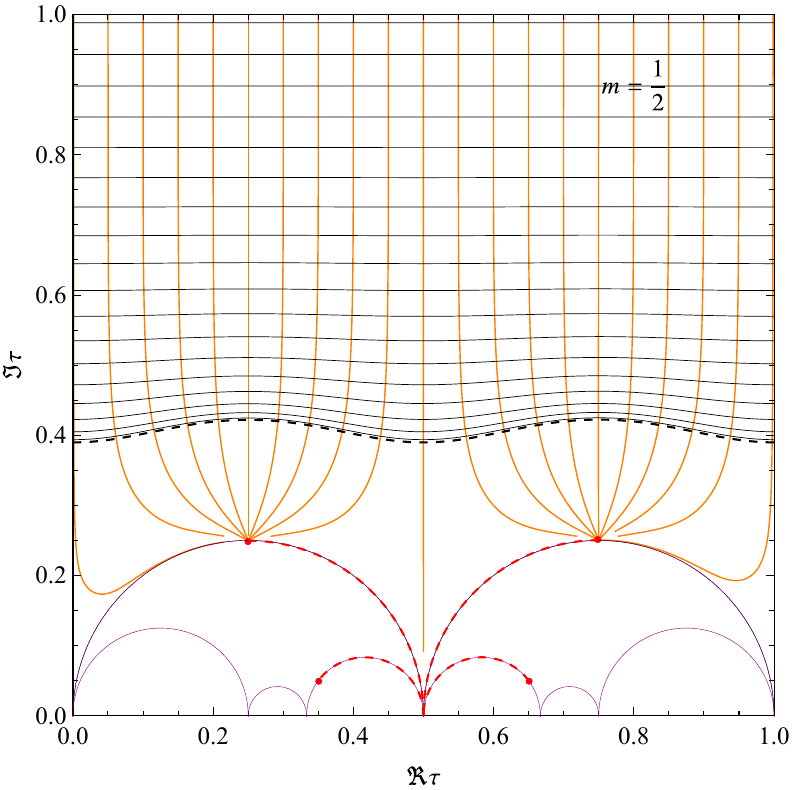}
    \caption{For $m=\frac12$, the contours with fixed $s=\tfrac{\Im T_D}{\Im T}$ in the strip $\tau_1\in (0,1)$, shown in orange, converge either to $\tau=\frac{1+\I}{4}$ or $\tau=\frac{3+\I}{4}$.
 The contours with fixed (and real) $t=\sqrt{2w-(s-\frac14)^2}$ stay well above the conifold
 and branch points
 In the region above the dashed line $t=0$, $\Pi$ stability is related to large volume stability 
by $GL(2,\IR)^+$.      \label{Contoursm1} }
\end{figure}

Due to  the branch points on $\cC$ and $\tilde\cC$, it now 
turns out that $s$ is discontinuous along the boundary of the fundamental domain, namely 
$s=0$ along the half-circle $\tilde\cC$  from $\tau=0$ up to $\tilde\tau_B$, $s=\frac12$ on the two arcs
in between  $\tilde\tau_B$ and $\tau_B$ and  $s=1$ along  $\cC$ from $\tau_B$ to $1$. Along these arcs, the central charge associated to $\cO(0,0), \cO(1,0)$ and $\cO(1,1)$ becomes real and negative,
respectively. Moreover,  along infinitesimal half-circles around 
 $\tilde\tau_B$   and $\tau_B$, $s$ increases continuously from $0$ to $\frac12$ and from $\frac12$ to $1$, respectively (see Fig.~\ref{Contoursm1} for the example of $m=\frac12$). Using 
 \eqref{TTD0}, \eqref{TTDhalf} and 
 the equality $w=s \Re T-\Re T_D$, one finds 
 that $(s,w)$ lies on the segment 
 $w=\left( \frac12 -m \right) s + \frac{m-1}{4}$ with $s\in (0,\frac12)$ on an infinitesimal half-circle around 
 $\tilde\tau_B$, and on the segment $w=\frac12 s - \frac{m+1}{4}$ with $s\in (\frac12,1)$
 on an infinitesimal half-circle around $\tau_B$, with the two  segments intersecting 
 at $(s,w)=(\frac12,-\frac{m}{4})$. On the other hand, the boundary circle arcs 
 map to vertical segments at $s\in \IZ/2$, ending at the images of the  
 conifold points $\tau=k$ and 
 dual conifold points $\tau=k+\frac12$ (see  \eqref{TTDk},  \eqref{TTDkhalf})
\be
(s,w) = \left(k, \tfrac12 k(k-m) \right)\quad \mbox{or} \quad
(s,w) = \left(k+\tfrac12, \tfrac12 k(k+1-m) \right)
\ee
 Hence, the union of the fundamental domains $\cF(k)$ is mapped to the
region above the piecewise linear curve (shown in magenta in Fig.~\ref{Contoursm1})
 joining the points
\be
\label{piecelc}
\left(k,\tfrac{k(k-m)}{2} + \tfrac{m-1}{4}\right) \to 
\left(k+\tfrac12, \tfrac{k(k+1-m)}{2}-\tfrac{m}{4}\right )  \to 
 \left(k+1,\tfrac{(k+1)(k+1-m)}{2} + \tfrac{m-1}{4}\right)  \to \dots
\ee

\begin{figure}[t]
    \centering
\includegraphics[height=8cm]{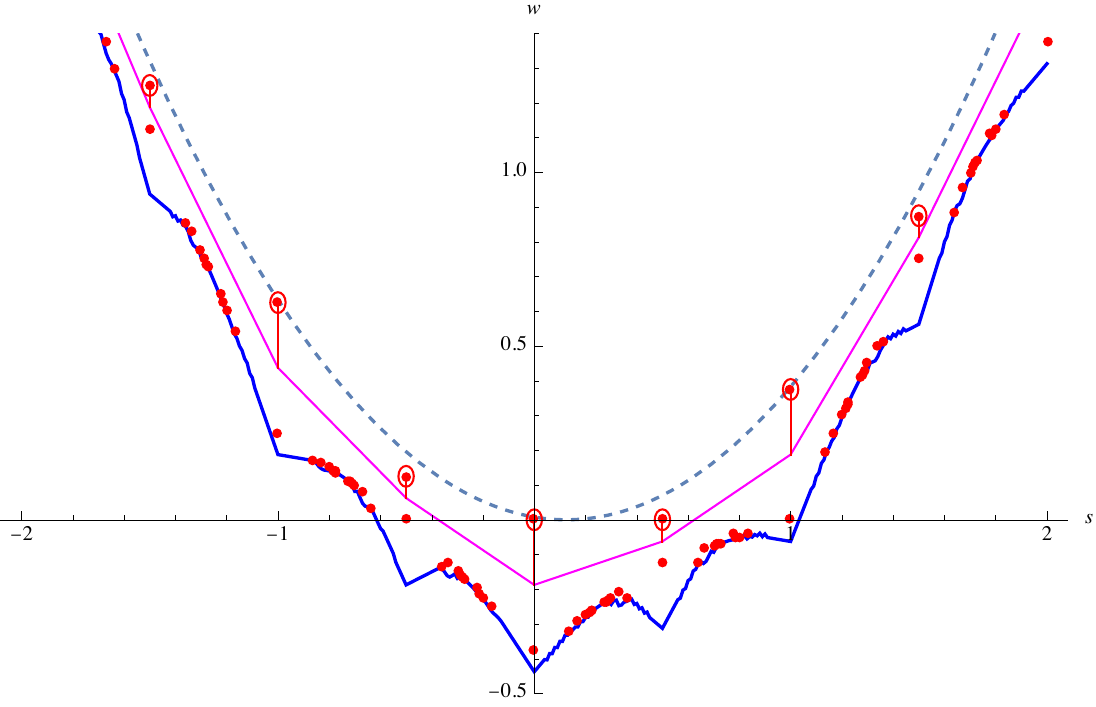}
    \caption{Image of the fundamental domain $\cF$ and its translates into the $(s,w)$ plane, shown for $m=1/4$. The dashed line is the parabola $w=\frac12 (s-\frac{m}{2})^2$, the red vertical segments are the images of the half-circles on the boundary of the fundamental domain and its translates
and the segments in magenta correspond to the values of $(s,w)$ on an infinitesimal circle around the branch points. The blue curve denotes the (continuous part of the) lower bound 
$\frac12 (s-\frac{m}{2})^2-\frac12\delta(s,1,m)$ on $w$ in the geometric domain \eqref{Ugeo}, 
and the red dots are the discontinuous values occurring at slopes of exceptional stable bundles. The circled ones are those which are massless at the conifold points $\tau=k$ and $\tau=k+\frac12$,
 lying along the shifted parabolas $w-\frac12 (s-\frac{m}{2})^2=-\frac{m^2}{8}$ and 
$-\frac{(1-m)^2}{8}$, respectively.   
 \label{fig_sw}}
\end{figure}

In order to apply the tilting construction explained below \eqref{defsw}, we need
to check that the condition \eqref{Ugeo} (with $\eta=1$)
\be
\label{condws}
w -\frac12 (s-\tfrac{m}{2})^2 > - \frac12 \delta(s,1,m)
\ee
is also satisfied. The lower bound is the blue jagged curve shown in Fig.~\ref{fig_sw}, along with
the red dots corresponding to exceptional bundles.
The left-hand side of \eqref{condws} is greater than $-\frac{1+\max(m,1-m)^2}{8}$, attained at the corners of the piecewise linear curve \eqref{piecelc}, while $\delta(s,1,m)\geq \frac12$ 
except when $s$ coincides with the slope of rank 1 or rank 3 exceptional bundles,
where inspection of Fig.~\ref{fig_sw} shows that the bound \eqref{condws} is still satisfied.  
Thus, every point in the interior of the fundamental domain and its translates is  endowed with a geometric stability condition, while the skyscraper sheaves become destabilized by $\cO(k,k),\cO(k+1,k)$  or $\cO(k+1,k+1)$ 
along the half-circles $\tilde\cC(k)$ and $\cC(k)$.

%

\subsubsection{$m$ complex}

When  $\Im m\neq 0$, there are several complications:  
\begin{itemize} 
\item[i)] The $GL(2,\IR)^+$-invariant parameters $M_1,\eta$ entering in the construction of the heart become non-trivial functions of $\tau$ (assuming, as we always do, that the parameter $m=M_1+(\eta-1) T$ is held fixed). This is already easily seen using the expansions~\eqref{TTDLV} for large $\Im \tau$, where (denoting $\tau=\tau_1+\I \tau_2, m=m_1+\I m_2$ 
\be
\label{sMetaLV}
M_1 \simeq \frac{ m_ 1\tau_2 - m_2\tau_1}{\tau_2-\frac{m_2}{2}}, \quad 
\eta\simeq\frac{\tau_2+\frac{m_2}{2}}{\tau_2-\frac{m_2}{2}}, \quad 
s\simeq \frac{\tau_1\tau_2 - \frac{m_1 m_2}{4}}{\tau_2-\frac{m_2}{2}}
\ee
up to worldsheet instanton corrections. As a result, we cannot map the fundamental domain in the $(s_1,w_1)$ plane without also keeping track of $(\eta,M_1)$.

\item[ii)]  The branch point $\tau_B$ moves inside the fundamental domain $\cF$, and the inequality $\Im T>0$ fails along a curve  extending from\footnote{To see this, one can use the expansion \eqref{Eichbranch} 
around $\tau_B$ to find that $\Im T=0$ along the infinitesimal segment
$\tau=\tau_B  - \frac{\I  \eta^8(2\tau_B)}
{\sin^2(\pi m/2) \eta^8(\tau_B) \eta^8(4\tau_B)} \epsilon + \cO(\epsilon^2)$
for small $\epsilon>0$, which points towards the real axis. } 
$\tau_B$ to $\tau=\frac34$. For the purpose of the current discussion, it is thus natural to choose this curve $\Im T=0$ 
as the branch cut. 
Moreover, the locus where $\Im Z(\cO(1,0))=0$
intersects the cut  at a point $\tau_b$ where $\Im T_D$ also vanishes,
and likewise the locus  $\Im Z(\cO(1,1))=0$  reaches the same point $\tau_b$ on
the other side of the cut
  (see Fig.~\ref{fig_geobound});

\item[iii)] Simultaneously,  the branch point $\tilde\tau_B$ and the loci where 
$Z(\cO(0,0))$ and $Z(\cO(1,0))$ are real and negative leave the fundamental domain $\cF$,  and move into its image $\tilde \cF$ under $M_S$. They intersect the locus $\Im T=0$ at a point $\tilde\tau_b$, but this locus lies below the arcs $\Im Z(\cO(0,0))=0$ and $\Im Z(\cO(1,0))=0$.
\end{itemize}


The last two bullet points show that the space of geometric boundary conditions cannot extend beyond the loci $s=0$, $s=\frac12$ and $s=\frac{1+\eta}{2}$,  along which skyscraper sheaves are destabilized by $\cO(0,0)$, $\cO(1,0)$ and $\cO(1,1)$, respectively, nor across the segment from $\tau_b$ to $\tau_B$ along the locus $\Im T=0$, along which skyscraper sheaves are destabilized by D2-branes. Let $\widehat\cF$ be the domain bounded by these loci in the strip $0\leq \Re(\tau_1)<1$.
Since $\cO(1,1)$ and $\cO(1,0)$ are mapped to $\cO(0,0)$ and $\cO(0,1)$ under $M_S$, 
$\widehat\cF$ 
is still a fundamental domain under $\Gamma_0(4)$, and it suffices to check that it satisfies the inequalities in \eqref{Ugeo}. We shall not attempt to prove it, but rather give some
numerical evidence that it does.

\begin{figure}[t]
    \centering
\includegraphics[height=8cm]{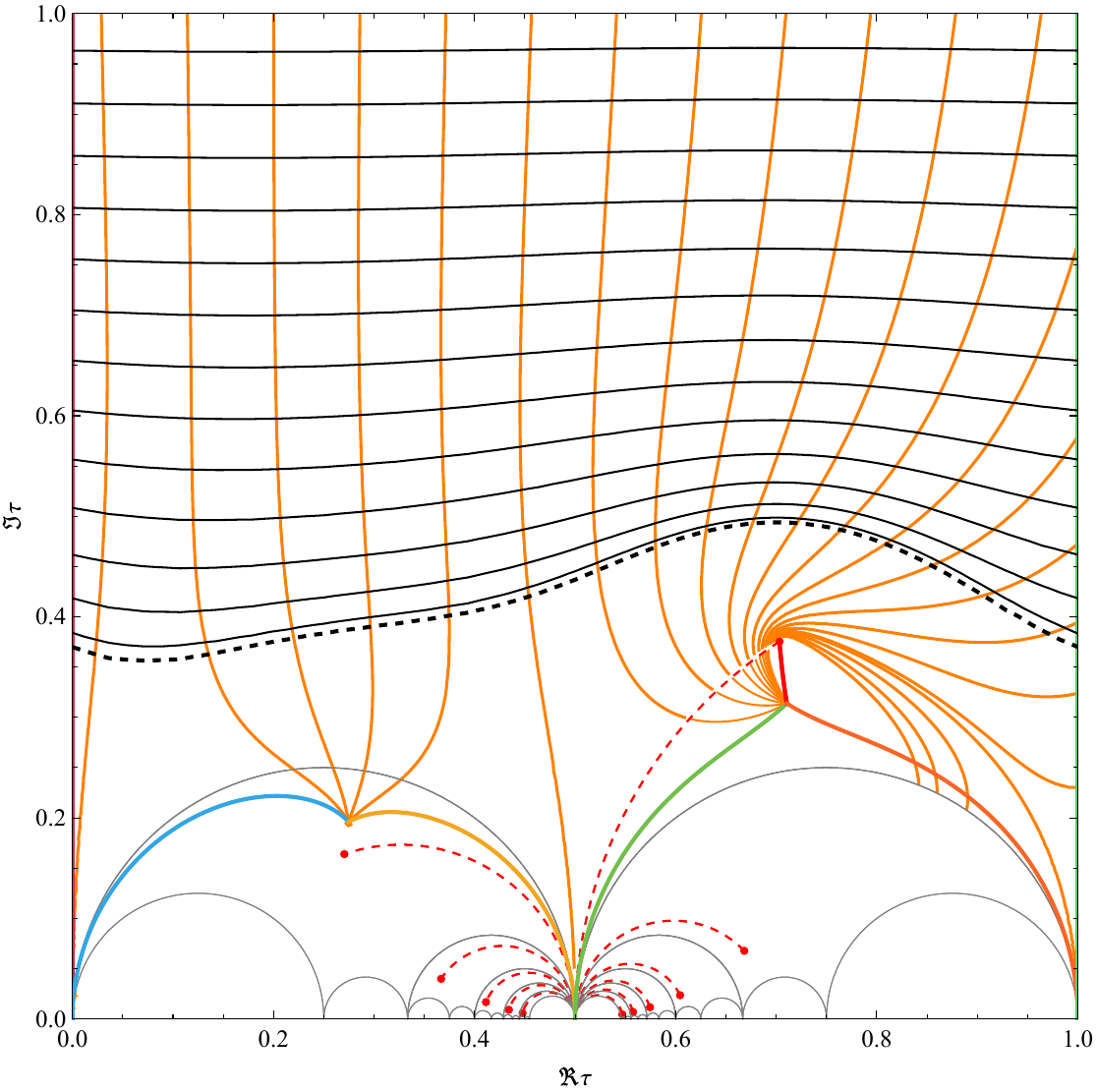}
    \caption{Contours with fixed $s$ (orange)
    or fixed $t=\sqrt{2w_1}$ (black) in the $\tau$ half-plane, for $m=0.4+0.3\I$.  
    In the region above the dashed line, corresponding to $t=0$, $\Pi$ stability is related to large volume stability 
by $GL(2,\IR)^+$. The blue, yellow, green, dark orange and red lines delimit (putatively)  the 
domain of geometric stability conditions, as in Fig.~\ref{fig_geobound}.
    \label{Contours1im} }
\end{figure}
\begin{figure}[t]
    \centering
\includegraphics[height=7cm]{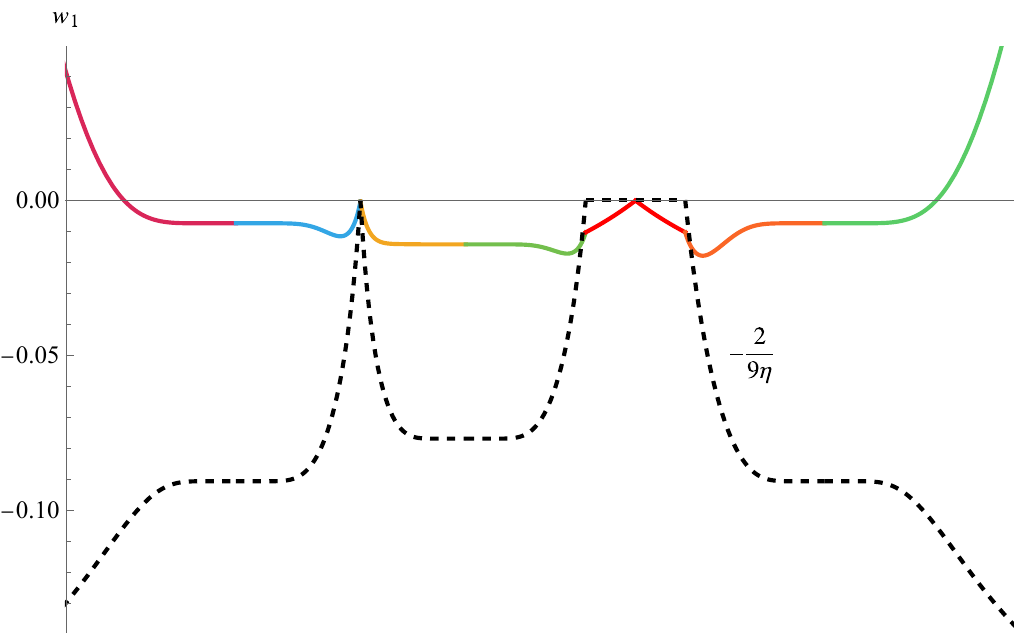} 
    \caption{
    Values of $w_1$ (defined in \eqref{defsw1}) along the various components of the boundary of the space of geometric stability conditions, including either side of the locus $\Im T=0$, compared to the sufficient lower bound $-2/(9\eta)$ (dashed line) 
    following from  $\delta(s,\eta,M_1)>\frac49$. The value of $m$ and the color coding for the boundary components 
    is the same as in Figs.~\ref{fig_geobound} and \ref{Contours1im}.
    \label{fig_walong}}
\end{figure}

In Fig.~\ref{fig_walong}, we show the value of $w_1$
along the boundary of the fundamental domain $\widehat\cF$, including the two sides of the 
locus $\Im T=0$.  This requires applying the monodromy $M_B(0)$ around $\tau_B$, acting as
\bea
(T,T_D) &\mapsto& (1-T,\tfrac{1+m}{2}-T_D) \nn\\
(s,w,\eta,M_1)& \mapsto& \left(s +\tfrac12 (1-\eta), -w +s - \tfrac{\eta+M_1}{2}, 2-\eta, M_1+ \eta-1 \right)
\eea
We first observe that the inequality $w_1>0$, which characterizes the large volume 
region, fails when $\Im\tau$ becomes small enough along the vertical boundaries at $\tau_1=0$ and 
$\tau_1=1$, consistent with Fig.~\ref{Contours1im}. Second, we observe that $w_1> -\frac2{9\eta}$ everywhere except on the two sides of the segment $\Im T=0$ on the boundary (as well as tiny segments along $\Im Z(\cO(1,0))=0$ and $\Im Z(\cO(1,1))=0$ in the vicinity of $\tau_b$). Since $\delta(s,\eta,M_1)\geq \frac49$ away from integer and half-integer slopes, this indicates that the inequality $w_1>-\delta(s,\eta,M_1)/(2\eta)$ is satisfied everywhere in $\widehat\cF$,
except perhaps inside a tiny sliver along the segment $\Im T=0$. We speculate that, using 
more refined information about $\delta(s,\eta,M_1)$, the inequalities \eqref{Ugeo}
could be shown to hold everywhere in the interior of the fundamental domain $\widehat{\cF}$, so that geometric stability conditions can be obtained using the tilting construction. 
If so, the intersection between the slice of $\Pi$-stability conditions and the 
space of geometric stability conditions would again coincide with 
the domain $\widehat\cF$ and its translates. 
Stability conditions on the full $\IZ^4$ cover of the Poincar\'e half-plane can then be obtained 
from those in $\cF$ by applying the autoequivalences $ST_{\cO(0,0)}, 
\cO(1,1) \otimes -$ and $M_B(0)$. 
 
\section{Quiver scattering diagram\label{sec_quiv}}

The derived category of (compactly supported) 
coherent sheaves on $K_{\IF_0}$ is equivalent to the derived category
of representations of quivers attached to cyclic strong exceptional collections~\cite{Bondal:1990,Aspinwall:2004bs}. In this section
we discuss the scattering diagram associated to the orbifold exceptional collection \eqref{extcoll},
also known as  phase II quiver, and to one of its mutations, sometimes known as phase I quiver,  
reproducing some of the results obtained in \cite[\S 4.1]{Beaujard:2020sgs}. 

\subsection{Basics of quiver scattering diagrams}
For completeness, we include a brief review of scattering diagrams for quivers with relations, heavily based on \S3.1 in our earlier work~\cite{Bousseau:2022snm}, with minor changes of notation.
\medskip

Let $(Q,W)$ be a quiver with potential. A representation $R$ of $(Q,W)$ is a set of finite dimensional
vector spaces $M_i$ for every node $i\in Q_0$, and linear maps $f_a\colon M_i\to M_j$ for every
arrow $(a\colon i\rightarrow j)\in Q_1$ such that for any arrow $a\in Q_1$, the element $\partial W/\partial a$ in the path algebra of $Q$ evaluates to zero on $R$.  The set of representations $R$ forms an Abelian category $\cA$  graded by the lattice $\IZ^{Q_0}$ spanned by the dimension vector $\vec N=\dim R=(\dim M_i)_{i\in Q_0}$. 
Given a choice of \textit{King stability parameters} $\zeta\in\IR^{Q_0}$, 
a representation $R$ is called $\zeta$-semi-stable if and only if  
$(\zeta,\vec N)=0$ and $(\zeta, \vec N')\leq 0$ for any subrepresentation $R'\subset R$ of dimension vector $\vec N'$. This coincides with the usual notion of $\sigma$-semistability
for a stability condition $\sigma=(Z,\cA)$ with central charge $Z(\vec N)=
\sum_{i\in Q_0} (-\zeta_i+\I\rho_i)N_i$ for any choice of $\rho_i>0$, provided one restricts to objects with $Z(E)\in \I \IR^+$.  We denote by $\cM_\zeta(\vec N)$
the moduli space of $\zeta$-semi-stable representations of dimension vector $\vec N$, by 
$\Omega_\zeta(\vec N)$ its refined Donaldson-Thomas invariant~\eqref{defOm} and by
$\bOm_\zeta(\vec N)$ its rational counterpart~\eqref{defOmb}.
These invariants vanish unless $\vec N$ belongs to the positive quadrant 
$\Gamma_+=\IN^{Q_0}$, in particular if all the $\zeta_i$'s have the same sign. 

The set of DT invariants is conveniently encoded in a stability scattering diagram $\cD_Q$ in the space $\IR^{Q_0}$ spanned by King stability parameters, defined as follows~\cite{bridgeland2016scattering}.
We first introduce the hyperplane orthogonal to $\vec N$, called \textit{geometric ray}, 
\be
\Rgeo_Q(\vec N)=\{\zeta : (\zeta,\vec N) = 0 \}
\ee
and define the \textit{active ray} $\Ract_Q(\vec N)$ as the subset  of $\Rgeo_Q(\vec N)$
\be
\Ract_Q(\vec N) = \{ \zeta :  (\zeta,\vec N) = 0,\quad \bOm_\zeta(\vec N)\neq 0\}
\ee
Furthermore, to each point  $\zeta\in \Ract_Q(\vec N)$ 
we associate an automorphism $\Uact_\zeta(\vec N)$ of the quantum torus algebra 
(or a suitable pronilpotent completion thereof)  generated by 
formal variables $\cX_{\vec N}$ for any $\vec N\in \Gamma_+$, subject to the relations 
\be
\label{qalg0}
\cX_{\vec N}\, \cX_{\vec N'} =(-y)^{\langle \vec N,\vec N'\rangle} \cX_{\vec N+\vec N'}
\ee
where 
\be
\langle \vec N,\vec N'\rangle=\sum_{a:(i\to j)\in Q_1} (n'_i n_j - n_i n'_j)
\ee
is the antisymmetrization of the Euler form $\chi_Q(\vec N,\vec N')=\sum_{i\in Q_0} n_i n'_i - \sum_{a:(i\to j)\in Q_1} n_i n_j'$. The automorphism $\Uact_\zeta(\vec N)$ then encodes the rational DT invariant $\bOm_\zeta(\vec N)$ via
\be
\label{defbUQ}
\Uact_\zeta(\vec N) =\exp\biggl( \frac{\bOm_\zeta(\vec N) \cX_{\vec N}}
{y^{-1}-y}\biggr)
\ee
The scattering diagram $\cD_Q$ is defined as the union of the active rays 
$\{\Ract_Q(\vec N) : \vec N\in\Gamma_+\}$ for all $\vec N$
equipped with the automorphism $\Uact_\zeta(\vec N)$
at each point. Since $\bOm_\zeta(\vec N)$  is invariant under 
rescaling $\zeta\to \lambda\zeta$ with $\lambda\in\IR_{>0}$, and since it can only jump on a finite set of hyperplanes $\Rgeo_Q(\vec N')$ corresponding to the destabilization by 
a subobject of dimension vector $\vec N'<\vec N$, $\cD_Q$ decomposes into 
a complex of convex rational polyhedral cones in $\IR^{Q_0}$.


The wall-crossing formula  ensures
that $\cD_Q$ is consistent in the following sense  \cite{bridgeland2016scattering}: for any 
generic closed path $\cP\colon t\in[1,0]\to\IR^{Q_0}$ (where generic means that
the intersection of $\cP$ with a ray $\Ract(\vec N_i)$ at $t=t_i$ is transverse
and does not meet any cone of codimension larger than one), 
 the  ordered product of automorphisms associated to each intersection is trivial, 
  \be
\label{constD0}
\prod_i  \cU_{\zeta(t_i)} (\vec N_i)^{\epsilon_i} = 1\ ,\quad 
\epsilon_i=\sign \left( \frac{\de \zeta}{\de t}, \vec N_i\right)
\ee
This consistency property  ensures that 
all rays can be deduced from the knowledge of the initial 
rays, defined as those rays $\Ract_\star(\vec N)$ which contain the
self-stability condition $\zeta_\star(\vec N):=\langle -,\vec N\rangle$ 
(such that $(\zeta_\star(\vec N),\vec N')=\langle \vec N',\vec N\rangle$ for
all $\vec N'$). We define the \textit{attractor invariant} $\Omstar(\vec N)$
as the index on the initial ray $\Ract_\star(\vec N)$. The attractor
tree formula of \cite{Alexandrov:2018iao} or its variant in \cite{Mozgovoy:2021iwz}
provide an algorithm to compute $\Omega_\zeta(\vec N)$ on the attractor invariants,
which is essentially equivalent to summing over all possible  scattering sequences of the initial rays~\cite{Arguz:2021zpx,Bousseau:2022snm}. 

For acyclic quivers (hence trivial potential), the only initial rays are those associated to the 
standard basis vectors $\vec N_i$, carrying $\Omstar(\vec N_i)=1$, and the construction of the scattering diagram is straightforward. For a general quiver with potential $(Q,W)$, determining the initial rays is a difficult problem. For quivers associated to non-compact Calabi-Yau threefolds, the case of interest in this work, it was conjectured in \cite{Beaujard:2020sgs,Mozgovoy:2020has,Descombes:2021snc} that 
$\Omstar(\vec N)=0$ unless $\vec N$ is a standard basis vector, or $\vec N$ belongs to the kernel of the antisymmetrized Euler form. See \cite[\S 3]{Bousseau:2022snm} for examples of scattering diagrams for Kronecker quivers, and for a definition of the scattering diagram $\cD_\psi$ in the full space of Bridgeland stability conditions.

\subsection{Orbifold quiver\label{sec_quivo}}
We  first consider the cyclic strong exceptional collection 
\be
\label{strongcoll2}
\cC = \left( \cO(0,0), \cO(1,0),\cO(1,1)\ ,\cO(2,1) \right) 
\ee
dual\footnote{In the sense that $\chi(E^i,E_j)=\delta^i_j$ for all $i,j$, see e.g.~\cite[\S 2.3,2.4]{Beaujard:2020sgs} for some background on exceptional collections. The collection 
\eqref{strongcoll2} appears in \cite{perling2003some} and arises by mutation from the more familiar
collection \eqref{strongcoll1} discussed in the next subsection.} to the Ext-exceptional collection 
\be
\label{dstrongcoll2}
\cC^\vee=\left(  \cO(0,0), \cO(-1,0)[1],\cO(1,-1)[1]\ ,\cO(0,-1)[2] \right) 
\ee
The Chern vectors of the objects $E^i$ in $\cC$ and dual objects $E_i$ in $\cC^\vee$ are 
\be
\begin{array}{ccl}
 \gamma^1 &=& \left[1,0,0,0\right] \\
 \gamma^2 &=& \left[1,1,0,0\right] \\
 \gamma^3 &=& \left[1,1,1,1\right] \\
 \gamma^4 &=& \left[1,2,1,2\right] \\
\end{array}
\qquad 
\begin{array}{ccl}
 \gamma_1 &=& \left[1,0,0,0\right] \\
 \gamma_2 &=& \left[-1,1,0,0\right] \\
 \gamma_3 &=& \left[-1,-1,1,1\right] \\
 \gamma_4 &=& \left[1,0,-1,0\right] \\
\end{array}
\ee
The corresponding quiver (sometimes called phase II quiver) is 
 \begin{center}
\begin{tikzpicture}[inner sep=2mm,scale=2]
  \node (a) at ( -1,1) [circle,draw] {$1$};
  \node (b) at ( 1,1) [circle,draw] {$2$};
  \node (c)  at ( 1,-1) [circle,draw] {$3$};
  \node (d)  at ( -1,-1) [circle,draw] {$4$};
 \draw [->>] (a) to node[auto] {$ $} (b);
 \draw [->>] (b) to node[auto] {$ $} (c);
 \draw [->>] (c) to node[auto] {$ $} (d);
 \draw [->>] (d) to node[auto] {$ $} (a);
\end{tikzpicture}
\end{center}
with quartic superpotential \cite{Feng:2000mi},\cite[(2.2)]{Feng:2001xr}
\be
\label{Worb}
W = \sum_{(\alpha\beta)\in S_2} \sum_{(\gamma\delta)\in S_2} 
\sign(\alpha\beta) \,  \sign(\gamma\delta) \,  \Phi_{12}^\alpha \, \Phi_{23}^\gamma \, \Phi_{34}^\beta\, \Phi_{41}^\delta\ .
\ee
A general Chern character $\gamma=[r,d_1,d_2,\ch_2]$ decomposes as $\gamma=\sum_i n_i \gamma_i$ with 
\bea
n_1 &=& d_1 + d_2 + \ch_2 + r \nn\\
n_2 &=& d_1 + \ch_2\nn\\
n_3 &=&\ch_2 \nn\\
n_4 &=& -d_2 + \ch_2
\eea
It is straightforward to check that the $\IZ_4$ cyclic symmetry $\gamma_i \to \gamma_{i+1}$ 
acts on the Chern vector as in \eqref{Z4aut}.

As in \cite{Beaujard:2020sgs}, we set $N_i=-n_i$ such that $N_i$ are all positive for large negative $\ch_2$. With these identifications, the expected dimension of the moduli space of Gieseker stable
sheaves coincides with the expected dimension of the moduli space of stable quiver representations
in the chamber $\zeta_1\geq 0,\zeta_4\leq 0$ where the arrows $\Phi_{41}^\alpha$ can be set to zero.
In order to match King stability with Gieseker stability for the canonical polarization $H=C+F$,  
one needs to take 
\be
\label{canchamber}
\zeta_1 = - \rho (d_1+d_2) - \ch_2, \quad 
\zeta_2=\rho (r + d_1+d_2) +\ch_2, \quad 
\zeta_3=\rho(d_1+d_2) + r+ \ch_2, \quad 
\zeta_4= -\zeta_2
\ee
with $\rho$ large and positive \cite[(4.27)]{Beaujard:2020sgs} (after fixing a sign mistake in loc.~cit.). 
We refer to the chamber \eqref{canchamber} as the canonical chamber\footnote{The canonical chamber is also called anti-attractor chamber, since the stability parameter \eqref{canchamber} is opposite to the attractor parameter
$\zeta_i = - \kappa_{ij} N_j$, with $\kappa_{ij}$ the number of arrows $i\to j$ minus the number of arrows $j\to i$.}.
With these identifications, the quiver index $\Omega_c(\vec N)$ in the canonical chamber
 is expected to agree with the Gieseker index $\Omega_1(\gamma)$  whenever $-r\leq d_1+d_2\leq 0$.
More generally, in order to reproduce Gieseker indices  $\Omega_\eta(\gamma)$ for polarization $H=C+\eta F$, the stability parameters should be chosen as \cite[(4.32)]{Beaujard:2020sgs}
\bea
\label{canchambereta}
\zeta_1 &=& - \mu_1 d_2 - \mu_2 d_1 - \ch_2
= \mu_1(N_3-N_4) + \mu_2(N_2-N_3) + N_3
\nn\\
\zeta_2 &=&  \mu_1 d_2 + \mu_2 (r+d_1) + \ch_2
= \mu_1 (N_4-N_3)+ \mu_2(2N_3-N_1-N_4) - N_3
\nn\\
\zeta_3 &=& \mu_1 (r+d_2) +\mu_2 (d_1-r) +r + \ch_2
= \mu_1 (N_2-N_1) + \mu_2( N_1+N_4-2N_2)+ N_2-N_1-N_4
\nn\\
\zeta_4 &=& - \mu_1 (r+d_2) - \mu_2 d_1 - \ch_2
=\mu_1 (N_1-N_2) + \mu_2 (N_2-N_3) + N_3\nn\\
\eea
with $\mu_1,\mu_2\gg 1$ such that $\eta=\mu_1/\mu_2$. These identifications ensure that 
\be
\sum_i N_i \zeta_i=0\ ,\quad
\sum_i N'_i \zeta_i= r ( \mu_1 d'_2 + \mu_2 d'_1) - r' (\mu_1 d_2+\mu_2 d_1) + r\ch'_2 - r' \ch_2
\ee
The conditions $\zeta_1\geq 0, \zeta_4\leq 0$ for the validity of the quiver description now translate into $-\eta r \leq d_1 + \eta d_2 \leq 0$.

\medskip

Since the kernel of the antisymmetrized Euler form is spanned by the dimension vectors $(1,0,1,0)$ and $(0,1,0,1)$, it is natural to consider  two-dimensional slices of the scattering diagram 
with fixed values of $\zeta_1+\zeta_3$ and $\zeta_2+\zeta_4$. Motivated by the values of the 
parameters $\zeta_i$ at the ramification point $\tau_B$ along the $\Pi$-stability slice, we choose 
\be
\label{quiver2param}
\zeta_1+\zeta_3 =\mu-1, \quad  \zeta_2 + \zeta_4 = -\mu
\ee
with $\mu$ a fixed real number. 
We can parametrize this slice by $(u,v)\in\IR^2$ such that
\be
\label{zetauvorb}
\zeta_1=v+\tfrac{\mu-1}{2}, \,\quad \zeta_2=u-\tfrac{\mu}{2}, \quad \zeta_3=-v+\tfrac{\mu-1}{2}, \quad \zeta_4=-u-\tfrac{\mu}{2}
\ee
The geometric rays $\cR_\gamma$ are then straight lines 
\be
(N_2-N_4) u + (N_1-N_3) v +\tfrac{\mu}{2} (N_1-N_2+N_3-N_4) -\tfrac{N_1+N_3}{2}= 0
\ee
oriented along the vector $v_{\gamma}=(N_3-N_1,N_2-N_4)$. Observe that the inner product 
$v_\gamma\wedge v_\gamma'$ is equal to the Dirac product $\langle \gamma, \gamma'\rangle$, 
up to a factor $-1/2$. 
Moreover, the angular momentum 
is constant along the rays, and positive for $0<\mu<1$,
\be
 (u,v)  \wedge v_\gamma = \tfrac{N_1+N_3}{2} -\tfrac{\mu}{2} (N_1-N_2+N_3-N_4)
\ee
so rays tend to rotate counterclockwise around the origin $u=v=0$. 

\medskip

By the same reasoning as in \cite[\S 5.2]{Bousseau:2022snm} one can show that
the only initial rays are those associated to $\gamma_i$, with unit BPS 
index $\Omstar(\gamma_i)=1$, and to the D0-branes (or skyscaper sheaves)
$k\delta$ with $\delta=(1,1,1,1)$. The indices $\Omega_c(\vec N)$ in the canonical chamber
for a selection of Chern vectors in the window $-r\leq d_1+d_2\leq 0$ are then given by the
following scattering sequences (or trees, for brevity):
\begin{equation*}
\begin{array}{|l|l|l|l|}
\hline
\left[r,d_1,d_2, \ch_2\right] & \vec N  &\mbox{Trees} & \Omega_c(\vec N)  
\\ 
\hline
\left[1,0,0,-1\right] &(0,1,1,1)& \{\gamma _2,\{\gamma _3,\gamma _4\}\} &y^2+2+1/y^2\ 
\\ 
\hline
\left[1,0,0,-2\right] &(1,2,2,2)&
  \begin{array}{@{}l@{}}
\{\{\{\gamma _1,2 \gamma _4\},2 \gamma _3\},2 \gamma _2\} \\
\{\{\{\{\gamma _1,\gamma _2\},2 \gamma _4\},2 \gamma
   _3\},\gamma _2\}\\
   \{\{\{\{\gamma _1,\gamma _4\},2 \gamma _3\},\gamma
   _4\},2 \gamma _2\}
    \end{array}
 &y^4+3 y^2 + 6 + 3/y^2 + 1/y^4 
 \\ 
  \hline
  \left[2,0,0,-2\right]&(0,2,2,2) & 
  \{2\gamma _2,\{2\gamma _3,2\gamma _4\}\}  & -y^5 - 2 y^3 -3 y -\dots
\\ 
 \hline
 \left[2,-1,0,-1\right] &(0,2,1,1) & \{2 \gamma _2,\{\gamma _3,\gamma _4\}\}
  & -y -1/y 
     \\ 
  \hline
 \left[2,-1,0,-2\right] &(1,3,2,2) & 
  \begin{array}{@{}l@{}}
 \{\gamma _1,\{3 \gamma _2,\{2 \gamma _3,2 \gamma _4\}\}\} \\
 \{\{\{\{\gamma _1,\gamma _2\},2 \gamma _4\},2 \gamma
   _3\},2 \gamma _2\}
    \end{array}
 & -y^5 - 3 y^3 -7 y -\dots 
  \\ 
  \hline
  \left[2,0,-1,-1\right] &(0,1,1,0) & \{\gamma _2,\gamma _3\} & -y -1/y 
\\  
 \hline
 \left[2,0,-1,-2\right] &(1,2,2,1) & 
   \begin{array}{@{}l@{}}
 \{\gamma _1,\{2 \gamma _2,\{2 \gamma _3,\gamma _4\}\}\} \\
 \{\gamma _1,\{\{2 \gamma _2,\{\gamma _3,\gamma _4\}\},\gamma
   _3\}\}\\
   \{\{\{\gamma _1,\gamma _4\},2 \gamma _3\},2 \gamma _2\}
   \end{array}
  &-y^5 - 3 y^3 -7 y -\dots
  \\ 
  \hline
 \left[2,-1,-1,-1\right] &(1,2,1,0) & 
 \{\gamma_1,\{\gamma_2,\gamma_3\}\}
  &-y^3- y -1/ y -1/y^3 
  \\ 
  \hline
  \left[3,0,0,-3\right] &(0,3,3,3) &
  \{3\gamma _2,\{3\gamma _3,3\gamma _4\}\}  
    &y^{10}+2 y^8+5   y^6 +8 y^4 + 9 y^2 + 10 +\dots
    \\ 
    \hline
  \left[3,0,-1,-2\right] &(0,2,2,1) &
  \begin{array}{@{}l@{}}
  \{2 \gamma _2,\{2 \gamma _3,\gamma _4\}\} \\
  \{\{2 \gamma _2,\{\gamma _3,\gamma _4\}\},\gamma _3\}
  \end{array}
   & y^4+2y^2 +4+\dots\\ \hline
  \left[3,-1,-1,-1\right] &(0,2,1,0) & \{2\gamma _2,\gamma _3\}  & 1\\ \hline
   \left[3,-1,-2,-2\right] &(2,3,2,0) &
     \begin{array}{@{}l@{}}
   \{2 \gamma _1,\{3 \gamma _2,2 \gamma _3\}\} \\
 \{\{   2 \gamma _1 ,\{\gamma _2,\gamma _3\} \} , \{ 
 2 \gamma _2 , \gamma _3 \} \} 
     \end{array}
    & y^8+2y^6+4y^4+4y^4+5+\dots\\ \hline
 \end{array}
\end{equation*}
This reproduces the
table in \cite[\S 4.1.1]{Beaujard:2020sgs}. Note that whenever $-r\leq d_1+d_2\leq 0$, the ray exits in the first quadrant $u>0, v>0$, and the index $\Omega_c(\vec N)$ in the canonical chamber coincides with the Gieseker index for canonical polarization.


\medskip

\begin{figure}[t]
    \centering
\includegraphics[height=10cm]{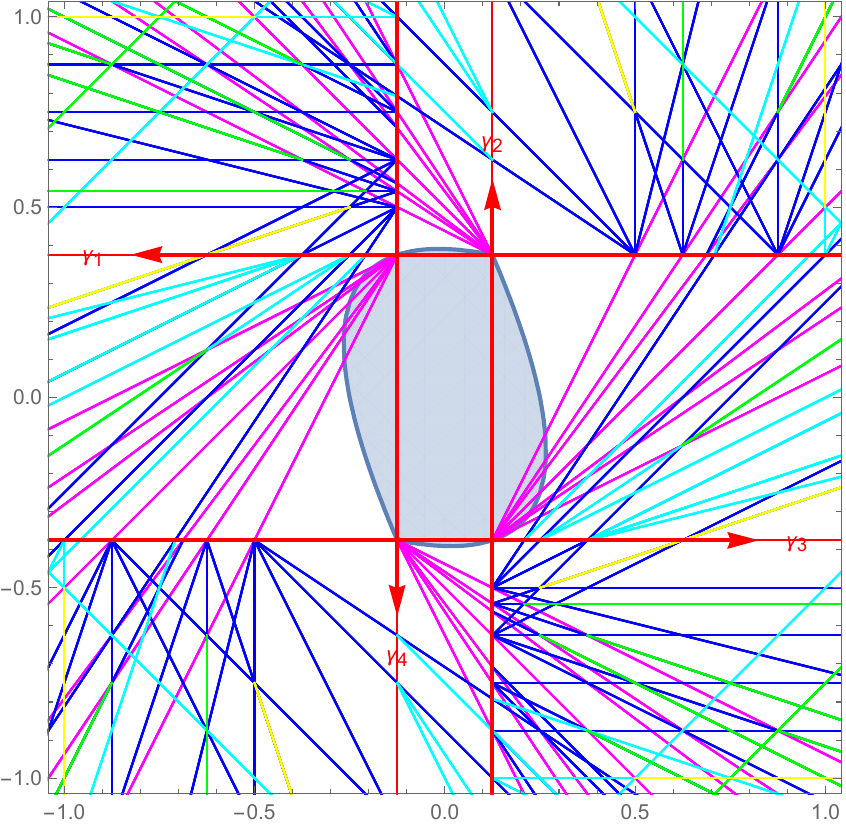}
    \caption{Scattering diagram $\cD^o_\mu$ for the orbifold quiver at $\mu=1/4$. 
    The shaded area corresponds
    to the generalized `collimation chamber' \eqref{delmontegen}.     \label{QuiverScattColl}}
\end{figure}

We further note that the BPS spectrum of this quiver has been studied in \cite{Closset:2019juk,Longhi:2021qvz,DelMonte:2021ytz,DelMonte:2022kxh}. In particular, in \cite{DelMonte:2022kxh}, elaborating on \cite[Appendix D]{Closset:2019juk}, 
the authors show that the spectrum simplifies drastically 
in the `collimation chamber' defined by\footnote{We rescale the 
central charges in   \cite{DelMonte:2022kxh} by a factor of $-R/(2\pi)$, such that the D0-brane
central charge is $-1$. Furthermore, we identify $\tau=1/\sqrt{\lambda}=e^{-\I\pi m}, \kappa=U/\sqrt\lambda$.}
\be
\label{delmonte}
Z(\gamma_3)=Z(\gamma_1)+\delta, \quad Z(\gamma_4)=Z(\gamma_2)-\delta, \quad
\arg Z(\gamma_1) > \arg Z(\gamma_2), \quad Z(\gamma_1+\gamma_2)<0
\ee
with $-\frac12<\delta<\frac12$. Namely, in this chamber 
the spectrum is the union of bound states of $\{\gamma_1,\gamma_2\}$, described by the Kronecker
quiver $K_2$, and bound states of $\{\gamma_3,\gamma_4\}$, described by the same quiver. 
More generally, the same conclusion holds when 
\be
\label{delmontegen}
Z(\gamma_3)=\rho Z(\gamma_1)+\delta, \quad Z(\gamma_4)=\rho Z(\gamma_2)-\delta, \quad
\arg Z(\gamma_1) > \arg Z(\gamma_2), \quad Z(\gamma_1+\gamma_2)<0
\ee
where $\rho>0$ and $-\frac{\rho}{1+\rho}< \delta < \frac{\rho}{1+\rho}$. When $\arg Z(\gamma_1) > \arg Z(\gamma_2)$, we similarly expect bound states of  $\{\gamma_4,\gamma_1\}$ and
 $\{\gamma_2,\gamma_3\}$ only. 
Central charges of the form \eqref{delmontegen} arise for
\be
T= -\frac{1}{1+\rho}, \quad T_D= \frac{(1+\delta- \mu)}{2(1+\rho)}
\ee
This  intersects the $\Pi$-stability slice  at the orbifold point where $\delta=0, \rho=1$. 
In terms of the King stability parameters 
\be
\label{zetafromZ}
\zeta_i=\frac{\Re(e^{-\I\psi} Z(\gamma_i))}{\cos\psi}
\ee
(with the phase $\psi$ chosen such that all central charges lie in the half-plane $\Im(e^{-\I\psi} Z(\gamma_i)) \geq 0$), the chamber \eqref{delmontegen}  lies along the locus
\be
\zeta_3 =\rho \zeta_1+\delta, \quad \zeta_4 = \rho \zeta_2-\delta
\ee
Further enforcing the conditions \eqref{quiver2param}, we get 
\be
u =\frac{\mu}{2} + \frac{\delta-\mu}{1+\rho}, \quad
v = \frac{1-\mu}{2} + \frac{\mu-1-\delta}{1+\rho}
\ee
Solving for $(\rho,\delta)$ in terms of $(u,v)$, we see that the region where $\rho>0, -\frac{\rho}{1+\rho}< \delta < \frac{\rho}{1+\rho}$ is bounded by the
two hyperbolae 
\be
(u+v)^2 \pm \mu (u+v) \mp u - \frac14 < 0
\ee
One may check that this region contains only rays coming from the primary scattering of $\{\gamma_1,\gamma_4\}$ and $\{\gamma_2,\gamma_3\}$, as predicted in \cite{Closset:2019juk}
(see Fig.~\ref{QuiverScattColl}).

\medskip

\subsection{Phase I quiver\label{sec_quivI}}

\begin{figure}
\begin{tikzpicture}[inner sep=2mm,scale=2]
  \node (a) at ( -1,0) [circle,draw] {$1'$};
  \node (b) at ( 0,1) [circle,draw] {$2'$};
  \node (c)  at ( 0,-1) [circle,draw] {$3'$};
  \node (d)  at ( 1,0) [circle,draw] {$4'$};
 \draw [->>] (a) to node[auto] {$ $} (b);
 \draw [->>] (a) to node[auto] {$ $} (c);
 \draw [->>] (b) to node[auto] {$ $} (d);
 \draw [->>] (c) to node[auto] {$ $} (d);
  \draw [->>>>] (d) to node[auto] {$ $} (a);
\end{tikzpicture}
\caption{Quiver associated to the Ext-exceptional collection \eqref{dstrongcoll1}, sometimes known as phase I quiver. This is accompanied
by the cubic superpotential $W$ in \eqref{WphaseI}.
\label{fig_quiverI}}
\end{figure}
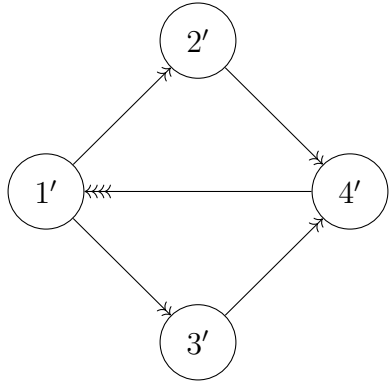

We now consider the cyclic strong 
exceptional collection \cite[Example 2.8]{bridgeland2010helices}
\be
\label{strongcoll1}
\cC=\left( \cO(0,0),\cO(1,0),\cO(0,1),\cO(1,1) \right)
\ee
dual to the Ext-exceptional collection 
\be
\label{dstrongcoll1}
\cC^\vee=\left( \cO(0,0), \cO(-1,0)[1], \cO(0,-1)[1], \cO(-1,-1)[2] \right)
\ee
This is obtained from the orbifold collection \eqref{dstrongcoll2} by applying a mutation with respect to node 4, and then exchanging the nodes $3$ and $4$. Note that $\cC^\vee$ is no longer invariant
under the order 4  autoequivalence $M_{\IZ_4}$, but it is invariant under fiber-base duality.
 The Chern vectors of the objects $E^i$ and dual objects $E_i$ are 
\be
\begin{array}{ccl}
 \gamma'^1 &=& \left[1,0,0,0\right] \\
 \gamma'^2 &=& \left[1,1,0,0\right] \\
 \gamma'^3 &=& \left[1,0,1,0\right] \\
 \gamma'^4 &=& \left[1,1,1,1\right] \\
\end{array}
\qquad
\begin{array}{ccl}
 \gamma'_1 &=& \left[1,0,0,0\right] =\gamma_1\\
 \gamma'_2 &=& \left[-1,1,0,0\right] = \gamma_2\\
 \gamma'_3 &=& \left[-1,0,1,0\right] = -\gamma_4 \\
 \gamma'_4 &=& \left[1,-1,-1,1\right]= \gamma_3+2\gamma_4 \\
\end{array}
\ee
The corresponding quiver, shown in Fig.~\ref{fig_quiverI},  is accompanied by the  cubic superpotential  \cite{Feng:2000mi}
\bea
\label{WphaseI}
W&=&\Phi_{12}^{1} \Phi_{24}^{1} \Phi_{41}^{4}-\Phi_{12}^{1} \Phi_{24}^{2} \Phi_{41}^{3}-\Phi_{12}^{2} \Phi_{24}^{1}
   \Phi_{41}^{2}+\Phi_{12}^{2} \Phi_{24}^{2} \Phi_{41}^{1} \nn\\
   &&-\Phi_{13}^{1} \Phi_{34}^{1} \Phi_{41}^{4}+\Phi_{13}^{1} \Phi_{34}^{2}
   \Phi_{41}^{2}+\Phi_{13}^{2} \Phi_{34}^{1} \Phi_{41}^{3}-\Phi_{13}^{2} \Phi_{34}^{2} \Phi_{41}^{1}
\eea
A general Chern character $\gamma=[r,d_1,d_2,\ch_2]$ decomposes as $\gamma=
\sum_i n'_i \gamma_i$ with 
\bea
n'_1 &=& d_1 + d_2 + \ch_2 + r \nn\\
n'_2 &=& d_1 + \ch_2\nn\\
n'_3 &=& d_2+ \ch_2 \nn\\
n'_4 &=& \ch_2
\eea
As in \cite{Beaujard:2020sgs}, we set $N'_i=-n'_i$ such that $N'_i$ are all positive for large negative $\ch_2$. The dimension vector $\vec N'$ with respect to $\cC'^\vee$ is related to the dimension vector $\vec N$ with respect to $\cC^\vee$ by $(N'_1,N'_2,N'_3,N'_4)=(N_1,N_2,2N_3-N_4,N_3)$. 

\medskip

With these identifications, the expected dimension of the moduli space of Gieseker stable
sheaves again coincides with the expected dimension of the moduli space of stable quiver representations
in the chamber $\zeta'_1\geq 0,\zeta'_4\leq 0$ where the arrows $\Phi_{41}^\alpha$ can be set to zero. This coincides with the canonical (or anti-attractor) chamber whenever $-2r\leq d_1+d_2\leq 0$.
In order to match King stability with Gieseker stability, one needs to take 
\be
\zeta'_1 = - \rho (d_1+d_2) - \ch_2, \quad 
\zeta'_2=\zeta'_3= \rho (r + d_1+d_2) +\ch_2, \quad 
\zeta'_4=-\rho(d_1+d_2+2r)+r-\ch_2, \quad 
\ee
with $\rho$ large and positive \cite[(4.11)]{Beaujard:2020sgs} (after fixing a sign mistake in loc.~cit.).
The
stability parameters for Gieseker stability with respect to $H=C+\eta F$ can be obtained from \eqref{canchambereta}
by equating $(\zeta'_1,\zeta'_2,\zeta'_3,\zeta'_4)=(\zeta_1,\zeta_2,-\zeta_4,\zeta_3+2\zeta_4)$. 

\medskip

Since the objects $E_1$ and $E_2$ are identical in the two collections $\cC^\vee$ and $\cC'^\vee$,
it is natural to choose the same two-dimensional slice as in \eqref{quiver2param}, 
\be
\label{quiver1param}
\zeta'_1=v+\frac{\mu-1}{2}, \quad \zeta'_2=u-\frac{\mu}{2}, \quad \zeta'_3=u+\frac{\mu}{2}, \quad \zeta'_4=-2u-v-\frac{\mu+1}{2}
\ee
 such that the rays $\cR(\gamma'_1)$ and 
$\cR(\gamma'_2)$ agree with the corresponding rays in the scattering diagram of the  orbifold quiver 
\eqref{quiver2param}. 
The geometric rays $\cR_\gamma$ are now straight lines 
\be
(N'_2+N'_3-2N'_4) u + (N'_1-N'_4) v + \frac{\mu}{2} (N'_1-N'_2+N'_3-N'_4)+ \frac{N'_1+N'_4}{2}= 0
\ee
oriented along $v_{\gamma}=(N'_4-N'_1,N'_2+N'_3-2N'_4)$. 

\medskip

As shown in \cite{mou2019scattering},
the scattering diagram for the mutated collection $\cC'^\vee$ coincides with the scattering diagram for 
the original collection $\cC^\vee$ on the half-space $\zeta'_4<0$. It follows that the Attractor 
Conjecture must also hold for the phase I quiver, namely the only initial rays are those associated to $\gamma'_i$, with BPS index $\Omstar(\gamma'_i)=1$, and to the D0-branes 
$k\delta$ with $\delta=(1,1,1,1)$. From these initial rays, we can compute the canonical indices
for the same Chern vectors as in the previous subsection, and check that the same result
arises, in general from different scattering sequences:  
\begin{equation*}
\begin{array}{|l|l|l|l|}
\hline
\left[r, d_1,d_2,\ch_2\right] & \vec N'  & \mbox{Trees} &\Omega_c(\vec N')  
\\ \hline
\left[1,0,0,-1\right] &(0,1,1,1)& \{\{\gamma'_3,\gamma'_4\},\gamma'_2\} & y^2+2+1/y^2 
\\  \hline
\left[1,0,0,-2\right] &(1,2,2,2)& 
\begin{array}{@{}l@{}}
\{ 2 \gamma' _2 , \left\{2 \gamma' _3,\left\{2 \gamma' _4\,\gamma' _1\right\}\right\} \} \\
\{ \gamma' _2 , \left\{2 \gamma' _3,\left\{2 \gamma' _4,\left\{\gamma' _1,\gamma'
   _2\right\}\right\}\right\} \} \\
\{ 2 \gamma' _2 , \left\{\gamma' _3,\left\{2 \gamma' _4,\left\{\gamma' _1,\gamma'
   _3\right\}\right\}\right\} \} \\
\end{array} &
y^4+3 y^2 + 6 + \dots
\\  \hline
\left[2,0,0,-2\right]&(0,2,2,2) & \{\{2 \gamma' _3,2 \gamma' _4\},2 \gamma' _2\}
 & -y^5 - 2 y^3 -3 y -\dots 
 \\  \hline
\left[2,-1,0,-1\right] &(0,2,1,1) &
\{ \{ \gamma' _3 , \gamma' _4\}, 2\gamma' _2\} & -y -1/y 
\\  \hline
\left[2,-1,0,-2\right] &(1,3,2,2) &
\begin{array}{@{}l@{}}
\{ \gamma' _1 , \left\{3 \gamma' _2,\left\{2 \gamma' _3,2 \gamma' _4\right\}\right\}  \} \\
\{ 2 \gamma' _2 , \left\{2 \gamma' _3,\left\{2 \gamma' _4,\left\{\gamma' _1,\gamma'
   _2\right\}\right\}\right\}  \} \\
\end{array}
 &  -y^5 - 3 y^3 -7 y -\dots  
 \\  \hline
 \left[2,0,-1,-1\right] &(0,1,2,1) &
\{ \{ 2\gamma' _3 , \gamma' _4\}, \gamma' _2\}& -y -1/y 
\\  \hline
\left[2,-1,-1,-1\right] &(1,2,2,1) & 
\{\gamma' _1,\{2 \gamma' _2,  \{2\gamma' _3, \gamma' _4\}\}\}
 & -y^3- y -1/ y -1/y^3 
 \\  \hline
\left[3,0,0,-3\right] &(0,3,3,3) & \{\{3 \gamma' _3 ,3 \gamma' _4\},3 \gamma' _2\} 
& y^{10}+2 y^8+5   y^6 +8 y^4 + 9 y^2 + 10 +\dots \\  \hline
\left[3,-1,0,-2\right] &(0,3,2,2) & 
\{3 \gamma' _2,\{2 \gamma' _3,2 \gamma' _4\} \}  
 & y^4+2y^2 +4+\dots 
 \\  \hline
\left[3,-1,-1,-1\right] &(0,2,2,1) &  \{ \gamma' _2, \{2 \gamma' _3, \gamma' _4\}\} 
& 1\\ \hline
\end{array} 
\end{equation*}
In all cases, 
the index $\Omega_c(\vec N')$ in the canonical chamber coincides with the Gieseker index.

%

\section{Large volume scattering diagram\label{sec_LV}}

\begin{figure}[t]
    \centering
\includegraphics[height=8cm]{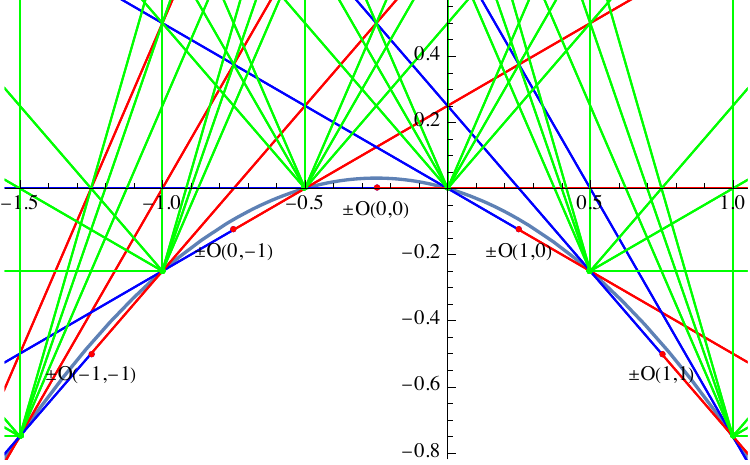}
    \caption{Scattering diagram in the $(x,y)$ plane at $m=\frac12$, including only primary scattering of initial rays.   \label{Scattxy_half}}
\end{figure}

In this section, we construct the scattering diagram for the  large volume stability condition
with central charge \eqref{ZLV}, which we repeat for convenience, 
\be
\label{ZLV0}
 Z^{\rm LV}_{T,m}(\gamma) = -r T(T+m) + d_1 T + d_2 (T+m) -\ch_2 
\ee
 Due to the identity (similar to  \cite[(4.26)]{Bousseau:2022snm})
\be
\label{Zpsi0}
\Re\left( e^{-\I\psi}  Z^{\rm LV}_{T,m}(\gamma)  \right) = 
\cos\psi\, \Re\left( Z^{\rm LV}_{T_\psi, m_\psi}(\gamma)  \right), \quad 
\begin{cases}
T_\psi :=& \Re T + \Im T \tan\psi  + \frac{\I \Im T}{\cos\psi}\\
m_\psi :=& \Re m +  \Im m \tan\psi + \frac{\I \Im m}{\cos\psi}
\end{cases}
\ee
there is no loss of generality in restricting attention to the scattering diagram 
for $\psi=0$, since the rays $\cR_\psi(\gamma)$ in the $T$-plane coincide with the rays $\cR_0(\gamma)$
in the $T_{\psi}$-plane after redefining $m \mapsto m_\psi$ (a trivial operation if $m$ is real). 
The identity \eqref{Zpsi0}
can be traced back to the fact both sides lead to the same values of the affine coordinates 
\be
\label{defxymu}
x=\frac{ \Re \left( e^{-\I\psi}T\right)}{\cos\psi}\ ,\quad y=-\frac{\Re\left( e^{-\I\psi} T_D\right)}{\cos\psi}, \quad \mu=\frac{ \Re \left( e^{-\I\psi}m\right)}{\cos\psi}\
\ee
such that the rays $\cR_\psi(\gamma)$ become straight segments in the $(x,y)$ plane,
\be
\label{straightray}
2r y + (d_1+d_2) x+ d_2 \mu-\ch_2 = 0
\ee
The region $\Im T>0$ in the $T$-plane maps to region above the parabola $y>-\frac12x(x+\mu)$, in the regime $\Im m>0$ that we consider (for $\Im m<0$ the large-volume region is $\Im(T+m)>0$, which has the same $(x,y)$ image), see Fig.~\ref{Scattxy_half}.

\medskip

For $\psi=0$, the real part of $T$ coincides with the coordinate $x$, and $m_1$ with $\mu$, so we set $T=x+\I t, m=\mu+\I m_2$. The coordinate $s=\frac{\Im T_D}{\Im T}$  determining the heart is $s=x+\frac{\mu}{2}+ \frac{m_2 x}{t}$,
so coincides with  $x+\frac{m}{2}$ when $m$ is real. The rays $\cR_0(\gamma)$ satisfy
\be
\Re Z(\gamma) = r (t^2-x^2) + (d_1+d_2-\mu r) x + \mu  d_2 - \ch_2 + m_2 r t = 0
\ee
For $r=0$, the ray is a vertical line at $x=\frac{\ch_2-\mu d_2}{d_1+d_2}$. For $r\neq 0$, it is
a branch of hyperbola centered at $(x,t)=(\frac{d_1+d_2-\mu r}{2r},-\frac{m_2}{2})$, intersecting the real axis $t=0$ at $x=\frac{d_1+d_2-\mu r \pm \sqrt{\Delta_{\mu}}}{2r}$ with
\be
\Delta_{\mu} = 4 ( d_1 d_2 -r \ch_2) + (d_2 -d_1 + \mu r)^2 
\ee
Note that the first term is proportional to the Bogomolov discriminant 
\eqref{slopedel}, and that  the same combination
appeared in \eqref{defdelmu} for $\eta=1, M_1=\mu$.  
For $\Delta_{\mu}>r^2 m_2^2$, the hyperbola consists of two branches intersecting
the axis $t=0$ at two distinct points, while for  $\Delta_{\mu}<r^2 m_2^2$, it consists
of a single branch intersecting the real axis twice. 

\begin{figure}[t]
    \centering
\includegraphics[height=8cm]{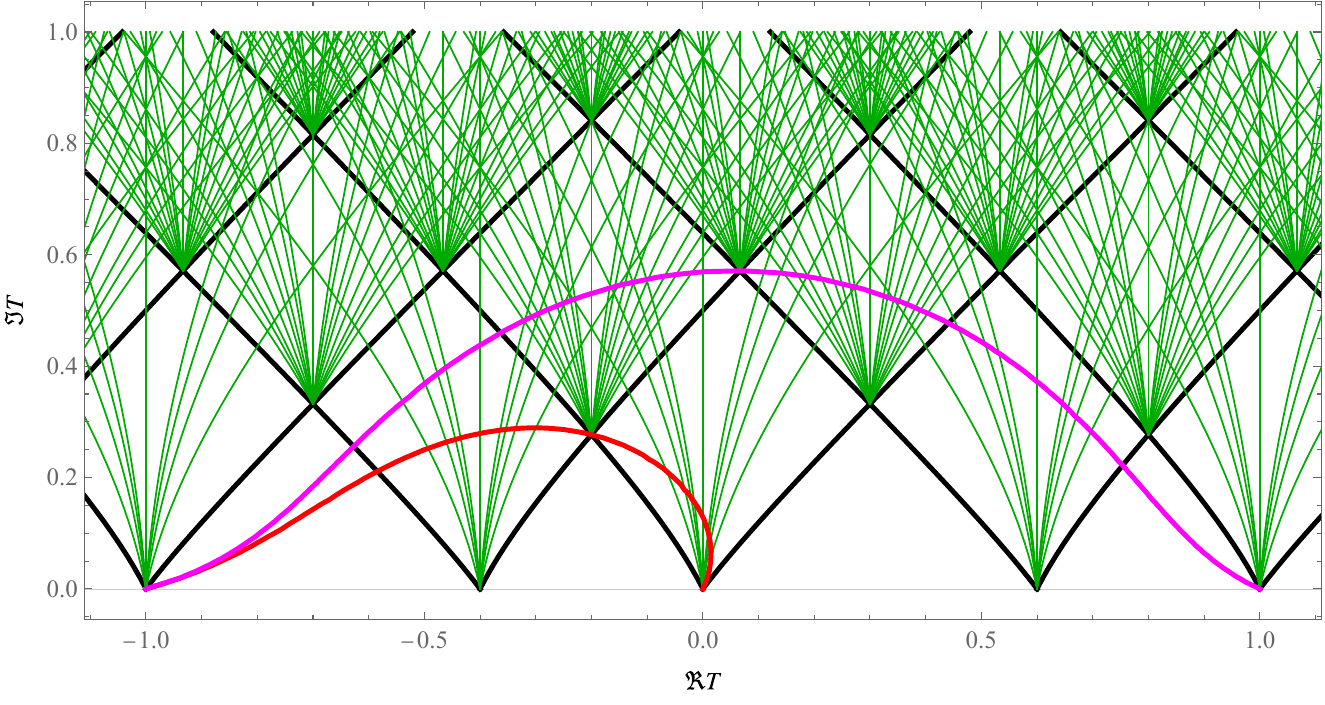}
    \caption{Initial rays for the large volume scattering diagram $\cD^\Lambda_{m,0}$ for fixed $m=0.4+ 0.3 \I$, and their primary scatterings. The  red and magenta lines denote the walls of marginal stability for  the binary trees $\{\cO(1,0), \cO(0,-1)[1]\}$
    and $\{\cO(1,1), \cO(0,-1)[1]\}$.
      \label{fig_LVmcx} }
\end{figure}

Walls of marginal stability where
$\Im Z(\gamma) \overline{Z(\gamma')}=0$ are in general cubic curves in the $(x,t)$ plane,
which reduce to half-circles centered on the real axis when $m_2=0$ (see Fig.~\ref{fig_LVmcx} for
an example of walls of marginal stability for $m_2>0$).
 For fixed charge $\gamma$, we expect a finite number of nested walls~\cite{arcara2013bridgeland,maciocia2014computing,arcara2015bridgeland}, outside which
the index $\Omega_{T,m}(\gamma)$ should reduce to a variant of the Gieseker index  
$\Omega_\eta(\gamma)$ with $\eta=1$, where the semi-stability condition for sheaves
with equal slope is in general different from the standard Gieseker prescription in footnote \ref{fooGieseker}. Indeed, for large $t$ the ratio 
\be
\frac{\Im Z(\gamma)}{\Re Z(\gamma)}
= \frac{\frac{d_1+d_2}{r}-m_1-2x}{t} + \frac{m_2(m_1+x-\frac{d_1}{r})}{t^2} + \cO(1/t^3)
\ee
is dominated by the slope $\mu_1(\gamma)=\frac{d_1+d_2}{2r}$, up to a charge independent constant; when the slopes are equal and $m_2\neq 0$, the subleading term requires comparing the ratios $\frac{d_1}{r}$ (or equivalently $\frac{d_2}{r}$), while $\ch_2$ enters only at sub-subleading order. When $m_2=0$, the subleading term vanishes and the sub-subleading term requires comparing $\frac{\ch_2-m_1 d_2}{r}$, which reduces to the Gieseker prescription for $m_1=0$ 
only (see \cite{Diaconescu:2007bf} for an early discussion of such discrepancies). 
We expect that these subtleties do not affect the index, at least for sufficiently generic charges.
 
\medskip

Focusing now on $\gamma=\pm \ch \cO(d_1,d_2)$,
the Bogomolov discriminant vanishes so the rays intersect the real axis at $x=d_1-\mu$ and $x=d_2$.
More precisely,  the ray $\cR_0(\cO(d_1,d_2))$ starts at $(\min(d_1-\mu,d_2),0)$ and bends to the left, while 
the ray $\cR_0(\cO(d_1,d_2))[1]$ starts at $(\max(d_1-\mu,d_2),0)$ and bends to the right. 
In the $(x,y)$ plane, the rays $\cR_\psi(\cO(d_1,d_2))$ and $\cR_\psi(\cO(d_1,d_2)[1]$ start at $x=d_1-\mu$ or $x=d_2$ along the parabola $y=-\frac12x(x+\mu)$, where the respective large volume central charges vanish. Note that the
ray $\cR_0(\cO(d_1,d_2))$ intersects $\cR_0(\cO(d'_1,d'_2))[1]$ if and only if $d_1\geq d_1'$ and $d_2\geq d_2'$.

\medskip

Similar to \cite[\S 6.2]{Bousseau:2022snm}, we regulate the scattering diagram using the cost function 
\be
\label{costF0}
 \varphi_x(\gamma)= d_1+d_2- r (2x+\mu)  = \frac{\Im Z^{\rm LV}(\gamma)}{\Im T} + m_2 (rx-d_2)
 \ee 
 which increases along any ray and is additive at each vertex.  Since the qualitative structure of the large volume scattering diagram $\cD^\Lambda_{m_1+\I m_2,\psi}$ turns out to be insensitive to $m_2$, we  henceforth set $m_2=0, m=\mu$ and $\psi=0$.

\subsection{\texorpdfstring{$m\notin \IZ$}{Non-integer m}}

\begin{figure}[t]
    \centering
\includegraphics[height=6cm]{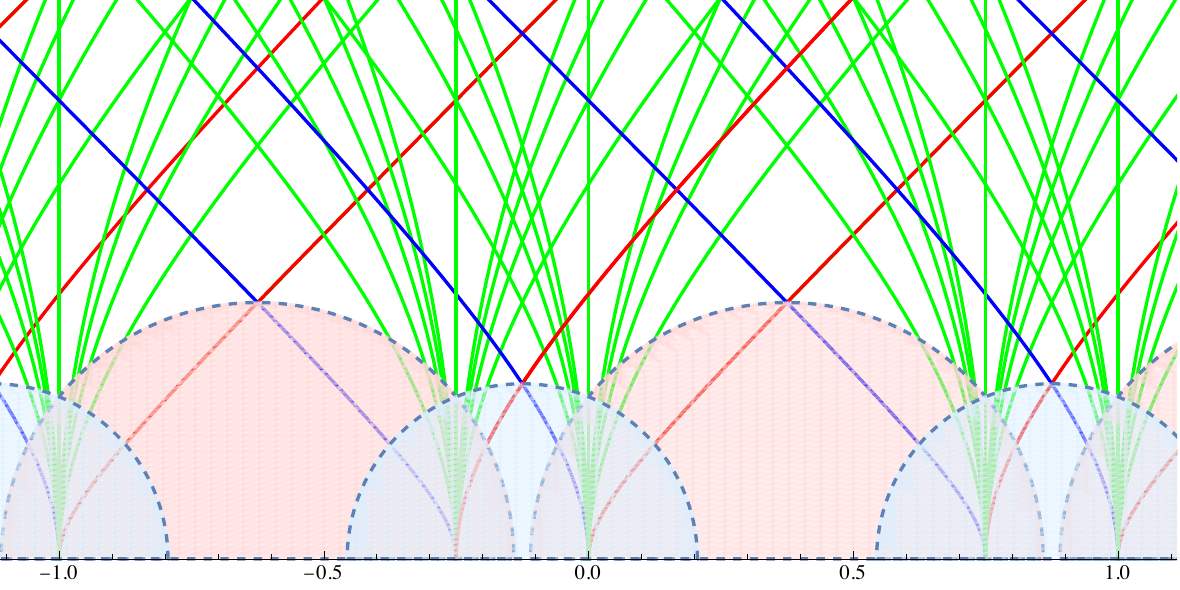}
    \caption{Initial rays for the large volume scattering diagram  in the $(x,t)$ plane at $m=1/4$. The shaded blue and red regions correspond to the validity of the  Ext-exceptional collection
    \eqref{dstrongcoll1} tensored with $\cO(k,k)$ and $\cO(k,k-1)$, respectively.  \label{LVScatt_quarter}}
\end{figure}

In view of the spectral flow invariance \eqref{Monm1m2},  we can assume $0<m<1$. One may check
that the region of validity (in the sense of footnote \ref{fooqvalid} on page \pageref{fooqvalid}) of the phase I quiver attached to the exceptional collection \eqref{strongcoll1}
 includes the interval $(-1,m-1)$ on the real axis, while the interval $(m-1,0)$ is covered by the
same exceptional collection \eqref{strongcoll1} 
tensored with $\cO(1,0)$ (see Fig.~\ref{LVScatt_quarter}). By applying a sequence of mutations, one finds that there is an infinite number of rays
emerging from $x=-m$ associated to $\cO(0,k)$ with $k>0$ and $\cO(0,-k)[1]$ with $k\geq 0$. Similarly, there is an infinite number of rays emerging from $x=0$, associated to $\cO(k,0)$ with $k>0$ and $\cO(-k,0)[1]$ with $k\geq 0$. The two set of rays emerging from $x=-m$ and $x=0$
are related by the combination of fiber/base duality $(x,m)\mapsto (x+m,-m)$ and derived duality $(x,m)\mapsto (-x,-m)$. For $m=1/2$, the diagram is symmetric with respect to the axis $x=-1/2$, see Fig.~\ref{LVScatt_half}. 

In fact, the infinite set of initial rays emerging from $x=-m$ can be 
can be understood as arising from the scattering of the two extremal rays $\cO(0,-1)[1]$
and $\cO(0,0)$ with charge 
$\gamma_1=[-1,0,1,0]$ and $\gamma_2=[1,0,0,0]$, described by the Kronecker quiver 
$K_2$ (also known as affine $A_1$ quiver).  Indeed, it is well known that 
the only populated dimension vectors 
for $K_2$ are of the form $(k,k+1), (k+1,k)$ and $(k,k)$, corresponding to outgoing rays 
\be 
k\gamma_1+(k+1)\gamma_2=\ch\cO(0,k), \quad 
(k+1)\gamma_1+k\gamma_2=\ch\cO(0,-k-1)[1]
\ee
with unit index, as well as $k\gamma_1+k\gamma_2 = [0,0,k,0]$ with index 
 $\Omstar(k\gamma_1+k\gamma_2)=-y-1/y$. For $k=1$, the latter is recognized 
 as the vertical ray $GV(0,1,0)$ associated the D2-D0 bound states with index $-y-1/y$  (see the first line in \eqref{D2D0F0}, with $d=0$). Similarly, the rays emerging from $x=0$ arise from the scattering of the two extremal rays $\cO(1,0)$ and $\cO(0,0)[1]$, with unit DT invariant, and also include the vertical ray $GV(1,0,0)$  with index  $-y-1/y$. 

Thus the full large volume scattering diagram $\cD^{\Lambda}_{m,0}$ 
can be obtained from the initial rays 
$\cO(k,k),$ $\cO(k,k-1)[1]$ starting from $x=k-m$ with $k\in \IZ$, slightly displaced into the lower 
half-plane, and 
$\cO(k+1,k), \cO(k,k)[1]$ starting from $x=k$,  slightly displaced into the lower half-plane. More precisely,
in the affine coordinates $(x,y)$, we can take the starting point for $\cO(k,k)$ and $\cO(k,k)[1]$
to be any point along the ray below the parabola $y=-\frac12 x(x+m)$, and similarly for 
$\cO(k+1,k)$ and $\cO(k+1,k)[1]$. A natural choice\footnote{This corresponds to the image in the $(x,y)$ plane 
of the points along the $\Pi$-stability slice where $\cO(k,k)$
and $\cO(k+1,k)$ are massless, for $m$ real and $\psi=0$. Along the large volume slice,
$\cO(k,k)$ and $\cO(k,k-1)[1]$ are instead massless at $x=k$ 
on the parabola $y=-\frac12 x(x+m)$, while $\cO(k,k)[1]$ and $\cO(k+1,k)$
are massless at $x=k-m$  along the same parabola.
} is the midpoint along the chord (see the red dots in Fig.  \ref{Scattxy_half}):
\be
\label{initraysxy0}
\begin{array}{lcl}
(x,y)=(k-\tfrac{m}{2},-\tfrac{k^2}{2}) &:& \cO(k,k) \ \mbox{and}\    \cO(k,k)[1]\\
(x,y)=(k+\tfrac{1-m}{2},-\tfrac{k(k+1)}{2}+\tfrac{m-1}{4}) &:& \cO(k+1,k)\  \mbox{and} \   \cO(k+1,k)[1]
\end{array}
\ee
Note that the initial points for $\pm \cO(k,k)$ lie
along the parabola $y=-\frac12 x(x+m)-\frac{m^2}{8}$ while the initial points for 
$\pm \cO(k+1,k)$ lie along the parabola $y=-\frac12 x(x+m)-\frac{(1-m)^2}{8}$.

More generally, when $m$ lies outside the interval $(0,1)$, 
decomposing $m=\lfloor m \rfloor + \Fr(m)$ and tensoring by $\cO(\lfloor m \rfloor,0)$, we get the initial rays 
\be
\begin{array}{lcl}
(x,y)=\left(k-\tfrac{ \Fr(m)}{2},-\tfrac{k^2}{2} - \tfrac{k}{2}
\lfloor m \rfloor + \tfrac14 \lfloor m \rfloor \Fr(m)
\right )&:& \pm  \cO(k+\lfloor m \rfloor ,k)
\\
(x,y)=\left(k+\tfrac{1- \Fr(m)}{2},
-\tfrac{k^2}{2} - \tfrac{k}{2} (1+ \lfloor m \rfloor )
-\tfrac14 ( 1- \Fr(m)) (1+\lfloor m \rfloor) \right) &:& 
\pm \cO(k+1+\lfloor m \rfloor ,k)
\end{array}
\label{initraysxy}
\ee
Their intersection on the parabola produces the initial rays at the boundary
of the upper half-plane,  
\be
\begin{split}
x=k:& \quad \cO(k+1+\lfloor m \rfloor+n,k), \quad \cO(k+\lfloor m \rfloor-n,k)[1], \quad  GV(1,0,k)  \\
x=k-\Fr(m):& \quad  \cO(k+\lfloor m \rfloor,k+n), \quad  \cO(k+\lfloor m \rfloor, k-n-1)[1], \quad GV(0,1,k+\lfloor m \rfloor)
 \end{split}
\ee
for all $n\geq 0$. Setting $s=x+\frac{m}{2}$ and shifting $k\mapsto k-\lfloor m \rfloor$ in the
second line, we recover the initial rays announced in \eqref{origLV}.

\medskip

It is now straightforward to determine the scattering sequences contributing to the Gieseker index for low enough Chern invariants, using as a cut-off the cost function \eqref{costF0}. 
As the ray crosses the parabola, or equivalently at $t=0$, one has 
\be
\begin{split}
 \varphi_{\rm in}(\pm \cO(k,k+\lfloor m \rfloor))=\Fr(m),  \quad 
  \varphi_{\rm in}(\pm \cO(k+1+\lfloor m \rfloor,k))=1-\Fr(m)
  \end{split}
\ee
Moreover, $\varphi_x(\gamma)$ vanishes along these rays (whose discriminant vanishes, namely $\ch_2 = d_1 d_2/r$) at
\be
(x,y) = \left( \frac{d_1+d_2}{2r} -m,  -\frac{d_1^2+d_2^2}{4r^2} + \frac{m(d_1-d_2)}{4r} \right)
\ee
such that $y+\frac12x(x+m) = -\frac18 ( m + \frac{d_2-d_1}{r})^2$. 
The same arguments 
as in \cite[\S4.2]{Bousseau:2022snm} guarantee that only a finite number of trees contribute to the
index, hence the  SAFC holds in the large volume slice.

\begin{table}
\begin{equation*}
\hspace*{-5mm}
\begin{array}{|l|l|l|l|}
\hline
\[r,d_1,d_2,\ch_2\]  & \mbox{Trees}  & \mbox{Range} & \Omega(\gamma) \\ \hline
 \[0,1,0,0\] & \{ \cO(0,0)[1],\cO(1,0) \} & \IR & K_2(1,1)=-y-1/y \\
 \hline
  \[0,1,1,0\] & \{ \cO(0,-1)[1],\cO(1,0) \} & [0,2] & K_4(1,1)=-y^3-y-\dots \\
 \hline
 \[0,1,2,0\] & \{ \cO(0,-2)[1], \cO(1,0) \} & [0,3] & \scalebox{.8}{$K_6(1,1)=-y^5-y^3-y-\dots$} \\
  \hline
 \[0,2,2,0\] &\begin{array}{@{}l@{}} \{ \cO(-1,-1)[1], \cO(1,1) \}  \\
       \{ \cO(0,-1)[1]^{\oplus 2}, \cO(1,0)^{\oplus 2} \}  \\
        \{ \cO(0,-2)[1], \cO(2,0) \}  
\end{array}  & 
\begin{array}{@{}l@{}} \[-2,2\] \\ \[0,2\] \\ \[0,4\] \end{array} 
& \scalebox{.8}{$ \begin{array}{@{}l@{}} K_4(2,2) +2 K_8(1,1) \\ = -y^9 - 3 y^7 - 4 y^5 \\
 - 4 y^3 - 4 y +\dots
\end{array}$}
\\ \hline
\[0,2,3,0\] &\begin{array}{@{}l@{}} \{ \{ \cO(-1,-1)[1], GV(0,1,0)\}, \cO(1,1) \}  \\
   \{ \cO(0,-3)[1], \cO(2,0) \}   \\
 \{  \{ \cO(0,-1)[1],\cO(1,0)^{\oplus 2}\},\cO(0,-2)[1]\} 
\end{array}  
&
\begin{array}{@{}l@{}} \[-\tfrac53,\tfrac52\] \\ \[0,5\] \\ \[0,\tfrac{10}{3}\] \end{array} 
&\scalebox{.7}{$  \begin{array}{@{}l@{}} (K_2(1,1)^2+ K_4(1,2)+ 1)  K_{10}(1,1)\\
=-y^{13} - 3 y^{11} - 8 y^9 - 10 y^7 \\ - 11 y^5 - 11 y^3 - 11 y+\dots
\end{array}$}
\\ \hline
\[1,0,0,-1\] & \{ \cO(0,-1), GV(0,1,-1)\}  & \IR^+ & K_2(1,1)^2 =y^2+2+\dots
\\ \hline
\[1,0,0,-2\] & 
\begin{array}{@{}l@{}}
\{ \cO(0,-1), GV(0,1,-2)\} \\
\{ \cO(-1,0), GV(1,0,-2)\} \\
\{ \cO(-2,-2)[1], \cO(-1,-1)^{\oplus 2} \} 
\end{array}
&\begin{array}{@{}l@{}} \[-1,\infty\] \\ \[-\infty,1\] \\ \[-1,1\] \end{array} 
& \scalebox{.8}{$\begin{array}{@{}l@{}} 2K_2(1,1)^2 + K_4(1,2) \\ =y^4+3y^2+6+\dots \end{array}$}
\\ \hline
\[1,0,0,-3\] & 
\begin{array}{@{}l@{}}
\{ \cO(0,-1), GV(0,1,-3)\} \\
\{ \cO(-1,0), GV(1,0,-3)\} \\
\{ \cO(-1,-1), \{ \cO(-1,-2), \cO(-2,-3)[1]\} \} \\
\{ GV(0,1,-2), \{ GV(1,0,-2), \cO(-1,-1)\}\}
\end{array}
&\begin{array}{@{}l@{}} \[-2,\infty\] \\  \[-\infty,2\] \\ \[-2,2\] \\ \[-\frac12,\infty\]  \end{array} 
&  \scalebox{.7}{$ \begin{array}{@{}l@{}}  2K_2(1,1)^2 + K_4(1,1)^2+ K_2(1,1)^4  \\
=y^6+3y^4+9y^2+14+\dots\end{array} $}
\\ \hline
\[2,0,0,-2\] & \{ \cO(0,-1)^{\oplus 2}, GV(0,2,-2)\} & \IR^+ & -y^5-2y^3-3y-\dots
\\ \hline
\[2,-1,0,-1\] & \{ \cO(0,-1), \cO(-1,1)\} & \IR^+ & K_2(1,1)=-y-1/y
\\ \hline
\[2,0,-1,-1\] & \{ 2\cO(0,-1), GV(0,1,-1) \} & \IR^+ & K_2(2,1)=-y-1/y
\\ \hline
\[2,-1,-1,-1\] & \{\cO(-1,-1)^{\oplus 3}, \cO(-2,-2)[1]\} & [-1,1] & K_4(3,1)=-y^3-y-\dots
\\ \hline
\[ 3,1,1,-1\] & \{ \cO(0, 0)^{\oplus 4}, \cO(-1,-1)[1] \} & [-1,1] & K_4(4,1)=1
\\ \hline
\[ 3,2,2,0\] & \{ \cO(1, 0)^{\oplus 2}, \cO(0,2) \} & \IR^+   & K_2(2,1)=1
\\ \hline
\[ 5,1,2,-2\] & \{ \cO(0, 0)^{\oplus 6}, \cO(-1,-2)[1] \} & [-1,2]&K_6(6,1)=1 
\\ \hline
\[ 5,3,4,0\] & \{ \cO(1, 0)^{\oplus 3}, \cO(0,2)^{\oplus 2} \} & \IR^+ &K_2(3,2)=1 \\
\hline
\end{array}
\end{equation*}
\caption{Trees contributing to the index $\Omega_{s,t}(\gamma)$ at $t\to \infty$, for 
$m$ in the range specified in the third column. As $m$ reaches the ends of the interval,
the last wall of marginal stability shrinks to zero radius. \label{tab_omLV}}

\end{table}

In Table \ref{tab_omLV}, we display the trees contributing to the index at large $t$, for
$m$ in an interval containing $m=\frac12$. For $m$ outside this range, the index is still the same
at large $t$, but BPS states arise from a different set of constituents. For example, 
\begin{itemize}
\item For $\gamma=[0,1,0,0]$, the ray $GV(1,0,0)$ exists for all $m$, but it arises as a bound state from different constituents: e.g $\{\cO(1,0),\cO(0,0)[1]\}$ when $0<m<1$, or more generally 
\be
\{\cO(\lfloor m \rfloor+1,0),\cO(\lfloor m \rfloor,0)[1]\}
\ee
 when $m\notin \IZ$.

\item For $\gamma=[0,1,1,0]$ and $\lfloor m \rfloor$ even, the contributing tree is 
\be
\{\cO(1+\tfrac12 \lfloor m \rfloor, -\tfrac12 \lfloor m \rfloor), \cO(\tfrac12 \lfloor m \rfloor, -1-\tfrac12 \lfloor m \rfloor)[1]\}
\ee
For odd $\lfloor m \rfloor$, the contributing tree is instead
\be
\{\cO(\tfrac12+\tfrac12 \lfloor m \rfloor, \tfrac12-\tfrac12 \lfloor m \rfloor), \cO(-\tfrac12+\tfrac12 \lfloor m \rfloor, -\tfrac12-\tfrac12 \lfloor m \rfloor)[1]\}
\ee
Writing $m=2d+p+\Fr(m)$ with $p\in\{0,1\}$, this can be rewritten in both cases as 
\be
\{ \cO(d+1,-d), \cO(d,-1-d)[1] \}
\ee
When $m$ varies in an interval $]2n,2n+2[$, the tree stays the same, but the circle of marginal stability shrinks as $m$ approaches the end of the interval. For $m$ even integer, the state arises as an initial ray. 

\item 
For $\gamma=[0,1,2,0]$ and $m=3d+p+\Fr(m)$ with $d=\lfloor \frac{m}{3} \rfloor$, the contributing tree is
\be
\{ \cO(d+1,-2d), \cO(d,-2-2d)[1] \}
\ee
This is independent of $p\in\{0,1,2\}$, but jumps as $d$ changes. 

\item For $\gamma=[1,0,0-1]$, we find the following trees, setting $d=\lfloor m \rfloor$ with $m>0$,
\be
\begin{cases}
0<m<1 :& 
\{\cO(0, -1), \{\cO(-1, -1), \cO(-1, -2)[1]\}\}
\\
1<m<2 :& 
\{\cO(0, -1), \{\cO(-1, -2), \cO(-1, -3)[1]\}\}
\\
m>2: & \{  \{ \cO(0,-d)^{\oplus d},   \cO(0,-d-1)[1]^{\oplus d+1} \},  \{ \cO(0,-d-1), \cO(-1,-d-2) \} \}
\end{cases}
\ee
whose first node is the same $\{\cO(0,-1), GV(0,1,-1)\}$ for any $m>0$. 
For $m<0$, we have the same trees upon setting $d=\lfloor -m \rfloor$ and applying fiber/base duality.
\end{itemize}

\subsection{\texorpdfstring{$m\in\IZ $}{Integer m}}

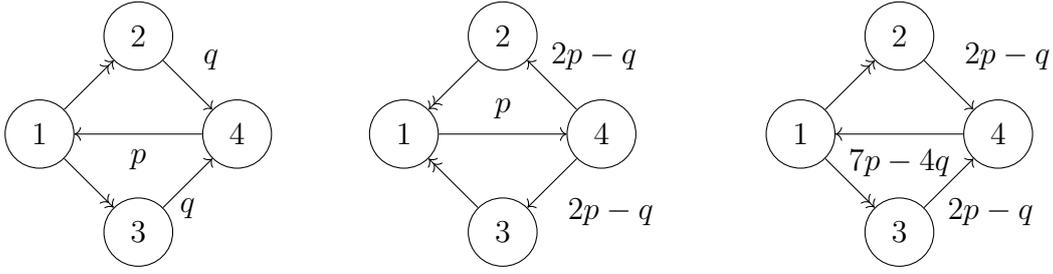
\begin{figure}
 \begin{center}
\begin{tikzpicture}[inner sep=2mm,scale=1.3]
  \node (a) at ( -1,0) [circle,draw] {$1$};
  \node (b) at ( 0,1) [circle,draw] {$2$};
  \node (c)  at ( 0,-1) [circle,draw] {$3$};
  \node (d)  at ( 1,0) [circle,draw] {$4$};
 \draw [->>] (a) to node[auto] {$ $} (b);
 \draw [->>] (a) to node[auto] {$ $} (c);
 \draw [->] (b) to node[auto] {$q$} (d);
 \draw [->] (c) to node[below] {$q$} (d);
  \draw [->] (d) to node[auto] {$p$} (a);
\end{tikzpicture}
\hspace*{1cm}
\begin{tikzpicture}[inner sep=2mm,scale=1.3]
  \node (a) at ( -1,0) [circle,draw] {$1$};
  \node (b) at ( 0,1) [circle,draw] {$2$};
  \node (c)  at ( 0,-1) [circle,draw] {$3$};
  \node (d)  at ( 1,0) [circle,draw] {$4$};
 \draw [->>] (b) to node[auto] {$ $} (a);
 \draw [->>] (c) to node[auto] {$ $} (a);
 \draw [->] (d) to node[above] {$\quad\quad\ \ 2p-q$} (b);
 \draw [->] (d) to node[auto] {$2p-q$} (c);
  \draw [->] (a) to node[auto] {$p$} (d);
\end{tikzpicture}
\hspace*{1cm}
\begin{tikzpicture}[inner sep=2mm,scale=1.3]
  \node (a) at ( -1,0) [circle,draw] {$1$};
  \node (b) at ( 0,1) [circle,draw] {$2$};
  \node (c)  at ( 0,-1) [circle,draw] {$3$};
  \node (d)  at ( 1,0) [circle,draw] {$4$};
 \draw [->>] (a) to node[auto] {$ $} (b);
 \draw [->>] (a) to node[auto] {$ $} (c);
 \draw [->] (b) to node[auto] {$2p-q$} (d);
 \draw [->] (c) to node[below] {$\quad\quad\ \  2p-q$} (d);
  \draw [->] (d) to node[auto] {$7p-4q$} (a);
\end{tikzpicture}
\end{center}
\caption{Images of the phase I quiver under mutations $\mu_1$ and $\mu_2\mu_3$.
\label{fig_itermut}
}
\end{figure}

\begin{figure}[t]
    \centering
\includegraphics[height=8cm]{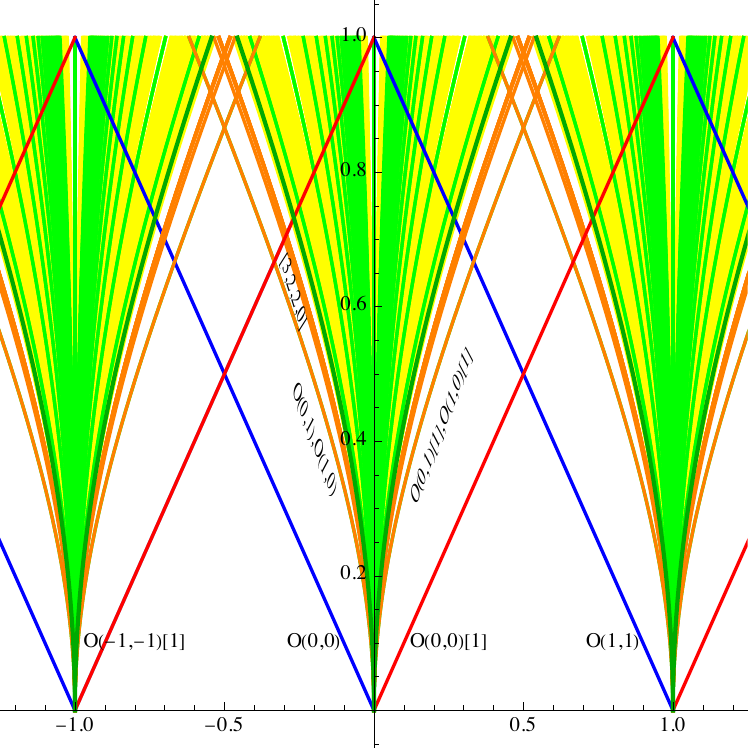}
    \caption{Initial rays in the $(x,t)$ plane for $m=0$. The orange rays are obtained from the phase I collection by iterating the mutations $\mu_1$ and $\mu_2\mu_3$. The dense yellow rays arise from the scattering of $\cR(\cO(1,0))$ and $\cR(\cO(0,-1)[1])$ with $|\langle\gamma_1,\gamma_2\rangle|=4$.  \label{LVScatt0}}
\end{figure}

When $m$ is an integer, the initial rays at $x\in \IZ$ and $x\in \IZ - \Fr(m)$ coalesce and
the structure of the scattering diagram becomes more complicated. In particular,
assuming $m=0$ without loss of generality, we see that four initial rays $\cO(0,0), \cO(0,0)[1],\cO(1,0), \cO(0,-1)[1]$ intersect at the origin $(x,y)=0$. In order to determine the outcoming rays, we start from
the phase I collection and apply mutations $\mu_1$ and $\mu_2 \mu_3$ alternatively
(see Fig.~\ref{fig_itermut}). Under these mutations, the charge 
$\gamma'_4$ attached to node 4 and the number of arrows $1\leftrightarrow 2,1\leftrightarrow 3$  remain unchanged (with the direction of the arrow being inverted at each step), but the 
charge vectors $\gamma'_{1,2,3}$ attached to nodes 1,2,3 and the numbers $p$ of arrows $4\leftrightarrow 1$ and $q$ of arrows $2\leftrightarrow 4,  3\leftrightarrow 4$, change according 
to 
\be
\begin{pmatrix} q \\ p \end{pmatrix} \mapsto 
\begin{pmatrix} -1 & 2 \\ -4 & 7 \end{pmatrix}\begin{pmatrix} q \\ p \end{pmatrix} , \quad 
\begin{pmatrix} \gamma'_1\\ \gamma'_2 \\ \gamma'_3 \end{pmatrix} \mapsto 
\begin{pmatrix} 7 & 2 & 2  \\ -2 & -1 & 0 \\ -2 & 0 & -1  \end{pmatrix}
\begin{pmatrix} \gamma'_1\\ \gamma'_2 \\ \gamma'_3 \end{pmatrix} 
\ee
Starting with the phase I collection \eqref{dstrongcoll1} with $(p,q)=(4,2)$ at $k=0$ and 
denoting by $(\alpha_{2k+1}, -\alpha_{2k})$ the charge vectors $(\gamma'_1,-\gamma'_3)$ after the $k$-th iteration, we find an infinite series of simple rays emitted at $(x,y)=(0,0)$, with charges
\be
\alpha_1=\cO(0,0), \ \alpha_2=\cO(0,1),\  \alpha_3=[3,2,2,0],\  \alpha_4=[5,4,3,0], \ 
\alpha_5=[17,12,12,0], \dots 
\ee
along with their images under fiber/base duality. Explicitly,
\be
r_k = \frac{(1 + \sqrt{2})^k + (1 + \sqrt{2})^{-k}}{2^{1+\Fr(k/2)}} , \qquad
(d_{1k},d_{2k}) = \frac{(1 + \sqrt{2})^k - (1 +
\sqrt{2})^{-k}}{2^{3/2+\Fr(k/2)}} \pm (-1)^{(k+1)/2} \Fr(k/2) .
\ee
such that $(d_1,d_2)/r_k$ tens to $1/\sqrt2$.
Similarly, there is a discrete set of rays $\tilde\alpha_k$ with negative rank emitted at $(x,y)=(0,0)$, related to $\alpha_k$ by derived duality. 
Beyond these discrete sets of rays, the diagram becomes dense and we shall not attempt to characterize it further,  see Fig.~\ref{LVScatt0}.

\medskip

\section{Scattering diagram in the \texorpdfstring{$\Pi$}{Pi}-stability slice \label{sec_Pi}}

In this section, we combine our understanding of the large volume and quiver scattering diagrams
with the action of $\Gamma_0(4)$ (or rather, the monodromy group $\Gamma$)
to elucidate the structure of the scattering diagram along the $\Pi$-stability slice.
In view of the complications introduced by the branch cuts, we do not aim for mathematical rigour, but rather propose a compelling physical picture supported by numerous numerical checks.
In \S\ref{ssec_safc} we outline the proof that this picture is correct, together with the proof of the Split Attractor Flow Conjecture, which go hand in hand.
Throughout, we assume $0<\Re m<1$ and $\Im m\geq 0$.

\subsection{Uncovering the scattering diagram}

By the same arguments as in \cite{Bousseau:2022snm}, we know that initial rays can only be emitted from conifold points along the boundary. By invariance under $\Gamma_0(4)$, this includes all rational points $\tau=\frac{p}{q}$ with $(p,q)$ coprime and $q\neq 0 \mod 4$, while rational points with $4|q$ correspond to images of the large volume point $\tau=\I\infty$. However, due to the existence of $\IZ_2$ ramification points at all $\Gamma_0(4)$ images of $\tau_B$, each fraction $\frac{p}{q}$ actually corresponds to a tower of conifold points indexed by $\IZ^4$. On the principal sheet,
obtained by analytic continuation from $\tau=\I\infty$ without crossing any cut, the initial rays 
at $\tau=0$ correspond to all homological shifts $\cO(0,0)[n]$ of the structure sheaf, while the initial rays at $\tau=\frac12$ correspond to $\cO(1,0)[n]$ or $\cO(0,1)[n]$, depending on the angular sector (see Fig.~\ref{fig_iniPi}). Initial rays on the principal sheet near other conifold points $\tau=\frac{p}{q}$
can be obtained by applying  the autoequivalences $ST_{\cO(0,0)}$ and $\cO(1,1)\otimes -$ generating $\Gamma_0(4)$. 

\medskip

For $|\psi|$ small enough (and $\Re m$ far enough from the endpoints of the interval
$0<\Re m<1$, depending on $\psi$), numerical analysis indicates that the only initial rays which can reach
the large volume point  $\tau=\I\infty$ are the rays $\cO(k,k)$ and $\cO(k,k)[1]$ starting 
from $\tau=k$ on the principal sheet, and the rays $\cO(k+1,k)$ and $\cO(k+1,k)[1]$ starting 
from $\tau=k+\frac12$ on the principal sheet. As discussed in \S\ref{sec_LV}, 
those are the same rays which generate the infinite collections of initial rays 
emanating from $x\in \IZ$ and $x\in \IZ-m$ in the large volume scattering diagram for $m$ real. 
In fact, using the affine coordinates $(x,y)$ and parameter $\mu$ introduced in \eqref{defxymu}, the part of the 
$\Pi$-stability scattering diagram which is connected to $\tau=\I\infty$ is identical
to the large volume scattering diagram $\cD^{\Lambda}_\mu$, except that the origin of the
initial rays is shifted to 
\be
\label{tauinitk}
\begin{split}
\tau=k :& \left( k+\cV_\psi , -\frac12 k(k+\mu) - k  \cV_\psi )\right) \\
\tau=k+\tfrac12 :& \left( k+\tilde\cV_\psi, 
-\frac12 k(k+\mu) - (k+\tfrac12)   \tilde\cV_\psi \right) 
\end{split}
\ee
where
\be
\cV_\psi := \frac{\Re[\I e^{-\I\psi} \cV(m)]}{\cos\psi}, \quad  \tilde\cV_\psi := \frac{\Re[\I e^{-\I\psi} \tilde\cV(m)]}{\cos\psi}
\ee
For $m$ real and $\psi=0$,  one has $(\cV_0,\tilde\cV_0)=(-\frac{m}{2},\frac{1-m}{2})$
so the initial points coincide with \eqref{initraysxy0}, namely the midpoint along the chord between $k-\mu<x<k$ and $k<x<k+1-\mu$  below the parabola. 
For later reference, we also note that the `jagged parabola' joining the successive points \eqref{tauinitk} is convex only if the following two conditions are obeyed
\be
\label{convexcond}
\cV_\psi ( 2 \cV_\psi + \mu +1 ) \leq \tilde \cV_\psi ( 2 \cV_\psi+\mu), \quad 
\tilde\cV_\psi ( 2 \tilde \cV_\psi + \mu -2 ) \leq \cV_\psi ( 2 \tilde\cV_\psi+\mu-1)
\ee
Indeed, the first inequality saturates when the points associated to $\tau\in\{k, k+\frac12,k+1\}$
become aligned, while the second saturates when the points  $\tau\in\{k-\frac12, k,k+\frac12\}$ become aligned. Furthermore, the two equations are exchanged under $(\cV_\psi,\tilde\cV_\psi,\mu)\mapsto 
(-\tilde\cV_\psi,-\cV_\psi,1-\mu)$.
We denote by $\cI_{\rm cvx}$ the range of $\psi$ such that the convexity conditions \eqref{convexcond} are satisfied. This range is obviously contained in the range where the initial rays \eqref{tauinitk} are ordered horizontally, namely 
$\cV_\psi<\tilde\cV_\psi<\cV_\psi+1$, and one may check (numerically) that it 
always contains $\psi=0$.

\medskip

\begin{figure}[t]
    \centering
\includegraphics[height=8cm]{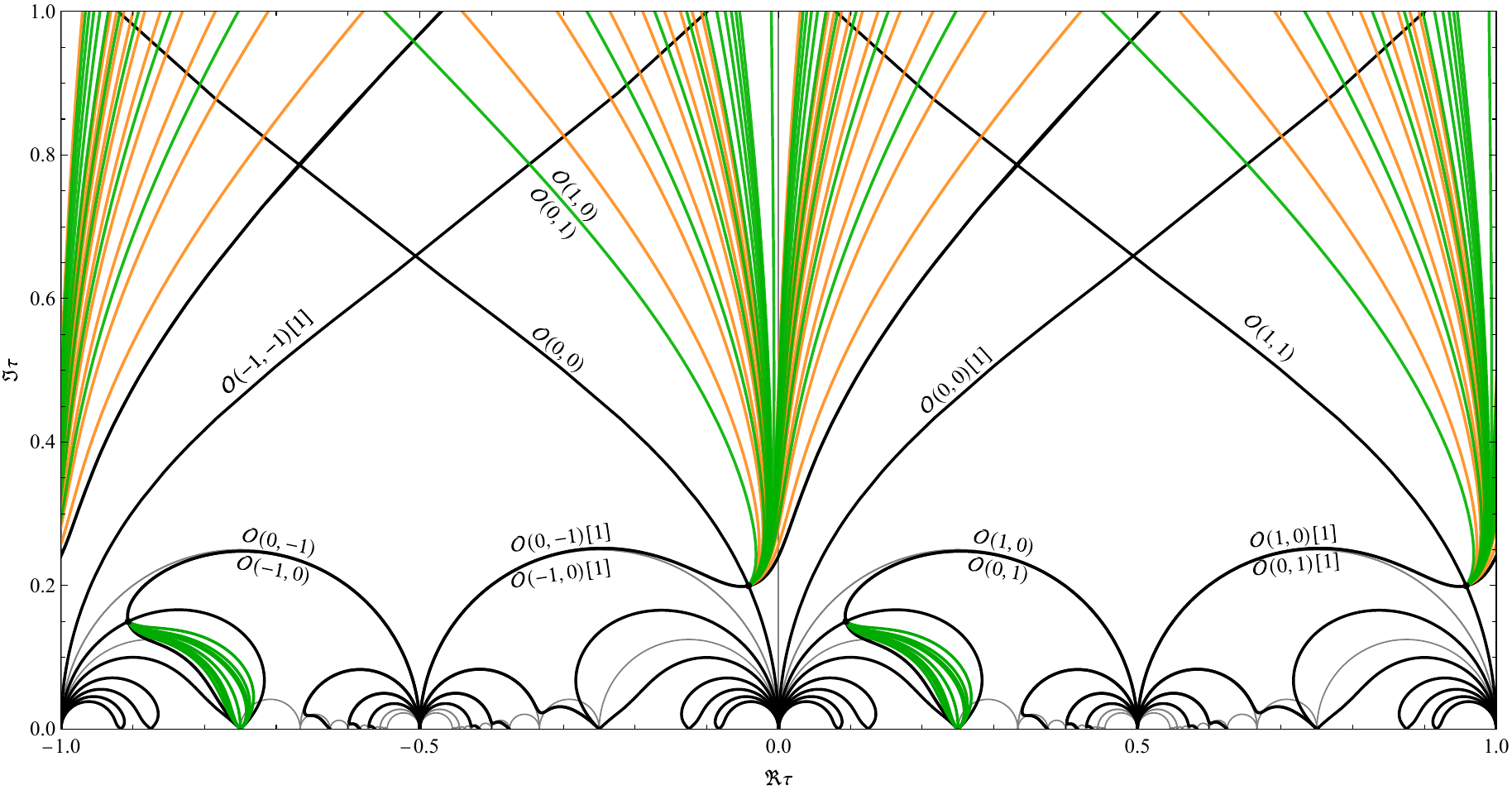}
    \caption{Scattering diagram  $\cD^\Pi_{m,\psi}$ for $m=0$, $\psi=0.01$, keeping only primary scatterings. Note that rays related by fiber-based duality are indistinguishable, with further degeneracies when $r=0$. As $\psi\to 0$, the intersection point between $\cO(0,-1)[1], \cO(-1,0)[1]$
    and $\cO(0,0)$ coalesces with $\tau=0$, similar to Fig. \ref{LVScatt0}.
   \label{fig_raypsim0} }
\end{figure}

As long as the initial
points stay on the aforementioned chords, i.e.\ when
\be
\label{ineqsimp}
-\mu<\cV_\psi<0 \quad \mbox{and} \quad 0<\tilde\cV_\psi<1-\mu
\ee
 the topology of the scattering diagram stays unaffected, and is identical
to the large volume diagram (see Figs.~\ref{fig_raypsi0} and \ref{fig_raypsi1}). This description clearly
breaks down for $m=0$ or $m=1$, where one of the ranges in \eqref{ineqsimp} becomes empty. 
Focusing for definiteness on the case $m=0$, we have $(\cV_\psi,\tilde\cV_\psi)=
(\frac{4G}{\pi^2} \tan\psi,\frac12)$ where $G$ is Catalan's constant. For arbitrarily small 
$\psi>0$, we find that the initial rays $\cO(0,1), \cO(1,0)$ starting from $\tau=\frac12$ approach $\tau=0$ before curling
back to another large volume point 
$\tau=\frac14$, while the initial rays $\cO(0,-1)[1], \cO(-1,0)[1]$ starting from $\tau=-\frac12$
approach $\tau=0$ and narrowly
 escape to $\tau=\I\infty$, crossing the ray $\cO(0,0)$ emanating from $\tau=0$ on their way (see Fig. \ref{fig_raypsim0}). The scattering of $\cO(0,-1)[1]^{\oplus n_1}$, $\cO(0,0)^{\oplus n_2}$ and   $\cO(-1,0)[1]^{\oplus n_3}$ is described by a 3-node quiver with $\kappa_{12}=\kappa_{32}=2$, leading to an infinite number of outgoing rays, all of them with $\ch_2=0$, which reproduce the dense set of rays in the large volume scattering diagram Fig. \ref{LVScatt0}. 
 
 \medskip

Returning to the case where $\Re m$ is not integer, we
find that as $|\psi|$ increases beyond certain critical values,
such that either of the inequalities \eqref{ineqsimp} get violated,
the initial points cross the primary intersection points at $x=k$ or $k-\mu$, and the topology of the scattering diagram changes: this occurs at phases $\psi^-_{\rm cr}<0<\psi^+_{\rm cr}$ where $\cV_\psi=-\mu$ and $\cV_\psi=0$ respectively, and phases $\tilde\psi^-_{\rm cr}<0<\tilde\psi^+_{\rm cr}$ where $\tilde\cV_\psi=0$ and $\tilde\cV_\psi=1-\mu$, respectively.
Explicitly,
\be
\label{psicritical}
\begin{aligned}
\psi^-_{\rm cr} & = \arctan\left( \frac{\Im[\cV(m)-\I m]}{\Re[\cV(m)-\I m]}\right),  \quad &
\tilde\psi^-_{\rm cr} & = \arctan\left( \frac{\Im\tilde\cV(m)}{\Re\tilde\cV(m)}\right),
\\
\psi^+_{\rm cr} & = \arctan\left( \frac{\Im \cV(m)}{\Re\cV(m)}\right),  \quad &
\tilde\psi^+_{\rm cr} & = \arctan\left( \frac{\Im[ \tilde\cV(m)+ \I(1-m)]}{\Re[\tilde\cV(m)+\I(1-m)]}\right)
\end{aligned}
\ee
For concreteness we consider positive $\psi$, and assume $\psi^+_{\rm cr}<\tilde\psi^+_{\rm cr}$, which holds for $m\in(0,1/2)$, and more generally for small positive $\Re m$. For $\psi>\psi^+_{\rm cr}$, the initial rays $\cO(k+1,k)$ starting from $\tau=k+\frac12$ no longer reach $\tau=\I\infty$ but instead curl back and escape to another large volume point $\tau=k+\frac14$, while the rays $\cO(k,k+1)[1]$ (which previously were curling back to $\tau=k+\frac34$) now escape to $\tau=\I\infty$ (see Fig.~\ref{fig_raypsi2}).  For $\psi>\tilde\psi^+_{\rm cr}$, a similar transition occurs where $\cO(k,k)$ crosses the cut and falls back on $\tau=k-\frac14$ (on a different sheet) and $\cO(k+1,k-1)[1]$ instead comes from $\tau=k$ on a different sheet, emerges into the principal sheet by crossing a cut, and escapes to $\I\infty$. At that point, the rays $\cO(0,0)$ and $\cO(-1,0)[1]$ (along with their images under the monodromy $M_B(-1)$) surround the ramification point at $\tau_B-1$ (see Fig.~\ref{fig_raypsi3}). Those are precisely the rays  associated to the objects in the orbifold collection \eqref{dstrongcoll2}, which we recall for convenience,
\be
\label{orbcoll1}
\gamma_1:\cO(0,0) , \quad  \gamma_2:\cO(-1,0)[1], \quad \gamma_3:\cO(1,-1)[1] , \quad \gamma_4:\cO(0,-1)[2]  
\ee
Note that their central charges at $\tau=\tau_B-1$ are equal in pairs, namely $\{\frac{m-1}{2}, -\frac{m}{2},\frac{m-1}{2},-\frac{m}{2}\}$, and become all equal to $-1/4$ when $m=1/2$. 
A portion of the orbifold scattering diagram embeds inside the exact scattering diagram by equating $\zeta_i$
with \eqref{zetafromZ}, or in terms of the affine coordinates~\eqref{zetauvorb}, 
\be
u=x-2y+\tfrac{\mu}{2}, \quad v=2y+\tfrac{1-\mu}{2}
\ee
The monodromy $M_{B}(-1)$ maps $(T,T_D)\mapsto (-1-T,\tfrac{1-m}{2}-T_D)$ hence $(u,v)\mapsto(-u,-v)$, so the branch point  maps to the origin in the $(u,v)$ plane. 
Unlike in the orbifold scattering diagram of \S\ref{sec_quivo}, the rays $\cR(\gamma_1)$ and $\cR(\gamma_2)$ no longer come from infinity, but start at $\tau=0$ and $\tau=-\frac12$ (approached on the principal sheet), which map to
\be\label{startuv-1}
\gamma_1:  (u,v) =\left( \tfrac{\mu}{2} + \cV_\psi,  \tfrac{1-\mu}{2} \right) \ , \qquad 
\gamma_2:  (u,v) = \left( \tfrac{\mu}{2} ,  \tfrac{1-\mu}{2}- \tilde\cV_\psi \right)
\ee
while the starting points for $\gamma_3,\gamma_4$ are the images under $(u,v)\mapsto(-u,-v)$. For $\psi=0$ and $m$ real,  the rays start at the midpoints on the rectangle bounded by the lines 
$u=\pm  \tfrac{\mu}{2}, v=\pm  \tfrac{1-\mu}{2}$, so do not intersect. The rays 
$\{\gamma_1,\gamma_2\}$ (and similarly 
$\{\gamma_3,\gamma_4\}$) start intersecting when $\cV_\psi\geq 0$, while the rays
 $\{\gamma_2,\gamma_3\}$    (and similarly 
$\{\gamma_4,\gamma_1\}$) start intersecting when $\tilde\cV_\psi\geq 1-\mu$.

\medskip

Similarly, the rays $\cO(1,0)$ and $\cO(1,-1)[1]$ (along with their two images under the
monodromy $\widetilde{M}_B(0)$) surrounding   the ramification point at $\tau=\tilde\tau_B=\frac{\tau_B-1}{4\tau_B-3}$
are recognized as those associated to the Ext-exceptional collection 
\be
\label{orbcoll2}
\gamma'_1:\cO(1,0), \quad \gamma'_2:\cO(1,-1)[1], \quad  \gamma'_3:\cO(0,1)[1] , \quad \gamma'_4:\cO(0,0)[2]  
\ee
related to \eqref{orbcoll1}  by applying fiber/base duality and then tensoring with $\cO(1,0)$. 
Affine coordinates $(u,v)$ defined by $\zeta'_1=v-\frac{\mu}{2}, \zeta'_2=u+\frac{\mu-1}{2}$ (related to \eqref{zetauvorb} by $\mu\mapsto 1-\mu$) are related to $(x,y)$ by
\be
u=-2y+ \tfrac{\mu-1}{2}, \quad v=x+2y+\tfrac{\mu}{2} .
\ee
The rays now originate from $\tau=\frac12$ (on the principal sheet) and $\tau=0$ (across the branch cut from $\tau=\tilde\tau_B$ to $\tau=\frac12$), which map to 
\be\label{startuv-2}
\gamma'_1:  (u,v) = \bigl( \tfrac{\mu-1}{2} +\tilde\cV_\psi , 
 \tfrac{\mu}{2} \bigr), \qquad 
\gamma'_2:  (u,v) = \bigl(  \tfrac{1-\mu}{2}, -\tfrac{\mu}{2} - \cV_\psi \bigr)
\ee
For $\psi=0$ and $m$ real, the rays start at the midpoints on the rectangle, so do not intersect. The rays $\{\gamma'_1,\gamma'_2\}$ start intersecting
for $\tilde\cV_\psi\geq 1-\mu$, while $\{\gamma'_2,\gamma'_3\}$ start intersecting for $\cV_\psi\geq 0$.\footnote{As one can see in Figure~\ref{fig_raypsi2}, in the regime where $\cV_\psi>0$ and $\tilde\cV_\psi<1-\mu$, the rays of bound states of $\gamma'_2$ and $\gamma'_3$ do not reach the $\tau=\I\infty$ large volume region.}

\begin{figure}[t]
    \centering
\includegraphics[height=6cm]{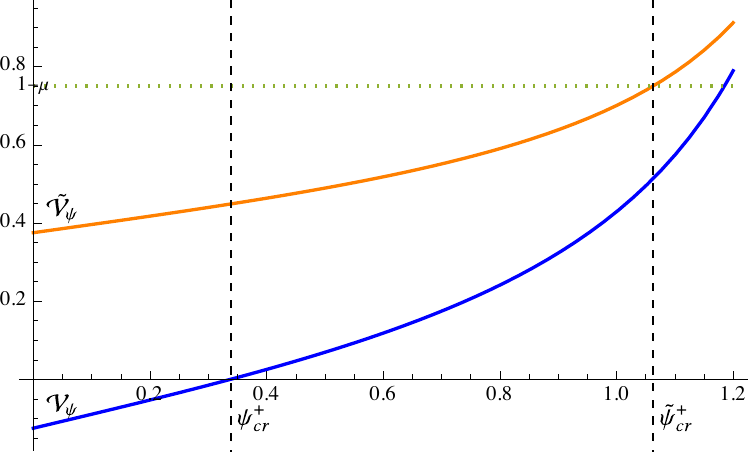}
    \label{PiScatt5-5}
    \caption{$\cV_\psi$ and $\tilde\cV_\psi$ as a function of $\psi$, for $m=1/4$. The values $\psi^+_{\rm cr}$ and $\tilde\psi^+_{\rm cr}$ correspond to the transitions where $\cV_\psi=0$ and $\tilde\cV_\psi=1-\mu$, respectively. For $m=1/2$, $\tilde\cV_\psi=\cV_\psi+\frac12$ and the two critical values coincide.}
\end{figure}

\medskip

We shall describe momentarily the scattering diagram for generic $\psi$ beyond those critical phases.
Let us first discuss the extreme value $\psi=\frac{\pi}{2}$ (the same happens {\it mutatis mutandis} 
for $\psi=-\frac{\pi}{2}$).
It follows from \eqref{ImZ} that the rays $\Im Z(\gamma)=0$ 
coincide with the loci $s=\frac{d_1+\eta d_2}{2r}$. For $m$ real, $\eta=1$ and these are
level sets of $s=\frac{\Im T_D}{\Im T}$. Hence they can only intersect at the branch point where $s$ becomes
ill-defined, hence there can be no wall-crossing between the branch point and $\tau=\I\infty$. 
For $\Im m>0$ however, $\eta$ becomes a function of $\tau$, which asymptotes to 1 as $\tau_2\to\I\infty$, see \eqref{sMetaLV}. 
Rays can intersect whenever $\frac{d_1+\eta d_2}{2r}=\frac{d'_1+\eta d'_2}{2r'}$
and $\langle\gamma,\gamma'\rangle \neq 0$, i.e.\ $\frac{d_1}{r} \neq \frac{d'_1}{r'}$ and $\frac{d_2}{r} \neq \frac{d'_2}{r'}$. Since $\langle\gamma,\gamma'\rangle=2(1-\eta)(r d_2'-r'd_2)$ must be a non-zero even integer, this can only happen when $\eta$ is a rational number such that $|(1-\eta)(r d_2'-r'd_2)|\geq 1$, hence for large charges if $\eta$ is close to one.

\subsection{Building the scattering diagram on the \texorpdfstring{$\Pi$}{Pi}-stability slice}

\begin{figure}[t]
    \centering
    \includegraphics[height=5cm]{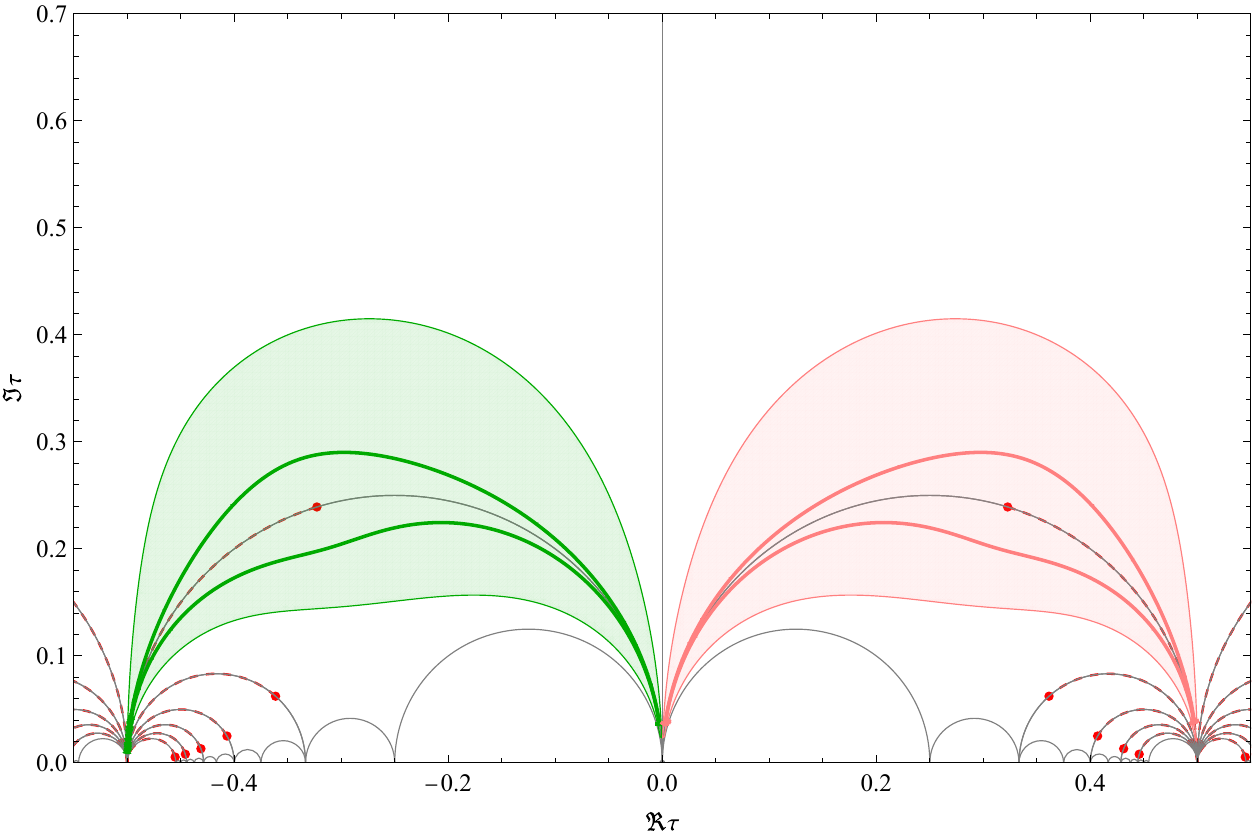}
    \includegraphics[height=5cm]{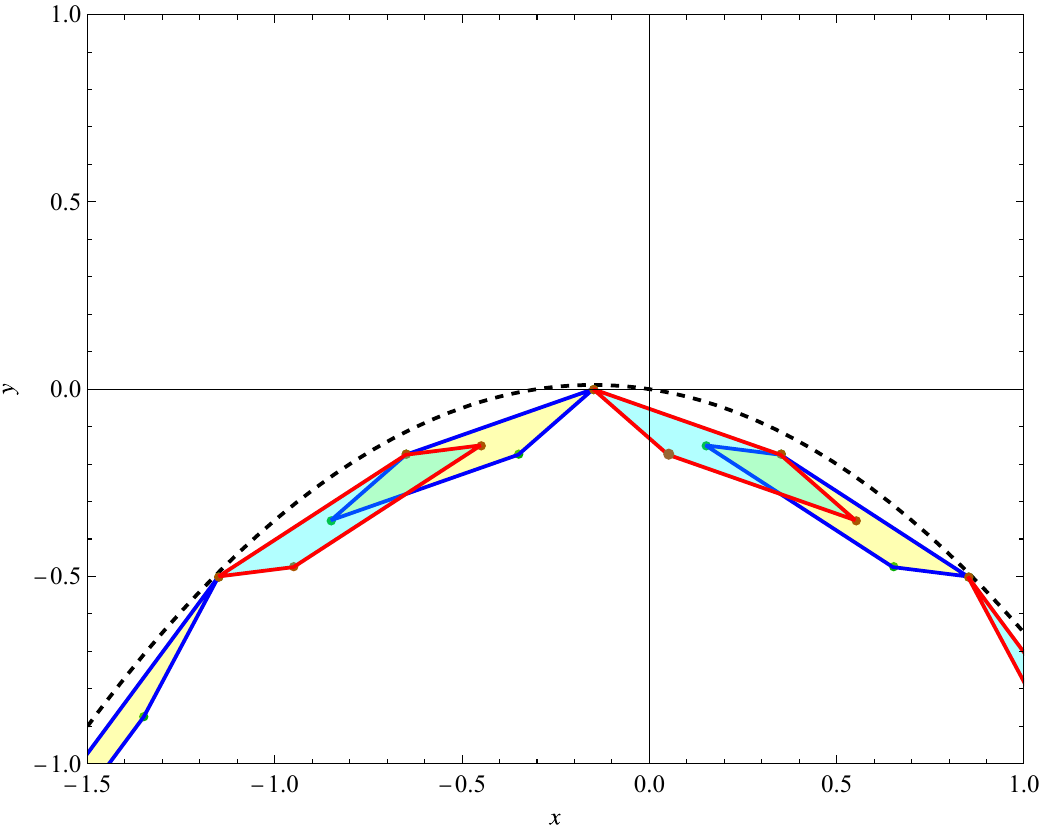}
    \caption{Left: Domain of validity of the quiver descriptions based on the 
    collections  \eqref{orbcoll1}, \eqref{orbcoll2}, for $m=0.3$. The solid green and red lines 
    are the boundaries of $\Delta_\psi$ and $\tilde\Delta_\psi$ and correspond to straight lines in the $(x,y)$ plane.
    Right: the same regions in the $(x,y)$ plane, together with some of their translates; the overlaps correspond to different stability conditions that map to the same values of $x,y$ coordinates.  The region above all of these translates is~$W_\psi$.\label{fig_diamond}}
\end{figure}

The orbifold diagram associated to the collection \eqref{orbcoll1} (resp.\ \eqref{orbcoll2}) is valid in the region (dubbed `orbifold region') around $\tau_B-1$ (resp.\ $\tilde\tau_B$) where all central charges $Z(\gamma_i)$ (resp.\ $Z(\gamma'_i)$) remain in the same half-plane.  In stark contrast to the local $\IP^2$ case of~\cite{Bousseau:2022snm}, these orbifold regions (depicted in  Fig.~\ref{fig_diamond}) and their $\tau\to\tau+1$ translates, together with the region where the central charge is $\GLt$-equivalent to the large-volume central charge, do not cover the whole fundamental domain $\cF$ for $\Gamma_0(4)$: the neighborhood of $\tau=\frac12$ is not covered.  Another issue is that the large volume description, while correctly counting BPS states through the large-volume scattering diagram, does not correctly describe the exact scattering diagram in the $\Pi$-stability slice, since the $\GLt$ transformation affects the phase of the central charge, hence affects the rays.

\medskip

To describe the exact scattering diagram we first define subregions $\Delta_\psi$ (resp.\ $\tilde\Delta_\psi$) of the orbifold regions near $\tau_B-1$ (resp.\ $\tilde\tau_B$), defined in the local affine coordinates $(u,v)$ adapted to the given exceptional collection as being the convex hull of the four initial points \eqref{startuv-1} (resp.\ \eqref{startuv-2}) of the rays, in other words, parallelograms in affine coordinates $(u,v)$ or $(x,y)$.  Each of their sides maps in the $\tau$-plane to a curve joining the conifold and dual conifold points $\tau=0,-\frac12$ (resp. $\tau=0,\frac12$) and opposite sides of the parallelogram map to the same curve since the monodromy around the branch point rotates the exceptional collection by two steps.  See Figure~\ref{fig_diamond}.\footnote{We checked numerically for a broad range of $m$ with $\Im m\geq 0$ and $0<\Re m<1$ that $\Delta_\psi$ and $\tilde\Delta_\psi$ are in the region of validity of the respective quiver descriptions.}
We also define\footnote{In \cite{Bousseau:2022snm} the large-volume components were called $\diamondsuit_\psi$ instead of $W_\psi$, but this could cause confusion since now the orbifold points are surrounded by four-sided regions.} $W_\psi$ as the connected component containing the large-volume point $\tau=\I\infty$ in the complement of $\Delta_\psi$, $\tilde\Delta_\psi$ and their $\Gamma_0(4)$ images. The image of its boundary $\partial W_\psi$ in the $(x,y)$ plane
is the `jagged parabola' introduced above \eqref{convexcond}.
 The $\Pi$-stability slice is partitioned in this way into $\Delta_\psi$, $W_\psi$ and their images under $\Gamma_0(4)$.  (In fact, $\tilde\Delta_\psi$ is one of the $\Gamma_0(4)$ images of $\Delta_\psi$.)

\medskip

The following picture emerges for the scattering diagram $\cD_{\psi,m}^{\Pi}$ along $\Pi$, for fixed $m$ and~$\psi$.
\begin{itemize}
\item For $|\psi|<\min(\psi^+_{\rm cr},\tilde\psi^+_{\rm cr})$, the region $\Delta_\psi$ and its $\Gamma_0(4)$ images do not contain any active\footnote{A ray is active if there exists a semi-stable object with that charge vector, namely if the corresponding index is nonzero. \label{fooactive}} ray so that the scattering diagram is a disjoint union of $\Gamma_0(4)$ images of the $W_\psi$ region, in which the scattering diagram coincides with the large-volume scattering diagram in affine coordinates $(x,y,\mu)$ given in \eqref{defxymu}.  The initial rays participating in that connected component are $\cO(k,k)$ and $\cO(k,k)[1]$ emerging at each integer (conifold) point $\tau=k$, and $\cO(k+1,k)$ and $\cO(k+1,k)[1]$ emerging at half-integer (dual conifold) points $\tau=k+\frac12$.

\item If $\psi^+_{\rm cr}<\tilde\psi^+_{\rm cr}$ then for $\psi^+_{\rm cr}<\psi<\tilde\psi^+_{\rm cr}$ the collision of the initial rays $\cO(k-1,k)[1]$ and $\cO(k,k)$ (emerging from $\tau=k-\frac12$ and $\tau=k$, respectively) produces an infinite set of bound state rays that exit from $\Delta_\psi$ into~$W_\psi$.  These bound states are described by the Kronecker quiver $K_2$, and they admit a large volume description\footnote{As  $\psi$ varies across $\psi^+_{\rm cr}$, this same collection of states goes from being described as bound states of $\cO(k,k)[1]$ and $\cO(k+1,k)$ for $\psi<\psi^+_{\rm cr}$ to the present description as bound states of $\cO(k-1,k)[1]$ and $\cO(k,k)$.  The dictionary involves shifting the label~$n$, which means for instance that $\cO(k+1,k)$ goes from being an initial ray for small~$\psi$ to a composite one for $\psi>\psi^+_{\rm cr}$, while conversely $\cO(k-1,k)[1]$ goes from composite to initial.} as $\cO(k-n-1, k)[1]$ and $\cO(k+n,k)$, and D2--D0 brane bound states with charge $[0,n,0,nk]$, for $n\geq 1$.  The scattering diagram within $W_\psi$ can then be built starting from these infinitely many rays together with $\cO(k,k-1)[1]$ emerging from $\tau=k-\frac12$ and $\cO(k,k)[1]$ emerging from $\tau=k$.  Altogether, the initial rays participating in the connected component of $\tau=\I\infty$ are $\cO(k,k)$ and $\cO(k,k)[1]$ emerging from $\tau=k$, and $\cO(k,k+1)[1]$ and $\cO(k+1,k)[1]$ emerging from $\tau=k+\frac12$.

\item If $\tilde\psi^+_{\rm cr}<\psi^+_{\rm cr}$ then for $\tilde\psi^+_{\rm cr}<\psi<\psi^+_{\rm cr}$ the situation is similar.  The initial rays $\cO(k,k)$ exit (translates of) $\Delta_\psi$ towards different images of the large volume region~$W_\psi$.  A new set of initial rays participate in the diagram in the region $\tau\to\I\infty$ of interest to us: $\cO(k+1,k-1)[1]$ emerges from $\tau=k$ on a different sheet, enters through the cut from $k+\tau_B$ to $k+\frac12$, collides with $\cO(k+1,k)$ emerging from $\tau=k+\frac12$, thus producing bound states with large-volume description as $\cO(k+1,k+n)$, $\cO(k+1,k-n-1)[1]$, and D2-D0 bound state with charge $[0,0,n,n(k+1)]$, for $n\geq 1$.
The initial rays participating in the connected component of $\tau=\I\infty$ are $\cO(k,k)[1]$ and $\cO(k+1,k-1)[1]$ emerging from $\tau=k$ on two different sheets, and $\cO(k+1,k)$ and $\cO(k+1,k)[1]$ emerging from $\tau=k+\frac12$.

\item For $\psi>\max(\psi^+_{\rm cr},\tilde\psi^+_{\rm cr})$ the scattering diagram is fully connected.  Within the region $k+\Delta_\psi$ (resp.\ $k+\tilde\Delta_\psi$), rays of the exceptional collection~\eqref{orbcoll1} (resp.\ \eqref{orbcoll2}) tensored with $\cO(k,k)$ interact according to the quiver scattering diagram.  Their bound state rays exit towards different $\Gamma_0(4)$ images of~$W_\psi$.  Those that enter~$W_\psi$ combine with the rays $\cO(k,k)[1]$ and $\cO(k+1,k)[1]$ emerging from $\tau=k,k+\frac12$ directly into the region~$W_\psi$ to produce the scattering diagram in that region.  Overall, the initial rays that participate to the scattering diagram in the large-volume region $W_\psi$ are $\cO(k,k)$, $\cO(k,k)[1]$, $\cO(k,k)[2]$, $\cO(k+1,k)$, $\cO(k+1,k)[1]$, $\cO(k,k+1)[1]$ from the principal sheet, $\cO(k+2,k)[1]$ and $\cO(k+1,k)[2]$ from behind a cut joining $k+\tau_B$ to $k+\frac12$, and finally $\cO(k+1,k-1)[1]$ and $\cO(k,k+1)[1]$ from behind a cut joining $k+\tilde\tau_B$ to $k+\frac12$.  Note that two different rays labeled $\cO(k,k+1)[1]$ and emerging from different points with the same coordinate $\tau=k+\frac12$ are involved, and likewise the two different rays\footnote{One of them is listed here as $\cO(k+2,k)[1]$, which simply amounts to shifing~$k$.} $\cO(k+1,k-1)[1]$ emerge from $\tau=k$ on different sheets.
\end{itemize}
For negative $\psi$ the situation is similar with different exceptional collections.  We only describe the case $\psi<\min(\psi^-_{\rm cr},\tilde\psi^-_{\rm cr})$, as the intermediate phases $\min(\psi^-_{\rm cr},\tilde\psi^-_{\rm cr})<\psi<\max(\psi^-_{\rm cr},\tilde\psi^-_{\rm cr})$ are simpler in that many of the rays mentioned below simply enter $W_\psi$~directly.  The rays $\cO(k,k-1)[1]$ and $\cO(k,k)[-1]$ on the principal sheet, together with rays $\cO(k-1,k)$ and $\cO(k+1,k-1)$ from across the cut joining $\tau_B-1$ to $\tau=\frac12$ interact to generate an orbifold scattering diagram in $k+\Delta_\psi$.  The rays $\cO(k,k+1)$ and $\cO(k,k)[1]$ on the principal sheet, together with rays $\cO(k+1,k)[-1]$ and $\cO(k+1,k-1)$ from across the cut joining $\tilde\tau_B$ to $\tau=\frac12$ interact to generate an orbifold scattering diagram in $k+\tilde\Delta_\psi$.  These two collections of four rays correspond to the exceptional collections \eqref{orbcoll1} and~\eqref{orbcoll2} tensored with $\cO(k,k)$ and
shifted by $[-1]$.  They generate infinitely many rays exiting from $k+\Delta_\psi$ and $k+\tilde\Delta_\psi$ into~$W_\psi$.  Together with the rays $\cO(k,k)$ and $\cO(k+1,k)$ that emerge directly into~$W_\psi$, they generate the scattering diagram in that region.

\subsection{Split Attractor Flow Conjecture\label{ssec_safc}}

An important claim underlying the above picture is that the only active  rays that enter into $\Delta_\psi$ (and likewise its $\Gamma_0(4)$ images) are the four initial rays participating in the exceptional collection, namely that collisions within $W_\psi$ do not produce any rays that can re-enter $\Delta_\psi$ and that could then participate in other large volume regions.  This ensures that portions of the scattering diagram in different $\Gamma_0(4)$ images of $W_\psi$ are completely independent.
The proof of this claim relies on the Split Attractor Flow Conjecture (SAFC), which itself relies on a closely related claim that attractor flow trees (see below) can only enter $\Delta_\psi$ and not exit it.
We discuss these related issues here and outline a proof of the SAFC, at least in 
the range $\psi\in \cI_{\rm cvx}$ defined below \eqref{convexcond},
 such that the initial points bound a convex region in the $(x,y)$ plane. 
We reiterate that we do not aim for mathematical rigour, hence will content ourselves with sketching the proof, letting the interested reader fill in the gaps.\footnote{A non-trivial gap, as in the local $\IP^2$ case, is to justify rigorously our claims about bijectivity properties of the map from the $\tau$ plane to the $(x,y)$ plane. Other minor gaps 
should be easy to fill by following the analogous proofs in~\cite{Bousseau:2022snm}.}

\medskip

The SAFC posits a mostly-combinatorical formula for the index $\Omega_\sigma(\gamma)$ for a given charge vector~$\gamma$ and $\Pi$-stability condition $\sigma$ as a sum over `attractor flow trees' \cite{Denef:2001xn,Denef:2007vg,Alexandrov:2018iao}.
Edges of these trees are paths in the space of $\Pi$-stability conditions defined by gradient flow of the central charge $|Z(\gamma_i)|$ for some charge vectors~$\gamma_i$ labeling the edge.
Nodes must obey charge conservation $\gamma_p=\gamma_1+\dots+\gamma_k$ when a parent edge with label $\gamma_p$ splits into children edges with labels $\gamma_1,\dots,\gamma_k$, and the central charges $Z(\gamma_i)$ involved at a given node must all have the same phase.
Finally, the first (root) edge of the tree starts at the stability condition~$\sigma$ and is labeled by~$\gamma$.
The contribution of a given tree is given by the product of indexes of the constituents (charge vectors of the leaves), multiplied by some combinatorial factors that account for multiplicities at each node of the tree.

\medskip

An important observation in the noncompact setting \cite{Bousseau:2022snm} is that the central charge $Z(\gamma)$ is a holomorphic function of $\tau$ and~$m$ (i.e.\ the normalizing factor $e^{-K/2}$ is absent).  The gradient flow with respect to its modulus $|Z(\gamma)|$ thus preserves its phase.  Since at each node the central charges $Z(\gamma_i)$ also have the same phase, all central charges involved in an attractor flow tree have the same phase as the central charge $Z_\tau(\gamma)$ at the initial point of the flow.  Moreover, 
 the metric determining the gradient flow is degenerate in such a way that the attractor flow in the $(\tau,m)$ space leaves $m$ invariant.
Thus, attractor flow trees lie in the scattering diagram $\cD_{\psi,m}^{\Pi}$, for fixed $m$ and~$\psi$.

\medskip

We are now ready to outline the proof of the SAFC for the scattering diagram $\cD_{\psi,m}^{\Pi}$. 
First, we should clear out a potential source of confusion, namely  that the attractor flow is oriented towards smaller values of $|Z(\gamma)|$, while when discussing scattering diagrams it is much more natural to think of rays as oriented in the opposite direction: exiting points where the object is massless, and pointing towards growing central charges.  We will mostly use the `scattering diagram direction', but it is sometimes necessary to mention the `attractor flow direction'.
We say that an oriented ray segment in $W_\psi$ (oriented in the scattering diagram direction) is \emph{outgoing} if the half-line in $(x,y)$ coordinates obtained by continuing the ray in that direction does not intersect $\partial W_\psi$, but the full line does.  Note that this definition excludes any ray that belongs to an $(x,y)$ line that does not intersect the `jagged parabola' $\partial W_\psi$ at all.
Note also that, at least when $\psi$ belongs to the range $\cI_{\rm cvx}$ introduced below \eqref{convexcond}, an outgoing ray cannot exit $W_\psi$, due to  the complement $\IR^2\setminus W_\psi$ in the $(x,y)$ plane being convex. It is possible that this property continues to be satisfied by active rays when the inequalities \eqref{convexcond} are no longer satisfied, but we shall not attempt to prove this. 

\medskip

The first part of the proof is to understand how attractor flow trees decompose into $W_\psi$, $\Delta_\psi$, $\tilde\Delta_\psi$ regions.
\begin{itemize}
\item In an attractor flow tree, by an induction going up from the leaves towards the root, we find that all ray segments that lie in $W_\psi$ are outgoing.  Indeed, rays that enter $W_\psi$ are necessarily outgoing, and then if several outgoing rays combine at a given node, then their bound state, whose ray lies in the convex hull of the children rays, must also be outgoing.

\item Thus, edges of attractor flow trees can only cross the boundary of $W_\psi$ one way: in the flow direction they can exit $W_\psi$ into one of the regions $k+\Delta_\psi$ or $k+\tilde\Delta_\psi$, but they can then never leave those regions.

\item In particular, any attractor flow tree with root in $W_\psi$ can be decomposed into `shrubs' in the regions $k+\Delta_\psi$ or $k+\tilde\Delta_\psi$ produced by the exceptional collections \eqref{orbcoll1} and \eqref{orbcoll2} and translates thereof, and a `trunk' in the region~$W_\psi$ that combines these shrubs together with additional lone branches $\cO(k,k)[1]$ and $\cO(k+1,k)[1]$ (for $\psi>0$) or $\cO(k,k)$ and $\cO(k+1,k)$ (for $\psi<0$).\footnote{While this step looks like a restatement of the structure of the scattering diagram proposed in the previous section, it concerns here all attractor flow trees, even if their branches are actually not active rays.  This phenomenon is ubiquitous and occurs due to cancellation between trees: for instance the SAFC applied to quivers produces trees for all non-negative dimension vectors even when the index happens to vanish.}
\end{itemize}

The main part of the proof is to make sense of the sum over attractor flow trees by proving that there are finitely many of them, as follows.
\begin{itemize}
\item In the $(x,y)$ plane the trunk of any attractor flow tree with root at a given point $(x,y)\in W_\psi$ must lie within the convex hull of $\partial W_\psi$ and $(x,y)$.  Trees can thus only involve shrubs from finitely many exceptional collections.

\item Then one exhibits a cost function
\be
\varphi_\tau(\gamma) = \begin{cases}
  d_1
+ d_2 + r\lfloor\cV_\psi-x\rfloor + r\lfloor\tilde\cV_\psi-x\rfloor& \text{in } W_\psi , \\
  c_\psi (N_1+N_2+N_3+N_4) & \text{in } \Delta_\psi , \\
  c_\psi (\tilde N_1+\tilde N_2+\tilde N_3+\tilde N_4) & \text{in } \tilde\Delta_\psi ,
\end{cases}
\ee
where the vectors $(N_1,N_2,N_3,N_4)$ and $(\tilde N_1,\tilde N_2,\tilde N_3,\tilde N_4)$ are dimension vectors of $\gamma$ with respect to the exceptional collections \eqref{orbcoll1} and~\eqref{orbcoll2}, and $c_\psi>0$ is a sufficiently small constant.

\item One checks that $\varphi_\tau(\gamma)\geq 1$ for the rays $\cO(k,k)[1]$ and $\cO(k+1,k)[1]$ near $\tau=k$ and $\tau=k+\frac12$ respectively, and for other initial rays one has $\varphi_\tau(\gamma)\geq c_\psi>0$ since dimension vectors are non-negative (and non-zero).

\item The cost function decreases monotonically along each branch of an attractor flow tree in the flow direction: in the large volume region, the variation of $x$ is appropriately correlated to the sign of~$r$, and one only needs to check the transition from $W_\psi$ to $\Delta_\psi$ or $\tilde\Delta_\psi$.
\end{itemize}
Note that this leads to explicit bounds hence an explicit algorithm for listing the trees for a given Chern vector.

As explained in \cite[\S 3.4--3.5]{Bousseau:2022snm}, once it has been shown that there are finitely many trees, a suitable choice of combinatorial factors (found in \cite{Alexandrov:2018iao,Arguz:2021zpx,Mozgovoy:2021iwz}) ensures that the index calculated by the Split Attractor Flow formula obeys the wall-crossing formula of \cite{ks}, or in other words that it defines a consistent scattering diagram.

\medskip

It remains to prove that the scattering diagram has the correct initial data and is uniquely characterized by consistency.  Large parts of the proof should mimic \cite[Appendix C]{Bousseau:2022snm}.
\begin{itemize}
\item The idea to compute $\Omega_\tau(\gamma)$ is to apply the KS wall-crossing formula starting from $\tau$ and tracking all relevant indices at every wall of marginal stability.

\item Concretely, one considers a notion of split attractor flow, which is a close cousin of attractor flow trees.  A split attractor flow starting from $\tau$, with charge vector $\gamma$ is obtained by the same gradient flow as for attractor flow trees, and the same type of splitting of a parent charge $\gamma_p=\gamma_1+\dots+\gamma_k$ into children charges, but now every edge has to be active (in the sense of footnote \ref{fooactive}), and the flow is not a priori required to reach initial rays in a finite number of steps.

\item The K\"ahler moduli space being compact except for large volume and conifold points, one finds that any such split attractor flow must in fact have finitely many branches, and must eventually reach the only non-compact points which are (dual) conifold points or the large volume point.  The large volume point is excluded by considering the asymptotics of the central charge.  Near the (dual) conifold points, suitable quiver descriptions force the split attractor flow to be given by initial rays that we have identified.

\item Altogether, repeatedly applying the wall-crossing formula to compute $\Omega_\tau(\gamma)$ in terms of simpler indices is a finite process.  It expresses the index in terms of split attractor flows, which are among attractor flow trees (with the added constraint that all rays be active).  The SAFC follows, after some bookkeeping to ensure that trees with some non-active rays do indeed cancel.
\end{itemize}

\subsection{Exploring the forest in the \texorpdfstring{$\Pi$}{Pi}-stability slice}

Given the complexity of the structure of initial rays, the above discussion of the SAFC only gives a somewhat impractical algorithm for determining the trees contributing to the index for an arbitrary Chern vector. As $\psi$ varies, the trees will in general jump, even if the index is constant.

To illustrate this phenomenon, we consider the simple cases of the Chern characters for the structure sheaves  $\cO_S$ and $\cO_C$, namely the pure D4-brane $\gamma=[1,0,0,0]$ and pure D2-brane $\gamma=[0,1,0,0]$. In this case, we find evidence that for fixed $\psi$, only one `shrub' contributes, even though the exceptional collection on which it is rooted varies as a function of $\psi$.  The tree gets more complicated as $|\psi|\to \frac{\pi}{2}$, albeit in a controlled fashion, as we now explain.

\begin{figure}[t]
    \centering
\includegraphics[height=6cm]{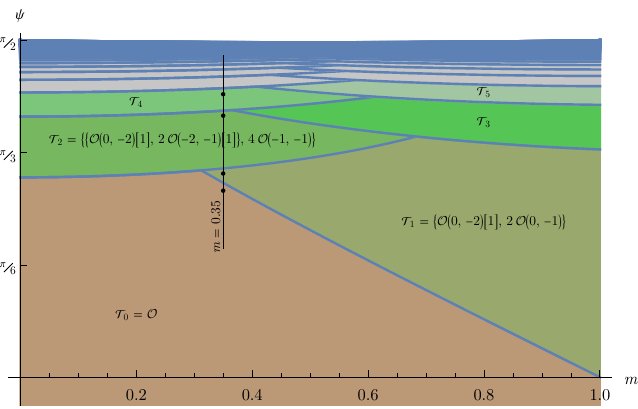}\\[5mm]
\includegraphics[width=12cm]{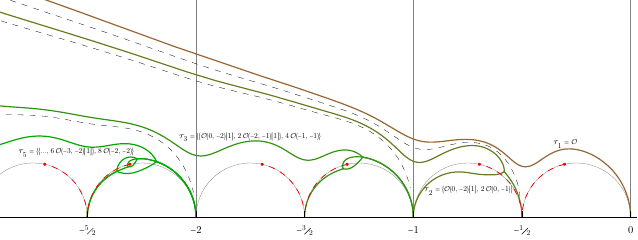}
    \caption{Top: Trees contributing to the structure sheaf index $\Omega([1,0,0,0])$
    as function of $m$ (real) and $\psi$. Bottom: Trees in $\tau$-plane 
    for $m=0.35$, varying $\psi$.  Note that the tree $\cT_4$ does not arise
    for this value. \label{fig_treeO}}
\end{figure}

\begin{figure}[t]
    \centering
\includegraphics[height=6cm]{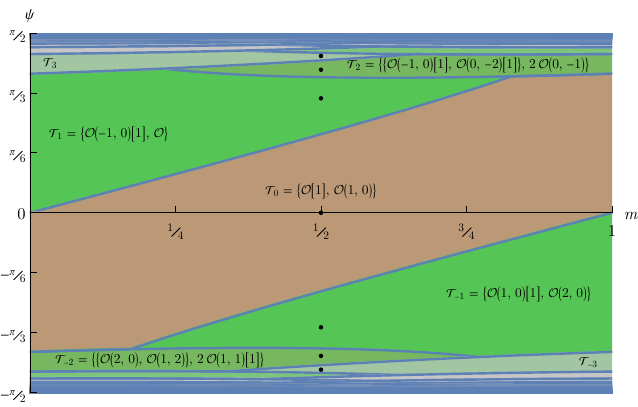}\\[5mm]
\includegraphics[width=12cm]{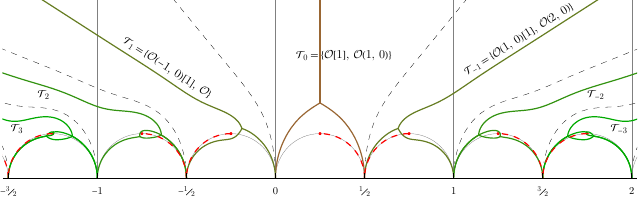}
    \caption{Top: Trees contributing to the D2 index $\Omega([0,1,0,0])$
    as function of $m$ (real) and $\psi$. Bottom: Trees in $\tau$-plane 
    for $m=0.5$, varying $\psi$.  \label{fig_treeD2}}
\end{figure}

For these Chern
characters, the dimension vectors  with respect to the collection \eqref{orbcoll1} tensored
with $\cO(-k,-k)$ (which we denote by $\cC(-k)$) are  given by 
\be
\cO_S: ( (k+1)^2, k^2+k, k^2,k^2-k)\ ,\quad
\cO_C: (k+1,k+1,k,k) 
\ee
while  the dimension vectors  with respect to the collection \eqref{orbcoll2} tensored
with $\cO(-k,-k)$  (which we denote by $\cC'(-k)$) are  given by 
\be
\cO_S: ( k^2+k, k^2, k^2-k,  (k-1)^2)\ ,\quad
\cO_C: (k+1,k,k,k-1)
\ee
In general  under shifting $k\mapsto k+1$, both dimension vectors transform as 
\be
\label{vecshiftk}
(N_1,N_2,N_3,N_4)\mapsto (4N_1-2N_2-3N_3+2N_4,N_4+2N_1-2N_3,N_1,N_2)
\ee
In fact, the dimension vector  along \eqref{orbcoll2} is mapped to the dimension vector along
\eqref{orbcoll1} via 
\be
\label{vecshiftk2}
(N_1,N_2,N_3, N_4)\mapsto (2N_1-2N_3+N_4,N_1,N_2,N_3)
\ee
and the transformation \eqref{vecshiftk} follows by acting with \eqref{vecshiftk2} twice. 
The transformation \eqref{vecshiftk2}
suggests that under shifting $\cC(-k)$ to $\cC'(-k+1)$, and then further to $\cC(-k+1)$, the
trees transform by a cyclic permutation $\sigma$ of the charges $\gamma_i\mapsto \gamma_{i+1}$, and adding an extra leg carrying $\langle \gamma,\gamma_2 \rangle$ units of charge $\gamma_1$,
\be
\label{TreeFlow2} 
T \mapsto \{ \langle \gamma,\gamma_2\rangle \gamma_1,  \sigma \cdot T \}
\ee
Of course, this only makes sense if $\langle \gamma,\gamma_2 \rangle>0$, and the full quiver scattering diagram could in principle contain additional trees. However, for the two cases
mentioned above, we find that this prescription correctly produces a tree contributing 
the same index (namely $1$ for $\cO_S$, and $-y-1/y$ for $\cO_C$) for increasing values of $\psi$,
see   Figures \ref{fig_treeO} and \ref{fig_treeD2}, so it is presumably the only contributing tree.

\clearpage

\appendix

\section{Central charge along the \texorpdfstring{$\Pi$}{Pi}-stability slice}
\label{sec_periods}

In this Appendix, we determine the central charge
at a general point 
along the slice of  $\Pi$-stability conditions, extending earlier studies of  mirror symmetry 
for local $\IF_0$~\cite{Katz:1996fh,Aganagic:2002wv,Aganagic:2006wq,Haghighat:2008gw,
Huang:2010kf,Huang:2011qx,Huang:2013yta} (or its close cousin $\IF_2$ \cite{Brini:2009nbd}). In particular, we express the coefficients $T,T_D$  as 
Eichler integrals of a certain weight 3 modular form under $\Gamma_1(4)$, 
and determine the asymptotic expansions
near the large volume, conifold, dual conifold and branch points.

\subsection{Mirror curve}
Local mirror symmetry \cite{Chiang:1999tz} relates type IIA strings on $K_{\IF_0}$ to type IIB strings on a conic bundle over
the genus one curve with equation
\be
\label{mcurvey}
y_1 + \frac{z_1}{y_1} + y_2 + \frac{z_2}{y_2}  + 1 = 0
\ee
where $y_1,y_2$ are coordinates in $\IC^\times \times \IC^\times$ and $z_1,z_2$ are complex structure parameters. Setting
\begin{align}
\lambda=\frac{z_1}{z_2}\,,\quad U=-\frac{1}{\sqrt{z_2}}
\, , \quad 	y_1=t\sqrt{z_1}\,,\quad y_2=w\sqrt{z_2}
\end{align}
we arrive at the form of the  mirror curve used in~\cite{Closset:2021lhd}:
\begin{align}
	\sqrt{\lambda}\left(\frac{1}{t}+t\right)+\frac{1}{w}+w-U=0\,,
	\label{eqn:cccurve}
\end{align}
Fiber-base duality acts by exchanging $z_1$ and $z_2$, or equivalently
\begin{align}
\lambda\mapsto \frac{1}{\lambda}\ ,\quad U\mapsto \frac{U}{\sqrt{\lambda}}
\end{align}
It will be convenient to define 
\be
\lambda= e^{2\pi\I m}
\ee
and choose $0\leq \Re(m)\leq \frac12$, which is always possible by fiber-base duality.  
Upon defining  $U = \frac{1}{Q_b}, \lambda=\frac{Q_f^2}{Q_b^2}$, 
or equivalently $(Q_f,Q_b)=(-\sqrt{z_1},-\sqrt{z_2})$, 
and rescaling $t$ and $w$,
one arrives at the representation in \cite{Banerjee:2020moh}
\be
\label{Qbf}
Q_b \left(w+\frac{1}{w}\right) + Q_f \left(t+\frac{1}{t}\right)-1=0
\ee

The curve \eqref{mcurvey} can be recast into the Weierstrass form 
\begin{align}
\label{mcurve}
	y^2=x^3+fx+g\,,
\end{align}
with coefficients given by
\begin{align}
	\begin{split}
	f=&-\frac{1}{48} \left(1-8 z_1+16 z_1^2-8 z_2-16 z_1 z_2+16 z_2^2\right)\,,\\
	g=&\frac{1}{864} \left(1-4 z_1-4 z_2\right) \left(1-8 z_1+16 z_1^2-8 z_2-40 z_1 z_2+16
   z_2^2\right)\,,
	\end{split}
\end{align}
Equivalently, one may write \eqref{mcurve} as a quartic \cite{Kim:2025pidkdi}
\be
\label{quarticcurve}
Y^2= (X^2+UX+1)^2 - 4 X^2 \cR^4\ ,\quad \cR=\sqrt{\lambda}, \quad U=-\frac{1}{\sqrt{z_2}}
\ee

\medskip

The curve \eqref{mcurve} becomes singular when the discriminant $\Delta$ vanishes, with \begin{align}
	\Delta=4f^3+27 g^2 = -\frac{1}{16} z_1^2 z_2^2 \left(1-8 (z_1+z_2)+16 (z_1-z_2)^2 \right)\,.
\end{align}
The discriminant locus $\Delta=0$ decomposes into 3 components, the two large volume limits
 $z_1=0$, $z_2=0$ and the conifold locus $\cC=\{1-8 (z_1+z_2)+16 (z_1-z_2)^2=0\}$ 
 (or equivalently $U=\pm 2(\cR^2\pm 1)$). 
 The symmetric slice $\{z_1=z_2=z\}$ (corresponding to $\lambda=1$) 
 intersects the discriminant locus at $z=0$ and $z=1/16$.
A generic slice $\{z_1=\lambda z_2\}$ with $\lambda \neq 1$ intersects  $\Delta$ in 3 points, 
for example $z=0, z=\pm \I/8$ when $\lambda=-1$. 

\medskip 
The Klein invariant of the elliptic curve \eqref{mcurve} evaluates to
\begin{align}
	J(z_1,z_2)=1728\frac{4f^3}{27g^2+4f^3}=
	\frac{\left[ (1-4z_1-4z_2)^2-48 z_1 z_2 \right]^3}
{z_1^2 z_2^2 \left(1-8 (z_1+z_2)+16 (z_1-z_2)^2 \right)}\,.
	\label{eqn:jz}
\end{align}
For fixed $\lambda=z_1/z_2$, the Klein invariant produces a 6:1 cover from the punctured
sphere parametrized by $z:=z_2$ to $\IH/PSL(2,\IZ)$. In order to trivialize this cover, we reduce
the modular group $SL(2,\IZ)$ to the subgroup $\Gamma_0(4)$, generated by 
integer matrices $\bigl(\begin{smallmatrix} a & b \\ c & d \end{smallmatrix}\bigr)$ with $ad-bc=1$ and $c=0\mod 4$.
\footnote{Note that  the authors of \cite{Closset:2021lhd,Aspman:2021vhs} consider instead 
$\Gamma^0(4)$, related by $\tau\mapsto \tau/4$. Moreover, $\Gamma_0(4)$ differs from $\Gamma_1(4)$ by the additional generator $\diag(-1,-1)$, which acts trivially on $\tau$.
}. The congruence subgroup 
$\Gamma_0(4)$ is an  index 6 subgroup of $SL(2,\IZ)$, with cusps at  
$\tau=\I\infty$ (and images thereof, $\tau=\frac{p}{q}$ with $q=0$ mod 4),  $\tau=0$ (and images thereof, $\tau=\frac{p}{q}$ with $q=1$ or $q=3$ mod 4) and  $\tau=\frac12$ (and images thereof, $\tau=\frac{p}{q}$ with $q=2$  mod 4). It will be convenient to choose the following fundamental domain of $\Gamma_0(4)$ in the Poincar\'e upper half-plane $\IH$ (see Fig.~\ref{figG04}) 
\be
\cF= \left\{ \tau\in \IH, 0\leq \Re\tau<1, \, |\tau-\tfrac14|>\tfrac14, \quad  |\tau-\tfrac34| \geq \tfrac14 \right\}
\ee
In particular, the lower boundary of $\cF$ consists of two half circles of radius 1/4, which we denote by
$\cC=\cC(\frac34,\frac14)$ and $\cC'=\cC(\frac14,\frac14)$.
The quotient $\IH/\Gamma_0(4)$ is in 1:1 correspondence with the punctured sphere $\IP^1
\backslash\{\infty, \pm 8\}$ via the   Hautpmodul 
\be
\label{defj4}
J_4(\tau) = 8 + \left(\frac{\eta(\tau)}{\eta(4\tau)}\right)^8 = \frac{1}{q}+20 q-62 q^3+216 q^5 - 641 q^7+\dots
\ee
which maps 
\be
J_4(\I\infty) = \infty, \quad J_4(0) =8\ ,\quad J_4(\tfrac12)=-8, \quad 
J_4(\tfrac{\I+1}{4})=J_4(\tfrac{\I+3}{4})=0 
\ee
Using the  relation between $J_4$ and the Klein invariant $J(\tau)=\frac{E_4^3}{\eta^{24}(\tau)}$
\be
\label{JJ4}
J(\tau)=\frac{(J_4^2(\tau) +240J_4(\tau) +2112)^3}{(J_4(\tau) -8)^4 (J_4(\tau) +8)} 
\ee
and setting $u=\frac{U^2}{4}-\lambda-1$, we find that 
\be
\label{uj4}
\frac{u}{2\sqrt{\lambda}} = \frac{J_4(\tau)+24}{J_4(\tau)-8}
\ee
This can be further simplified by introducing the Fricke transform
\be
\begin{split} 
J_4'(\tau) = & J_4\left(-\frac{1}{4\tau}\right) 
=  8+256  \left(\frac{\eta(4\tau)}{\eta(\tau)}\right)^8 
= 8+\frac{256}{J_4(\tau) -8} \\
= &  8 + 256 q + 2048 q^2 + 11264 q^3 +  \dots
\end{split}
\ee
Indeed, \eqref{uj4} now becomes $\frac{u}{\sqrt{\lambda}} = \frac14 J_4'(\tau) $
or  equivalently 
\be
z = \frac{1}{\sqrt{\lambda} J'_4(\tau) + 4 (\lambda+1)} 
\ee
In order to map the large volume limit $z\to 0$ to $\I\infty$, we shall henceforth trade $\tau$ for its
Fricke transform $\tau'=-1/(4\tau)$ (i.e.\ perform the isogeny $\tau\mapsto 4\tau$ followed by the inversion) and write 
\be
\label{zj4h}
z = \frac{1}{\sqrt{\lambda} J_4(\tau) + 4 (\lambda+1)} 
\ee
This relation gives a 1:1 map from $\IH/\Gamma_0(4)$ to the punctured line  parametrized by $z$,
\be
\IP^1  \backslash \{ 0 , \frac{1}{4(1+ \sqrt{\lambda})^2}, \frac{1}{4(1- \sqrt{\lambda})^2}, \infty \} 
\ee
corresponding to the images of $\tau=\I\infty, 0,\frac12, \tau_B$, respectively, where 
$\tau_B$ is the point in the fundamental domain $\cF$ defined by 
\be
\label{J4B}
J_4(\tau_B) = -4 (\lambda^{1/2} + \lambda^{-1/2} ) = -8 \cos \pi m
\ee
Circling around the point $\tau_B$, $z$ is single valued but $U=-1/\sqrt{z}$ changes sign.
Thus, for generic values of $\lambda$, the $U$-line is a double cover of $\IH/\Gamma_0(4)$.
We choose the branch cut starting from
$\tau_B$ and ending at $\tau=\frac12$ along a hyperbolic geodesic (equivalently, we require that 
its image under the Atkin-Lehner 
involution $\tau''=\frac{\tau}{1-2\tau}$ mapping $(\I\infty, 0, \frac12)$ to $(-\frac12,0,\I\infty)$
lies along a vertical line with fixed $\Re \tau''$), and extend this choice to the full 
Poincar\'e upper half-plane by $\Gamma_0(4)$ invariance. We observe the following
simplifications at special values of $\lambda$:
\begin{itemize}
\item
For $|\lambda|=1$ (or $m$ real),
 the branch point $\tau_B$ lies along the half circle $\cC:=\cC(\frac34,\frac14)$ 
while its image under $\tau\mapsto \frac{\tau-1}{4\tau-3}$ lies along the half circle $\tilde\cC:=\cC(\frac14,\frac14)$, see Fig.~\ref{figG04}. 
\item For $\Re(m)=0$, $\tau_B$ lies along the vertical axis $\Re\tau_B=1/2$. 
For $\Re(m)=\frac12+\I 0^+$ (respectively $\Re(m)=\frac12-\I 0^+$), the branch point $\tau_B$
lies on the vertical axis $\Re(\tau)=\frac14$ (respectively  $\Re(\tau)=\frac34$)
\item For  $\lambda\to 1$ (or rather $m\to 0$), the branch point $\tau_B$ collides with the cusp at $\tau=\frac12$, so the branch cut disappears. The coordinate $U$ then becomes single-valued 
on the doubled fundamental domain $\cF \cup (\cF+1)$, which can be viewed as 
a fundamental domain of $\Gamma_0(8)$ after rescaling $\tau\mapsto \tau/2$ \cite{Closset:2021lhd}. 
\item For $\lambda\to -1$ (or rather $m\to 1/2$)
we have $J_4(\tau_B)\to 0$ so $\tau_B$ coincides with the $\IZ_4$-orbifold point $\frac{\I+3}{4}$,
and its image with $\frac{\I+1}{4}$. The auto-equivalence $M_{0,1}\cdot M_{FB}$ from \eqref{Monm1m2} gives an extra symmetry $\tau\mapsto \tau+\frac12$, which extends 
$\Gamma_0(4)$ to $\Gamma_0(2)$, now acting on $2\tau$. Indeed, for $\lambda=-1$ 
the Coulomb branch parameter $U=-1/\sqrt z$ can be written in terms of the Hauptmodul for 
$\Gamma_0(2)$
\be
J_2(\tau) = \left(\frac{\eta(\tau)}{\eta(2\tau)}\right)^{24}+24 = \frac{1}{q}
+276 q - 2048 q^2 + 11202 q^{3}+ \dots
\ee
as 
\be
U^4 = - (J_2(2\tau)+40 )
\ee
where we used $J_2(2\tau)=J_4^2(\tau)-40$. Thus, the $U$-line becomes a $4:1$ cover of $\IH/\Gamma_0(2)$~\cite{Closset:2021lhd}. 
\item For $\lambda\to+\infty$ (or $m\to-\I\infty$), corresponding to the 4D limit (see \S\ref{sec_4d}), the branch point $\tau_B$ moves off to infinity
\end{itemize}

\subsection{Picard-Fuchs equations and mirror map}
The dependence of the coefficients $T_D,T_1,T_2$  on the K\"ahler parameters $z_1,z_2$
 is dictated by mirror symmetry. The Picard-Fuchs (PF) system annihilating the periods is generated by the two operators~\cite{Katz:1996fh}
\begin{align}
\label{PF0}
	\begin{split}
	\mathcal{D}_1=&\theta_1^2-2z_1(\theta_1+\theta_2)(2\theta_1+2\theta_2+1)\,,\\
	\mathcal{D}_2=&\theta_2^2-2z_2(\theta_1+\theta_2)(2\theta_1+2\theta_2+1)\,.
	\end{split}
\end{align}
Setting $z=z_2, \lambda=\frac{z_1}{z_2}$, 
the first operator is rewritten as 
\be
\label{PF1}
\mathcal{D}_1 = (\lambda\partial_\lambda)^2 - 2 \lambda z (2 \theta+1 ) \theta
\ee
where $\theta=z\partial_z$, while the second operator can be traded for a third order differential operator in $z$~\cite{Huang:2013yta} 
\begin{align}
\label{D3}
	\mathcal{D}=\left[\theta ^2-2 (1+\lambda) z  (2 \theta+1 )^2+4 (1-\lambda)^2 z^2  (2 \theta+1 ) (2 \theta+3 )\right]\theta\,,
\end{align}
where $\lambda$ plays the role of an external parameter. This operator has 4  singularities at 
\be
\label{zsing}
z=0,\ \quad z=\frac{1}{4(1+ \sqrt{\lambda})^2}, \quad z=\frac{1}{4(1- \sqrt{\lambda})^2},
\quad  z=\infty
\ee
corresponding to the large volume point $\tau=\I\infty$, conifold point $\tau=0$, `dual conifold' point $\tau=\frac12$ and branch point $\tau_B$, respectively. 
The constants 
\be
\varpi_0=1\,,\quad \varpi_1=\log\lambda = 2\pi\I m 
\ee
are manifestly annihilated by $\cD$ and $\cD_1$, while non-constant solutions of $\cD_1=0$
have a logarithmic or doubly logarithmic singularities at $z=0$, 
\begin{align}
\label{pi23}
\begin{split}
\varpi_2=&\log z + 
  2 \sum_{k+\ell>0} \frac{(2k+2\ell-1)!}{(k!)^2 (\ell!)^2} z^{k+\ell} \lambda^{\ell} \\
		&=\log z +2z(1+\lambda) +(3+12\lambda+3\lambda^2) z^2
		+\frac{20}{3} (1+9\lambda+9\lambda^2+\lambda^3) z^3+\ldots\,,\\
\varpi_3=&-\log^2 z 
  +  \left( 2\log z + \log\lambda\right) \varpi_2
   - 2   \sum_{k+\ell>0} \tfrac{(2k+2\ell)!}{(k!)^2 (\ell!)^2 (k+\ell)} 
  \left( H_k + H_\ell - 2 H_{2k+2\ell} + \tfrac{1}{k+\ell} \right) 
  z^{k+\ell} \lambda^{\ell}\\
&=-\log^2 z +\varpi_2\left[\log(\lambda)+2\log(z)\right]+4z(1+\lambda)+(13+40\lambda+13 \lambda^2)z^2+\ldots\,.
	\end{split}
\end{align}
where $H_k=1+\frac12+\dots+\frac{1}{k}$ are the harmonic numbers.
Note that $\varpi_2$ is invariant under fiber-base duality, $(z,\lambda)\mapsto (z\lambda,1/\lambda)$,
whereas $\varpi_3$ transforms by a shift proportional to $\varpi_1$. 
On the other hand, the period $\varpi_1$ is independent of the complex structure of the mirror curve, 
odd under fiber-base duality, and is the residue of the meromorphic differential. 

Guided by the large volume limit of the D4-brane central charge 
$T_D$, we define the period $T$ and dual period $T_D$ (or 
$(2a,\frac12 a_D)$ in the notations of ~\cite{Closset:2021lhd}) via
\begin{align}
T=\frac{1}{2\pi i}\varpi_2\,,\quad T_D=\frac12\frac{1}{(2\pi i)^2}\varpi_3+\frac{1}{12}\,,\quad
\label{defTTDm}
\end{align} 
Setting $Q=e^{2\pi\I T}$, 
we can obtain the mirror map order by order in $Q$,
\begin{align}
	\begin{split}
	z(Q)=&Q-2(1+\lambda)Q^2+3(1+\lambda^2)Q^3\\
	&-4(1+\lambda+\lambda^2+\lambda^3)Q^4+5(1-5\lambda^2+\lambda^4)Q^5+\mathcal{O}(Q^6)\,,
	\end{split}
\end{align}
The prepotential $F_0$ is obtained by  integrating the relation $T_D=\partial_T F_0$,
\begin{align}
	F_0=\frac16T^3+\frac14 m T^2+ \frac{T}{12} + \frac{1}{(2\pi\I)^3}
	\left[ 2(1+\lambda)Q + \frac14(1+16\lambda+\lambda^2)Q^2+\mathcal{O}(Q^3) \right]\,.
\end{align} In the large volume limit, we get the expected
large volume expression of the D4-brane central charge, including the Todd-class correction $\chi(\IF_0)/48=1/12$,
\be
T_D =\frac12 T(T+m)+\frac1{12}+\cO(Q) 
\ee
The modular parameter of the mirror curve is given by the second derivative of the prepotential, 
\begin{align}
	\tau=\partial_T T_D = \partial_T^2F_0
\end{align}
such that
\begin{align}
	q=e^{2\pi i\tau}=Q\sqrt{\lambda}\left[1+2(1+\lambda)Q+(3+20\lambda+3\lambda^2)Q^2+\mathcal{O}(Q^3)\right]\,,
\end{align}
After inverting the series, we can express $Q$ a in terms of $\tau$, 
\be
	\label{eqn:modmap}
	\sqrt{\lambda}Q = q-2\left(\lambda^{-\frac12}+\lambda^{\frac12}\right)q^2+\left(5\lambda^{-1}-4+5\lambda\right)q^3+\mathcal{O}(q^4) 
\ee
and therefore $z$  in terms of $\tau$
\be
z = \frac{q}{\sqrt\lambda} - 4 \frac{1+\lambda}{\lambda} q^2 
+ \frac{4(4+3\lambda+4\lambda^2)}{\lambda^{3/2}} q^3
- \frac{32(2+\lambda+\lambda^2+2\lambda^3)}{\lambda^{2}} q^4
+\dots
\ee
which agrees with the $q$-expansion of \eqref{zj4h}. 

\subsection{Explicit solutions at \texorpdfstring{$\lambda=\pm 1$}{lambda = 1 or -1}}
When $\lambda=1$ (more precisely $m=0$), one of the singularities \eqref{zsing} moves off to infinity  and the PF equation \eqref{D3}
admits explicit solutions in terms of Meijer G functions, 
\be
\begin{split}
\label{MeijerG}
\varpi_2(z) =& -\frac{1}{\pi} G_{3,3}^{2,2} \left( 
\begin{array}{ccc} \frac12 & \frac12 & 1 \\ 0& 0 & 0 \end{array} 
\vert \, -16 z \right) - \I\pi\ ,\quad  \\
\varpi_3(z) =&  -\frac{2}{\pi} G_{3,3}^{3,2} \left( 
\begin{array}{ccc} \frac12 & \frac12 & 1 \\ 0& 0 & 0 \end{array} 
\vert \, 16 z \right) - \frac{4\pi^2}{3}
\end{split}
\ee
such that 
\be
\tau=\frac{z\partial_z T_D}{z\partial_z T}
=  \frac{\I}{2} \frac{_2F_1( \tfrac12, \tfrac12; 1; 1-16 z)}{_2F_1( \tfrac12, \tfrac12; 1; 16 z)},
\quad z=\frac{1}{J_4(\tau)+8}
\ee
One easily checks that the expansion near $z=0$ (assuming $\Im z<0$) reproduces \eqref{pi23}, 
while the value
at the  conifold point $z=1/16$, corresponding to $\tau=0$ or $U=\pm 4$ yields
\bea
\varpi_2= -\frac{8G}{\pi}  ,\quad \varpi_3=\frac{2\pi^2}{3}
\eea
where $G=\Im\Li_2(\I)\simeq 0.915966$ is Catalan's constant, hence
\be
\label{TTDcon}
m=0: \qquad T(0)=\I \frac{4G}{\pi^2}, \quad T_D(0)=0
\ee
The expansion near $z=\infty$ (or $U\to \infty$) is more subtle, due to the branch cut in the Meijer G-function. For $\Im z<0$, the Meijer G-function goes to zero and $\tau\to -\frac12$, yielding
$(T,T_D)=(-\tfrac12, \tfrac14)$. For $\Im z>0$, $\tau$ instead goes to $+\frac12$ and we obtain the 
image of the previous result under spectral flow,  
\be
\label{TTDcond}
m=0: \qquad  T(\tfrac12) =\tfrac12, \quad T_D(\tfrac12) =\tfrac14
\ee
We shall recompute these values below \eqref{Lf1f2} using $L$-series. 
By using analytic properties of the Meijer-G functions, we find the monodromies
of the reduced period vector $(1,T,T_D)^t$ around  $z=0$, $z=1/16$ and $z=\infty$,
\be
\label{monDG}
M_{LV}=\begin{pmatrix}
1 & 0 & 0 \\
1 & 1 & 0 \\
\frac12 & 1 & 1 
\end{pmatrix}, \quad
M_{C}=\begin{pmatrix}
1 & 0 & 0 \\
0 & 1 & -4 \\
0 & 0 & 1 
\end{pmatrix}, \quad
 M_{o} =\begin{pmatrix}
1 & 0 & 0 \\
1 & 1 & -4 \\
\frac12 & 1 & -3 
\end{pmatrix}
\ee
This reduces to the matrices in  \cite[(5.22-23)]{Closset:2021lhd}
upon ignoring the mixing with the constant solution and translating notations.
The corresponding actions on the $\tau$ parameter are the $\Gamma_0(4)$
transformations 
\be
\tau\mapsto\tau+1,\quad 
\tau\mapsto -\frac{\tau}{4\tau-1},\quad 
 \tau\mapsto \frac{3\tau-1}{4\tau-1}
\ee

\medskip

For $\lambda=-1$, the PF equation \eqref{D3} becomes invariant under $z\to -z$, 
with singularities at $z\in \{0,\pm\frac{\I}{8},\infty\}$, and 
also admits explicit solutions in terms of Meijer G functions. Matching the  behavior 
\eqref{pi23} near $z=0^+$, we find 
\cite[(5.69)]{Closset:2021lhd}, 
\be
\label{Mpi23p}
\begin{split}
\varpi_2(z) =& -\frac{1}{2\pi\sqrt2} \, G_{3,3}^{2,2} \left( 
\begin{array}{ccc} \frac14 & \frac34 & 1 \\ 0& 0 & 0 \end{array} 
\vert \, 64 z^2 \right) \ ,\quad  \\
\varpi_3(z) =&  -\frac{1}{2\pi\sqrt2} \, G_{3,3}^{3,2} \left( 
\begin{array}{ccc} \frac14 & \frac34 & 1 \\ 0& 0 & 0 \end{array} 
\vert \, -64z^2 \right) - \frac{\pi^2}{3}
\end{split}
\ee
such that 
\be
\tau=\frac{\I}{2\sqrt2} \frac{_2F_1( \tfrac14, \tfrac34; 1; 1+64 z^2)}{_2F_1( \tfrac14, \tfrac34; 1; -64 z^2)},
\quad z=\frac{\I}{J_4(\tau)}
\ee
In the limit $z\to \infty$, corresponding to $\tau\to \tau_B$, 
the Meijer G functions go to 0, so that 
\be
\label{TTDorb12}
m=-\frac12: \quad T(\tau_B)=0, \quad T_D(\tau_B) =\frac18 
\ee
with $\tau_B=\frac{\I+3}{4}$. 
If instead we match the expansion near $z=0^-$, we find
\be
\label{Mpi23m}
\begin{split}
\varpi_2(z) =& -\frac{1}{2\pi\sqrt2} \, G_{3,3}^{2,2} \left( 
\begin{array}{ccc} \frac14 & \frac34 & 1 \\ 0& 0 & 0 \end{array} 
\vert \, 64 z^2 \right) + \I \pi  \\
\varpi_3(z) =&  -\frac{1}{2\pi\sqrt2} \, G_{3,3}^{3,2} \left( 
\begin{array}{ccc} \frac14 & \frac34 & 1 \\ 0& 0 & 0 \end{array} 
\vert \, -64z^2 \right) +2\pi\I \varpi_2(z)
- \frac{\pi^2}{3}
\end{split}
\ee
leading to 
\be
\label{TTDorb12m}
m=\frac12: \quad T(\tau_B)=\frac12, \quad T_D(\tau_B) =\frac38 
\ee
At the conifold points $z=\pm \I/8$, the period $T_D$ vanishes, while $T$ is given (in agreement with the results for the $E_7$ model in \cite[(3.57)]{Mohri:2000kf}) by
\bea
\label{TMohri}
T&=&\mp \frac14 + \frac{3}{64\pi\I} \,_3F_4(1,1,\tfrac54,\tfrac74;2,2,2;1) -\frac{\log8}{2\pi\I} 
\simeq \pm \frac14+  \I\times 0.305131
\eea
where the upper (lower) sign corresponds to \eqref{Mpi23p} and \eqref{Mpi23m}, respectively.
These correspond to the values at $\tau=\frac12$ and $\tau=0$, respectively.

\subsection{Eichler integral representation} 


We shall now establish an Eichler integral representation for the periods, similar to the one obtained in \cite{Bousseau:2022snm}\footnote{With hindsight, such a representation could have been extracted from~\cite{Aganagic:2006wq}, Sections 6 and 7.}
 for local $\IP^2$. For this, we need to compute $C(\tau,m):=\partial_{\tau} T=1/\partial_T {\tau}=1/C_{TTT}$, the inverse of the Yukawa coupling. Defining $e(\tau,m)=z\partial_z\varpi_2$
and using
\be
\label{qdqJ4}
q\partial_q J_4 = - 16 \frac{\theta_3^4(2\tau) \theta_4^4(2\tau)}{\theta_2^4(2\tau)}  
\ee
we easily compute the ratio
\be
\label{qdz}
\frac{C}{e} = \frac{q}{z} \frac{\partial z}{\partial q} = 16 \frac{\theta_3^4(2\tau) \theta_4^4(2\tau)}{\theta_2^4(2\tau)} 
 (J_4+ 4\lambda^{1/2} + 4\lambda^{-1/2})^{-1}
\ee
The l.h.s. of \eqref{qdz} should be identified as $C/e$, where $C=1/C_{zzz}$ is the inverse Yukawa coupling. On the other hand, the ratio $C/e^3$ is equal to the discriminant,
\be
\frac{C}{e^3} = \frac{1}{(2\pi\I z \partial_z T)^3 C_{TTT}} =  \frac{1}{(2\pi\I z)^3 C_{zzz}}=
1 - 8 z (1 + \lambda) + 16 z^2 (\lambda-1)^2 
\ee
It follows that $e$ and $C$ are given separately by 
\bea
\label{eCgen}
e(\tau,m) &=&  \frac{ 4\theta_3^2(2\tau)\,\theta_4^2(2\tau)\, z^{1/2}\lambda^{1/4}}
{\theta_2^2(2\tau)\sqrt{1 - 8 z (1 + \lambda) + 16 z^2 (\lambda-1)^2}} = 
 \frac{4\theta_3^2(2\tau) \theta_4^2(2\tau)}{\theta_2^2(2\tau)} \sqrt{\frac{J_4+ 4\lambda^{1/2} + 4\lambda^{-1/2}}{J_4^2-64}} \nn\\
 C(\tau,m)&=&  
\frac{64\theta_3^6(2\tau) \theta_4^6(2\tau)\, z^{3/2}\lambda^{3/4}}
{\theta_2^6(2\tau)\sqrt{1 - 8 z (1 + \lambda) + 16 z^2 (\lambda-1)^2}}
=  \frac{64\theta_3^6(2\tau) \theta_4^6(2\tau)}
{\theta_2^6(2\tau) \sqrt{(J_4^2-64)(J_4+ 4\lambda^{1/2} + 4\lambda^{-1/2})}}
\nn\\
\eea
For $\lambda=1$, the branch cuts disappear and $e,C$ become simple ratios of 
Dedekind eta functions, in turn equal to Eisenstein series of weight $1$ and $3$ under $\Gamma_1(4)$, respectively,
\bea
e(\tau,0) &=& \frac{\eta(2\tau)^{10}}{\eta(\tau)^4\eta(4\tau)^4}
=1+4\sum_{n\geq 1} \frac{\chi_4(n) q^n}{1-q^n}    
	=1+4q+4q^2+4q^4+8q^5+\mathcal{O}(q^6)\,,
	\nn \\
	C(\tau,0) &=& \frac{\eta(\tau)^4 \eta(2\tau)^6}{\eta(4\tau)^4}
=1-4\sum_{n\geq 1} \frac{n^2 \chi_4(n) q^n}{1-q^n}    
=1 - 4 q - 4 q^2 + 32 q^3 - 4 q^4 - 104 q^5 +\mathcal{O}(q^6) \nn\\
\eea
Using these expressions, we can simplify the expression
of $e$ and $C$ for arbitrary $\lambda$, 
\bea
e(\tau,m) &=& \frac{\eta(2\tau)^{10}}{\eta(\tau)^4\eta(4\tau)^4} \sqrt{\frac{J_4+ 4\lambda^{1/2} + 4\lambda^{-1/2}}{J_4+8}} 
\nn\\
&=&1+2q(\lambda^{\frac12}+\lambda^{-\frac12})-2q^2(\lambda-4+\lambda^{-1})+4q^3(\lambda^{\frac32}-\lambda^{\frac12}-\lambda^{-\frac12}+\lambda^{-\frac32})+\ldots\,,
\nn\\
C(\tau,m) &=& \frac{\eta(\tau)^4 \eta(2\tau)^6}{\eta(4\tau)^4} 
\sqrt{\frac{J_4+8}{J_4+ 4\lambda^{1/2} + 4\lambda^{-1/2}}}  
\nn\\
&=&1-2q(\lambda^{\frac12}+\lambda^{-\frac12})+q^2(6\lambda-16+6\lambda^{-1})
+q^3(-20\lambda^{\frac32}+36\lambda^{\frac12}+36\lambda^{-\frac12}-20\lambda^{-\frac32})+\ldots
\nn\\
\eea
We determine the sign of the square root in the same way as described below \eqref{J4B}, namely
we choose a branch cut extending from $\tau=\tau_B$ to $\tau=\frac12$ along a geodesic circle.
Moreover, the product $e C$ is independent of $\lambda$,  and is modular of weight 4 under $\Gamma_0(2)$ after rescaling $\tau\to\tau/2$,
\be
e(\tau,m) C(\tau,m) = \frac{\eta(2\tau)^{16}}{\eta(4\tau)^8}
\ee
For $\lambda=-1$, using the identities 
\be
\label{th234doub}
\theta_2(\tau)=\frac{2\eta^2(2\tau)}{\eta(\tau)}\ ,\quad
\theta_3(\tau)=\frac{\eta^5(\tau)}{\eta^2(\tau/2) \eta^2(2\tau)}\ ,\quad
 \theta_4(\tau)=\frac{\eta^2(\tau/2)}{\eta(\tau)}\ ,\quad 
 \frac{\theta_3\theta_4}{\theta_2}=\frac{\eta^5(\tau)}{2\eta^4(2\tau)}
\ee
we find the simpler expressions
\be
\label{eCm12}
e(\tau,\tfrac12)=\frac{1}{\sqrt{2}} \sqrt{ \theta_3^4(2\tau)+ \theta_4^4(2\tau)}\ ,\quad 
C(\tau,\tfrac12)= \frac{\sqrt2\,  \theta_4^8(4\tau)}{\sqrt{ \theta_3^4(2\tau)+ \theta_4^4(2\tau)}}
\ee

From the fact that $\partial_\tau (T,T_D)=C (1,\tau)$, we conclude that 
\be
\label{Eichlercst}
\begin{pmatrix} T \\ T_D  \end{pmatrix}(\tau)
= \begin{pmatrix} T(\tau_0)  \\ T_D(\tau_0) \end{pmatrix} 
+  \int_{\tau_0}^{\tau} \begin{pmatrix} 1 \\u \end{pmatrix} \, 
\, C(u) \de u
\ee
for any point $\tau_0$ where $(T,T_D)$ takes a finite value. It is natural to choose $\tau_0=\tau_B$,
the branch point where the factor $J_4+4\lambda^{1/2}+4\lambda^{-1/2}$ in the expression \eqref{eCgen} for $C$ vanishes. Although we have not been able to derive this directly, we have computed numerically the periods at $\tau=\tau_B$ by integrating \eqref{Eichlercst} to large values of $\Im\tau$, and found overwhelming confirmation for the following formulae for 
$\Im m>0$, $0<\Re m<1$, and $\Im\tau\geq \Im\tau_B$ 
\be
\label{Eichlerl}
 \begin{pmatrix} T \\ T_D  \end{pmatrix}
= \begin{pmatrix} \frac12 \\ \frac{1+m}{4}  \end{pmatrix} 
+  \int_{\tau_B}^{\tau} \begin{pmatrix} 1 \\u \end{pmatrix} \, 
\, C(u) \de u
\ee
where the contour runs from $\tau_B$ to $\tau$ without crossing the cut. 
For $\Im m<0$ and $0<\Re m<1$ and $\Im\tau\geq \Im\tau_B$, one finds instead
\be
\label{Eichlerlm}
 \begin{pmatrix} T \\ T_D  \end{pmatrix}
= \begin{pmatrix} \frac12-m \\ \frac{1-m}{4}  \end{pmatrix} 
+  \int_{\tau_B}^{\tau} \begin{pmatrix} 1 \\u \end{pmatrix} \, 
\, C(u) \de u
\ee
with the same choice of contour. These formulae can then be used to analytically continue $T,T_D$
into the Poincar\'e upper half-plane, although the result will of course depends on the path, due to
the square root branch cuts in $C$. 
We note that the two values for opposite signs of $\Im m$ are related by fiber-base duality \eqref{FB}.

Moreover,  the values at $\tau_B$ are 
invariant under the order 2 monodromies
$M_{B}(0)$ and $ \widetilde{M}_B(0)$ respectively, where
\be
\label{MBk}
\begin{split}
M_{B}(k) 
= & \begin{pmatrix}
 1 & 0 & 0 & 0 \\
 0 & 1 & 0 & 0 \\
 2 k+1 & 0 & -1 & 0 \\
 k^2+k+\frac{1}{2} & k+\frac{1}{2} & 0 & -1 \\
\end{pmatrix}\ ,\quad M_{B,k}^2=1
\\
\widetilde{M}_{B}(k)
=& \begin{pmatrix}
 1 & 0 & 0 & 0 \\
 0 & 1 & 0 & 0 \\
 2 k+1 & -2 & -1 & 0 \\
 k^2+k+\frac{1}{2} & -\frac{1}{2}-k & 0 & -1 \\
\end{pmatrix}\ ,\quad \widetilde{M}_{B,k}^2=1
\end{split}
\ee
It is worth noting that $M_B(k)$ and $\widetilde{M}_B(k)$ are themselves the square of order 4 matrices,
\be
\label{MZ4}
\begin{split}
M_{\IZ_4}(k) =& \begin{pmatrix}
 1 & 0 & 0 & 0 \\
 1 & -1 & 0 & 0 \\
 -2 k (k+1) & 2 k+1 & 4 k+3 & -4 \\
 -2 k (k+1)^2 & \frac{1}{2} (2 k+1)^2 & 4 k^2+6 k+\frac{5}{2} & -4 k-3 \\
\end{pmatrix}
\\
\widetilde{M}_{\IZ_4}(k) =& 
\begin{pmatrix}
 1 & 0 & 0 & 0 \\
 1 & -1 & 0 & 0 \\
 -2 k^2 & 2 k+1 & 4 k+1 & -4 \\
 -k^2 (2 k+1) & 2 k^2+k+\frac{1}{2} & 4 k^2+2 k+\frac{1}{2} & -4 k-1 \\
\end{pmatrix}
\end{split}
\ee
which preserve the values $(\tau,m)=(\frac{3+k+\I}{4},\frac12)$ and 
 $(\frac{1+k+\I}{4}, \frac12)$, respectively. For $k=-1$ the matrices $M_{\IZ_4}(-1)$ 
 and $M_B(-1)$ coincide with
 $M_{\IZ_4}$ and $M_B$ in \eqref{MConZ4} and \eqref{MBm1}. 

\subsection{Expansions around singular points}
We shall now obtain the expansion of the periods near $\tau=\I\infty$, $\tau=0$, $\tau=\frac12$
and $\tau=\tau_B$. The expansion near all other singular points in the Poincar\'e upper half-plane
are then determined by the action of $\Gamma_0(4)$. 

\subsubsection{Large volume expansion}  
At large $\Im\tau$ the argument of the square root behaves as $1+q(1-\cos\pi m)+\cO(q)$ as $\tau\to\I\infty$ so the path does not cross the cut. Upon 
Fourier expanding 
\be
C = \sum_{n\geq 0} c_n(m) \, q^n\ ,\quad \lambda=e^{2\pi\I m}
\ee
and integrating term by term,  
we easily obtain expansions near $\tau\to\I\infty$, 
\bea
\label{TTDLV}
T_1 &=&T = \tau-\frac12m+\frac{1}{2\pi\I}  f_1(\tau,m)  \nn\\
T_2 &=& T+m = \tau+\frac12m+\frac{1}{2\pi\I} f_1(\tau,m) \\
T_D &=&\frac12 \tau^2 -\frac18 m^2 + \frac{1}{12}+ \frac{\tau}{2\pi\I} f_1(\tau,m) +
\frac{1}{(2\pi\I)^2} f_2(\tau,m)  \nn
\eea
with 
\be
\label{f12}
f_1(\tau,m) = \sum_{n\geq 1} \frac{c_n(m)}{n} q^n, \quad 
f_2(\tau,m) = -\sum_{n\geq 1} \frac{c_n(m)}{n^2} q^n
\ee
The radius of convergence is fixed by the distance to the nearest singularity, so 
the series  $f_1$ and $f_2$ converge uniformly when $\Im\tau>\Im\tau_B$. 
Their value at $\tau_B$ is determined by consistency with the Eichler integral representation 
\eqref{Eichlerl}, e.g.\ for $\Im m>0$
\be
\begin{split}
\tau_B+ \frac{f_1(\tau_B,m)}{2\pi\I} =& \frac{m+1}{2} \\
\frac12 \tau_B^2+ \frac{\tau_B}{2\pi\I} f_1(\tau_B,m) + 
 \frac{ f_2(\tau_B,m)}{(2\pi\I)^2} =& \frac{m^2}{8} + \frac{m}{4}+\frac16
 \end{split}
\ee

Since $f_i(\tau-\frac12, m\pm 1)=f_i(\tau,m)$, we find that the periods $(T,T_D)$ transform 
according to \eqref{Monm1m2} under tensoring with $\cO(1,1)$, $\cO(-1,0)$, $\cO(0,-1)$ and
$\cO(-1,1)$, respectively, 
\bea
T(\tau+1,m) = T(\tau,m)+1, \quad  && 
T_D(\tau+1,m)= T_D(\tau,m)+T(\tau,m)+ \tfrac{m+1}{2} \nn\\
T(\tau-\tfrac12,m-1) = T(\tau,m), \quad  && 
T_D(\tau-\tfrac12,m-1)= T_D(\tau,m)-\frac12 T(\tau,m) \nn\\
T(\tau-\tfrac12,m+1) = T(\tau,m)-1, \quad &&
T_D(\tau-\tfrac12,m+1)= T_D(\tau,m)-\frac12(T(\tau,m)+m) \nn\\
T(\tau,m-2) = T(\tau,m)+1, \quad  && 
T_D(\tau,m-2) = T_D(\tau,m)+\tfrac{m-1}{2} \nn\\
\label{monoLV}
\eea
Note however that these formulae no longer hold when $\Im\tau<\Im\tau_B$, due to the presence of branch cuts. 
Moreover, from \eqref{TTDLV} and the fact that $c_n(m)=\overline{c_n(\bar m)}=c_n(-m)$ 
we see that $T,T_D$ satisfy the reality conditions
\be
\label{TTDreal}
T(-\bar\tau,-\bar m)  = - \overline{T(\tau,m)} \ ,\quad 
T_D(-\bar\tau, -\bar m)  = \overline{T_D(\tau,m)}
\ee
As a result, the central charge along the $\Pi$-stability slice has the same transformation
property as the large volume central charge \eqref{derduality}, 
\be
\label{derdualityPi}
Z_{-\overline{\tau},-\overline{m}}(-r,d_1,d_2,-\ch_2) = - \overline{ Z_{\tau,m}(r,d_1,d_2,\ch_2)  }
\ee

As in \cite{Bousseau:2022snm}, we can use these expansions and zeta function regularization to compute the 
values of the periods at the conifold
points when $m=0$. For this purpose, we compute the $L$-series associated to $f_1$ and $f_2$,
\be
\label{Lf1f2}
\begin{split}
L_{f_1}(s) \coloneqq& \sum_{n\geq 1} \frac{c_n(0)}{n^{1+s}} = -4 \zeta(s+1)\, L(\chi_4,s-1)\\
L_{f_2}(s) \coloneqq& -\sum_{n\geq 1} \frac{c_n(0)}{n^{2+s}} = 4 \zeta(s+2)\, L(\chi_4,s)
\end{split}
\ee
where $L(\chi_4,s)\coloneqq\sum_{m\geq 1} \chi_4(m) m^{-s}=4^{-s}\left( \zeta(s,\frac14)-\zeta(s,\frac34) \right)
$ is the Dirichlet $L$-series. Therefore
\be
\label{f12L0}
\begin{split}
f_1(0,0) =& -4 \lim_{s\to 0}  \zeta(s+1)\, L(\chi_4,s-1) = -\frac{8G}{\pi}  \\
f_2(0,0) =& 4 \lim_{s\to 0} \zeta(s+2)\, L(\chi_4,s) = \frac{\pi^2}{3}
\end{split}
\ee
where we used $L(\chi_4,0)=1/2$. Similarly, the periods at $\tau=\tfrac12$ can be computed by noting that the insertion of a sign $(-1)^n$ replaces $\zeta(s)$ by 
$\zeta_2(s):=\sum_{k\geq 1} (-1)^k/ k^s = (2^{1-s}-1)\zeta(s)$, leading to 
\be
\label{f12Lhalf}
\begin{split}
f_1(\tfrac12,0) =& -4 \lim_{s\to 0}  \zeta_2(s+1)\, L(\chi_4,s-1) = 0  \\
f_2(\tfrac12,0) =& 4 \lim_{s\to 0} \zeta_2(s+2)\, L(\chi_4,s) =- \frac{\pi^2}{6}
\end{split}
\ee
Inserting these results in \eqref{TTDLV}, we reproduce the values \eqref{TTDcon} and 
\eqref{TTDcond}. Unfortunately, we do not know how to reproduce the values \eqref{TMohri}
at $m=\frac12$ from the $L$-series associated to \eqref{eCm12}.

\subsubsection{Expansion near $\tau=0$}
We can now obtain the expansion $\tau=0$ by performing a Fricke transform $\tau\mapsto -1/(4\tau)$,
$u\mapsto -1/(4u)$  
in the Eichler integral representation \eqref{Eichlerl}. This involution maps the fundamental domain $\cF$
to its shift $\cF(-1)$.  Since $J_4(-1/(4\tau)) = 8+\frac{256}{J_4(\tau)-8}$, we get (assuming $\Im m>0$)
\be
\label{Eichlerl0}
 \begin{pmatrix} T \\ T_D  \end{pmatrix} \left(-\frac{1}{4\tau}\right)
= \begin{pmatrix} \frac12 \\ \frac{1+m}{4}  \end{pmatrix}
+8\I \int_{-1/(4\tau_B)}^{\tau} \begin{pmatrix} 4u \\-1 \end{pmatrix} \, 
\, C'(u) \de u
\ee
with 
\bea
C'(\tau) &=& \frac{C(-1/(4\tau))}{128\I \tau^3} \nn
\\
&=& \frac{\eta(4\tau)^4 \eta(2\tau)^6}{\eta(\tau)^4} 
\sqrt{\frac{4(J_4+8)}{2J_4+48+(J_4-8) (\lambda^{1/2}+\lambda^{-1/2})}} \nn\\
&=& 
 \frac{2q}{\lambda^{1/4}+\lambda^{-1/4}} + 8
\frac{3 \sqrt\lambda-2 + 3/\sqrt{\lambda}}{(\lambda^{1/4}+\lambda^{-1/4})^3} q^2 
   + 16 \frac{9 \lambda-44\sqrt\lambda+86-44/\sqrt{\lambda}+9/\lambda}{(\lambda^{1/4}+\lambda^{-1/4})^5}   q^3 + 
 \dots \\ \nn
\eea
It follows that 
\be
\label{TTDp}
T(-1/(4\tau)) = -\frac{8\I}{\pi^2} f'_2(\tau,m) + {\rm cte}\ ,\quad 
T_D(-1/(4\tau)) = -\frac{4}{\pi} f'_1(\tau,m) + {\rm cte}
\ee
where $c'_n$ are the Fourier coefficients of $C'(\tau,m)$, 
\be
f'_1(\tau,m) = \sum_{n\geq 1} \frac{c'_n}{n}  q^n, \quad
f'_2(\tau,m) =  2\pi\I \tau  f'_1(\tau,m) - \sum_{n\geq 1} \frac{c'_n}{n^2}  q^n
\ee
In particular, under $\tau\to \tau+1$, the period vector $\Pi^t=(1,m,T,T_D)^t$ transforms into
$M_C \Pi^t$, where $M_C$ was defined in \eqref{MConZ4}.
This is the expected action of the Seidel-Thomas twist \eqref{STtwist} with respect to the spherical object $\cO(0,0)$, which is massless at $\tau=0$, the fixed point of the action $\tau\mapsto \frac{\tau}{1-4\tau}$ of $M$ on $\tau$. 

To determine the additive constants in \eqref{TTDp}, we solve the PF equation \eqref{D3} around 
$z=\frac{1}{4(1+\sqrt\lambda)^2}$. It is convenient to use the variable 
$z'=\frac{1}{\sqrt{\lambda}J_4'+4(\lambda+1)}$ obtained by 
Fricke involution, which vanishes at that point, 
\be
z= \frac{1}{4(1+\sqrt\lambda)^2} - \frac{16\lambda z'}{(1+\sqrt\lambda)^4-4z'(1-\lambda)^2(1+6\sqrt\lambda+\lambda)}
\ee
The regular and logarithmic solutions are given by
\be
\label{periods23c}
\begin{split}
\varpi'_2 =& \frac{\lambda^{3/4} z'}{\sqrt{\lambda}+1}+\frac{\lambda^{3/4} \left(2 \sqrt{\lambda} \left(2 \sqrt{\lambda}+7\right)  (\lambda+1)+4\right)
   z'^2}{(\sqrt{\lambda}+1)^3}+ \dots\\
\varpi'_3 =& \varpi'_2 \log (z'\sqrt{\lambda}) +
\frac{\I}{8}\left( \Li_2(\I \lambda^{1/4}) - \Li_2(-\I \lambda^{1/4}) \right) + 
 \frac{\lambda^{3/4} z'}{\sqrt{\lambda}+1} \\&
 +\frac{\lambda^{3/4} \left(8 \lambda^2+ 25 \lambda^{3/2}+10 \lambda+25 \sqrt{\lambda}+8\right)
   z'^2}{(\sqrt{\lambda}+1)^3}+ \dots 
   \end{split}
\ee
where the $y$-dependent normalization and constant terms are fixed by \eqref{PF1}. Rewriting this in terms of $q'$, we find
\be
\frac{\partial_{z'} \varpi'_3 }{\partial_{z'} \varpi'_2}=2+\log q'
\ee
and recognize
\be
\varpi'_2 =\frac12 f'_1, \quad 
\varpi'_3 = f'_1 + \frac12 f'_2 + \frac{\I}{8} \left( \Li_2(\I \lambda^{1/4}) - \Li_2(-\I \lambda^{1/4}) \right)  
\ee
Hence, we conclude that,  possibly up to linear terms in $m$,
\be
\label{TTDzero}
\begin{split}
 T(-1/(4\tau)) =&\frac{16\I}{\pi^2} (2\varpi'_2-\varpi'_3) = -\frac{8\I}{\pi^2}  f'_2(\tau) 
+\I \cV(m)  \\
 T_D(-1/(4\tau)) =& -\frac{8}{\pi} \varpi'_2 = - \frac{4}{\pi} f'_1(\tau) 
\end{split}
\ee
where
\be
\label{defVm}
\cV(m) = \frac{2\I}{\pi^2}  \left( \Li_2(-\I e^{\I \pi m/2}) - \Li_2(\I e^{\I \pi m/2}) \right)  
\ee
In particular, as $\tau\to \I\infty$ we get
\be
\label{TTD0}
T(0,m) = \I \cV(m), \quad T_D(0,m) = 0
\ee
so that \eqref{Eichlerl} may be written as 
\be
\label{Eichlerl2}
 \begin{pmatrix} T \\ T_D  \end{pmatrix} \left(-\frac{1}{4\tau}\right)
= \begin{pmatrix} \I \cV(m)  \\0  \end{pmatrix}
+8\I \int_{\I\infty}^{\tau} \begin{pmatrix} 4u \\-1 \end{pmatrix} \, 
\, C'(u) \de u
\ee
We have checked  the absence of additional linear terms in $m$ by numerical integration. 
Remarkably, the value of $T(0)$ for $m=0$ and $m=1/2$ reproduces \eqref{TTDcon} and \eqref{TMohri}, respectively. Note that $\cV(m)$ is anti-periodic under $m\mapsto m+2$,
in fact using the expansion near $\tau=0$ we find
\be
T(\tau,m-2) = -T(\tau,m), \quad T_D(\tau,m-2) = -T_D(\tau,m),
\ee 
in contrast to the periodicity at large volume \eqref{monoLV}. 
For $m$ real, using Joncqui\`ere's identity we get 
\be
\label{ImcV}
\Im \cV(m) = 2 B_2\left( {\rm Fr}( \tfrac{m-1}{4}) \right) - 2 B_2\left( {\rm Fr}( \tfrac{m+1}{4}) \right)
\ee
where ${\rm Fr}(x)=x-\lfloor x \rfloor$ and $B_2(x)=x^2-x+\frac16$ is the second Bernoulli polynomial.
For $m\in [-1,1]$, this evaluates to $\Im \cV(m)=\frac{m}{2}$, while for $m\in [1,3]$, $\Im \cV(m)
=1-\frac{m}{2}$. The real part oscillates between $\pm 4G/(\pi^2)$. By using the second equation \eqref{monoLV} , we get, for $k\in\IZ$ and $0<\Re (m)<1$,
\bea
\label{TTDk}
T(k,m) &=& k+\I \cV(m), \quad T_D(k,m)=\frac{k(k+m)}{2}+ k \I \cV(m) 
\eea
such that $\cO(k,k)$ is massless at $\tau=k$.

\subsubsection{Expansion near $\tau=\frac12$}
Let us now define
\be
J_4''(\tau)=  J_4\left(\frac{\tau}{2\tau+1}\right)\ ,\quad 
C''_1(\tau)=- \frac{1}{4(2\tau+1)^3} C_1\left(\frac{\tau}{2\tau+1}\right)
\ee
In particular, setting $\tau'=-1/4\tau, \tau''=\frac{\tau}{1-2\tau}$ we have
\be
J_4(\tau) = J'_4(\tau') = J''_4(\tau''), \quad \tau'=-\frac{1}{4\tau''}-\frac12
\ee
Note that when $\tau$ lies in the standard fundamental domain in the interval $(0,1)$,
$\tau''$ belongs  to the standard fundamental domain in the interval $(-1,0)$. The large volume
point is now at $\tau''=-\frac12$, while the conifold and shifted conifold points are now at $\tau''=0$
and $\tau''=\I\infty$.
Using a sequence of $S$ and $T$ transformations we find
\bea
\eta\left(\frac{\tau}{2\tau+1}\right) &=&-(-1)^{5/6} \sqrt{2\tau+1} \, \eta(\tau) \nn\\
\eta\left(\frac{2\tau}{2\tau+1}\right) &=&-(-1)^{11/12} \sqrt{2\tau+1} \, \eta(2\tau) \\
\eta\left(\frac{4\tau}{2\tau+1}\right) &=&\frac{1}{\sqrt2} \sqrt{2\tau+1} \, \eta(\tau-\tfrac12) \nn
\eea
We further note the relation
\be
\eta(\tau-\tfrac12) = e^{-\frac{\I\pi}{24}}\frac{\eta^3(2\tau)}{\eta(\tau)\eta(4\tau)}
\ee
Using these relations, we get 
\be
J_4''(\tau) 
= -8 + \frac{256}{J_4(\tau)+8}
= 8 - \frac{16 \eta^{16}(\tau)  \eta^{8}(4\tau)}{\eta^{24}(2\tau)}
=-8 + 256 q - 2048 q^2+\dots
\ee
and 
\be
C''_1(\tau)= (-1)^{1/3}  \frac{\eta^6(2\tau)\eta^4(\tau)} 
{\eta^4(\tau-\tfrac12) }
= \frac{\eta^{8}(\tau)\eta^4(4\tau)}{\eta^6(2\tau)}
= \sqrt{q} - 8 q^{3/2} + 26 q^{5/2} - 48 q^{7/2} + \dots
\ee
More generally, for $\lambda\neq 1$, 
\be
\label{Cppgen}
\begin{split}
C''(\tau,m) =&  \frac{\eta^{8}(\tau)\eta^4(4\tau)}{\eta^6(2\tau)} \frac{8}{\sqrt{48-2J_4+(J_4+8)(\lambda^{1/2}+\lambda^{-\frac12})}} \\
=& \frac{8q}{\lambda^{1/4}-\lambda^{-1/4}}- \frac{32(3\lambda^{1/2}+2+3\lambda^{-1/2}) q^2}{(\lambda^{1/4}-\lambda^{-1/4})^3} +\dots
\end{split}
\ee
Note that the leading power jumps from $q^{1/2}$ at $\lambda=1$ to $q$ at $\lambda\neq 1$. For $\lambda=-1$, we have $\lambda^{1/4}-\lambda^{-1/4}=\I \sqrt2$ hence
\be
C''(\tau,\pm \tfrac12) \to -4\I\sqrt2 \left( q + 4 q^2 + 136 q^3 + 2576 q^4+\dots \right)
\ee
Moreover, note the relation between the Fourier coefficients of $C'$ and $C''$,
\be
c'_n(m-1) = \frac{\I}{4} (-1)^{n+1} c''_n(m)
\ee

Similarly as above,  we solve the PF equation \eqref{D3} around 
$z=\frac{1}{4(1-\sqrt\lambda)^2}$, by using the variable 
$z''=\frac{1}{\sqrt{\lambda}J_4''+4(\lambda+1)}$ obtained by 
Fricke involution, which vanishes at that point, 
\be
z= \frac{1}{4(1-\sqrt\lambda)^2} - \frac{16\lambda z''}{(1-\sqrt\lambda)^4-4z''(1-\lambda)^2(1-6\sqrt\lambda+\lambda)}
\ee
The regular and logarithmic solutions are given by 
\be
\begin{split}
\varpi''_2 =& \frac{\lambda^{3/4} z''}{1-\sqrt{\lambda}}+\frac{\lambda^{3/4} 
\left(4-14 \sqrt{\lambda} + 4 \lambda -14\lambda^{3/2} + 4 \lambda^2\right)
   z''^2}{(1-\sqrt{\lambda})^3}+ \dots\\
\varpi''_3 =& \varpi''_2 \log (z''\sqrt{\lambda}) +
\frac{1}{8}\left( \Li_2(\lambda^{1/4}) - \Li_2(-\lambda^{1/4}) \right) + 
 \frac{\lambda^{3/4} z''}{1-\sqrt{\lambda}} \\&
 +\frac{\lambda^{3/4} \left(8 \lambda^2- 25 \lambda^{3/2}+10 \lambda-25 \sqrt{\lambda}+8\right)
   z''^2}{(1-\sqrt{\lambda})^3}+ \dots 
   \end{split}
\ee
This agrees with the continuation of \eqref{periods23c} under $\sqrt{\lambda}\to - \sqrt{\lambda}$. 
Moreover,
\be
\frac{\partial_{z''} \varpi''_3 }{\partial_{z''} \varpi''_2}=2+\log q'' \ ,\quad 
q''\partial_q'' \varpi''_2
= -\frac18 C''
\ee
Setting
\be
f''_1(\tau,m) = \sum_{n\geq 1} \frac{c''_n(m)}{n}  q^n, \quad
f''_2(\tau,m) =  2\pi\I \tau  f''_1(\tau,m) - \sum_{n\geq 1} \frac{c''_n(m)}{n^2}  q^n
\ee
we recognize 
\be
 \varpi''_2= -\frac18 f''_1, \quad 
  \varpi''_3 = -\frac18 ( 2f''_1+ f''_2) +\frac{1}{8}\left( \Li_2(\lambda^{1/4}) - \Li_2(-\lambda^{1/4}) \right) 
\ee
In particular, under $\tau\to \tau+1$, the period vector $\Pi^t=(1,m,T,T_D)^t$ transforms into
$M_{\tilde C} \Pi^t$ where $M_{\tilde C}$ is the  Seidel-Thomas twist \eqref{STtwist} with respect to the spherical object $\cO(1,0)$. The latter is indeed massless at $\tau=\frac12$, the fixed point under the action $\tau\mapsto 
\frac{\tau-1}{4\tau-3}$  of $M_{\tilde C}$ on the upper half-plane.

By demanding consistency with the Eichler integral representation
\be
\label{Eichlerm1pp}
 \begin{pmatrix} T \\ T_D  \end{pmatrix} \left( \tfrac{\tau}{2\tau+1},m \right) 
= \begin{pmatrix} a'' \\ b''  \end{pmatrix} 
+  4 \int_{\I\infty}^{\tau} \begin{pmatrix} 2u+1 \\ u \end{pmatrix} \, 
\, C''(u,m) \de u
\ee
we conclude that
\be
\label{TTDhalfexp}
\begin{split}
T \left( \tfrac{\tau}{2\tau+1},m\right) =& \frac{16}{\pi^2}  ( \varpi''_3-2\varpi''_2)
 + \frac{16\I}{ \pi} \varpi''_2 = -\frac{2}{\pi^2} f_2'' -\frac{2\I}{\pi} f_1'' 
  +\I \tilde\cV(m)  \\ 
T_D \left( \tfrac{\tau}{2\tau+1}, m \right) =& \frac{8}{\pi^2} (\varpi''_3- 2\varpi''_2)
=  -\frac{1}{\pi^2} f_2''  + \frac{\I}{2} \tilde\cV(m)
\end{split}
\ee
up to an additive constant and additive multiple of $\log\lambda$. In particular, as $\tau\to\I\infty$,
\be
\label{TTDhalf}
T(\tfrac12,m) = 2T_D(\tfrac12,m) = 
\I \tilde\cV(m)
\ee
with 
\be
\label{defVtm}
\tilde\cV(m):= 
\frac{2\I}{\pi^2} \left( \Li_2(-\lambda^{1/4}) - \Li_2(\lambda^{1/4}) \right) = \cV(m-1)
\ee
This determines the constants $a'',b''$ in \eqref{Eichlerm1pp}, possibly up to linear terms in $m$,
\be
\label{Eichlerm1pp2}
 \begin{pmatrix} T \\ T_D  \end{pmatrix} \left( \tfrac{\tau}{2\tau+1}, m \right) 
= \begin{pmatrix} \I \tilde\cV(m) \\ \frac12 \I \tilde\cV(m)   \end{pmatrix} 
+ 4  \int_{\I\infty}^{\tau} \begin{pmatrix} 2u+1 \\ u \end{pmatrix} \, 
\, C''(u,m) \de u
\ee
We have checked by numerical integration that there are no additional linear terms. By spectral flow,
we find the periods at the conifold points $\tau=k+\frac12$,
\be
\label{TTDkhalf}
T(k+\tfrac12,m) = k+\I \tilde\cV(m), \quad T_D(k+\tfrac12,m)=\frac{k(k+m)}{2}+
 (k+\tfrac12) \I \tilde\cV(m) 
\ee
such that $\cO(k+1,k)$ is massless at $\tau=k+\frac12$.

\subsubsection{Expansion near $\tau=\tau_B$}

For completeness, we finally solve the Picard-Fuchs equations near the branch point $z=\infty$. 
Choosing the Ansatz $\varpi= \sum_{n\geq 0} a_n(\lambda) z^{-n-\frac12}$ and expanding
in $1/z$, one finds that the equation $\cD\varpi=0$ determines all coefficients in terms of 
$a_0(\lambda)$ and $a_1(\lambda)$ via
\be
-8 n (2 n^2-n-1) (1 - \lambda)^2 a_n(\lambda) + 
 4 (n - 1)^2 (2 n - 1) (1 + \lambda) a_{n - 1}(\lambda) - 
 \frac18 (2 n - 3)^3 a_{n - 2}(\lambda)=0
 \ee
  in particular
\bea
a_2(\lambda) &=& -\frac{a_{0}(\lambda)-96 (\lambda+1) a_{1}(\lambda)}{640 (\lambda-1)^2}
\nn\\
a_{3}(\lambda) &=&
   \frac{3 \left(23 \lambda^2+82 \lambda+23\right) a_{1}(\lambda)-(\lambda+1) a_{0}(\lambda)}{2688
   (\lambda-1)^4}, \\
a_{4}(\lambda)  &=&  \frac{384 \left(11 \lambda^3+85 \lambda^2+85 \lambda+11\right) a_{1}(\lambda)-\left(71
   \lambda^2+242 \lambda+71\right) a_{0}(\lambda)}{884736 (\lambda-1)^6} \nn
\eea   
Moreover, the equation $\cD_1\varpi=0$ implies that $a_1$ is determined from $a_0$, 
which must satisfy the second order differential equation
\be
a_1(\lambda)=\frac16 ( a'_0 + \lambda a''_0 ) \ ,\quad 
4\lambda(1-\lambda) a''_0 - 4(2y-1) a'_0 - a_0 =0
\ee
where the prime denotes the derivative with respect to $\lambda$.
The solutions are linear combinations of elliptic functions $K(\lambda)$ and $K(1-\lambda)$.
The regular solution at $\lambda=1$ is obtained by choosing $a_0(\lambda) =\tfrac{2}{\pi} K(1-\lambda)$,
\be
\begin{split}
\varpi_2^B =& \frac{2K(1-\lambda) }{\pi \sqrt{z}} + 
\frac{(1+\lambda) K(1-\lambda)-2 E(1-\lambda)}{12\pi (\lambda-1)^2 z^{3/2}}
 + \dots 
 \\
=&  \frac{1}{\sqrt{z}}\left( 1+\frac14(1-\lambda)+\frac{9}{64}(1-\lambda)^2+\dots\right) 
\\ & +  \frac{1}{192z^{3/2}}\left( 1+\frac34(1-\lambda)+\frac{75}{128}(1-\lambda)^2+\dots\right)
 +\dots
 \end{split}
\ee
The logarithmic solution is obtained by choosing $a_0(\lambda) =\frac{4\log 4}{\pi} K(1-\lambda) - 2 K(\lambda))$,
\bea
\varpi_3^B &=& 
 \left(  \frac{4\log 4}{\pi} K(1-\lambda) - 2 K(\lambda) \right) z^{-1/2} + \nn\\
 &&  + \frac{  \pi(1-\lambda) K(\lambda) -2\pi E(\lambda) +2\log 4 
 \left( (1+\lambda) K(1-\lambda) -2 E(1-\lambda)\right)}{12\pi(1-\lambda)^2}
z^{-3/2} + \dots 
 \nn\\
  &=&  \varpi_2^B \log(1-\lambda) + 
 \frac{1}{\sqrt{z}}\left( \frac12(1-\lambda)+\frac{21}{64}(1-\lambda)^2+\dots\right) \nn\\&&
 +  \frac{1}{z^{3/2}} \left( 
 -\frac{1}{6(1-\lambda)^2}+\frac{1}{24(1-\lambda)}+
 \frac{5}{384}
  +\frac{5}{512}(1-\lambda)+\frac{255}{32768}(1-\lambda)^2+\dots\right)
 +\dots \nn\\ 
\eea
One may check that the expansions around $z=\infty$ and $\lambda=1$ satisfy the recursion
in \cite[\S 6.3]{Aganagic:2002wv}, and agree with   \cite[(5.24)]{Haghighat:2008gw} at low orders. 

Notice that both $\varpi_2^B$ and $\varpi_3^B$ vanish at $z=\infty$. The periods at the branch point are therefore linear combinations of $1$ and $\varpi_1^B=-\log\lambda$. In order to fix 
 the relevant combination, we can evaluate the Eichler integral \eqref{Eichlerl} in the vicinity of the branch point using \eqref{qdqJ4} and \eqref{th234doub},
  \be
 z =\frac{1}{\sqrt\lambda( J_4(\tau)- J_4(\tau_B))} \sim
- \frac{1}{2\pi\I\sqrt\lambda}\frac{\eta^{16}(4\tau_B)}{\eta^{20}(2\tau_B)(\tau-\tau_B)}
 \ee
 \be
 C(\tau,m)\sim \frac{\eta^4(\tau_B) \eta^6(2\tau_B)}{\eta^4(4\tau_B)} 
\sqrt{\frac{8-4\lambda^{1/2}-4\lambda^{-1/2}}{(\tau-\tau_B) \partial_\tau J_4(\tau_B)}}
= \pm \frac{2 \eta^4(\tau_B)  \eta^4(4\tau_B) (\lambda^{1/4}-\lambda^{-1/4})}
{\eta^4(2\tau_B) \sqrt{(2\pi\I)(\tau-\tau_B)}}
 \ee
 hence
 \be
 \begin{split}
 \label{Eichbranch}
 \int_{\tau_B}^{\tau} \begin{pmatrix} 1 \\u \end{pmatrix} \, 
\, C(u,m) \de u \sim&
\pm \begin{pmatrix} 1 \\ \tau_B   \end{pmatrix} 
\frac{4(\lambda^{1/4}-\lambda^{-1/4}) \eta^4(\tau_B) \eta^4(4\tau_B)
\sqrt{(\tau-\tau_B)}}
{\eta^4(2\tau_B)\sqrt{2\pi\I}}
\\
\sim  & 
\pm \begin{pmatrix} 1 \\ \tau_B   \end{pmatrix}
\frac{2\eta^{12}(4\tau_B)\eta^4(\tau_B)}{ \eta^{14}(2\tau_B)}
\frac{1-\lambda^{-1/2}}{\pi\sqrt{z}}
\end{split}
 \ee 
Moreover, we can use the fact that 
for $k=\frac{\theta_2^4(4\tau)}{\theta_3^4(4\tau)}$, 
\be
\label{ellid}
K(k) = \frac{\pi}{2}\theta_3^2(4\tau), \quad 
\quad 
E(k)=\frac{\theta_3^4(4\tau)+\theta_4(4\tau)^4}{3\theta_3^4(4\tau)} K(k) + \frac{\pi^2 E_2(4\tau)}{12 K(k)}\ ,\quad 
4\tau = \frac{\I K(1-k)}{K(k)} 
\ee
The identification $k=\lambda$ is justified by the observation
that the branch point where $J_4(\tau_B)+4(\lambda^{1/2}+\lambda^{-1/2})=0$ satisfies 
\be
\left( \frac{\theta_2(2\tau_B)}{\theta_3(2\tau_B)} \right)^4 = \frac{4\sqrt{\lambda}}{(1+\sqrt\lambda)^2} 
\ \Rightarrow \left( \frac{\theta_2(4\tau_B)}{\theta_3(4\tau_B)} \right)^4 = \lambda
\ee
where the second equality follows from the doubling identities 
\be
\theta_2(4\tau)=\frac{1}{\sqrt2}\sqrt{\theta_3^2(2\tau)-\theta_4^2(2\tau)}\ ,\quad 
\theta_3(4\tau)=\frac{1}{\sqrt2}\sqrt{\theta_3^2(2\tau)+\theta_4^2(2\tau)}
\ee
and holds whenever $\tau_B$ is in the fundamental domain of $\Gamma_0(4)$ (if not the r.h.s seems to evaluate to $1/\lambda$ rather than $\lambda$).

 To evaluate the precise linear combination, we evaluate the coefficient of $1/\sqrt{z}$ at $\lambda=-1$.
 Using the Meijer-G representation  \eqref{MeijerG} above, we expect
 \be
 \varpi_2 = -\frac{\Gamma(\tfrac14)^2}{2\sqrt2 \pi^{3/2}}z^{-1/2} + \dots, \quad 
 \varpi_3 = -\frac{\pi^2}{3} + 2 K(2)\, z^{-1/2} + \dots
 \ee 
 where we used (or observed)  
 \be
 \begin{split}
 \Gamma(\tfrac54) =& \frac14 \Gamma(\tfrac14), \quad 
 \Gamma(\tfrac34)= \frac{\pi \sqrt2}{\Gamma(\tfrac14)}
\\
 K(2)=&\frac{1-\I}{8} \sqrt{\frac{2}{\pi}}\Gamma(\tfrac14)^2\ ,\quad
 E(2)=\frac{1+\I}{2} \sqrt{\frac{2}{\pi}}\Gamma(\tfrac34)^2\ 
\\
 K(-1) =& \frac{1+\I}{2} K(2), \quad 
 E(-1) = \frac{8 \pi ^2+\Gamma \left(\frac{1}{4}\right)^4}{4 \sqrt{2 \pi } \Gamma
   \left(\frac{1}{4}\right)^2}
   \end{split}
 \ee
 On the other hand,
 the eta functions in \eqref{Eichbranch} evaluate to 
 \be
 \eta(\tfrac{1+\I}{4}) =  \frac{e^{-\I \pi/24} \Gamma(\tfrac14) }{2^{5/8}\pi^{3/4}}\ ,\quad 
  \eta(\tfrac{1+\I}{2}) =  \frac{e^{\I \pi/24} \Gamma(\tfrac14) }{2^{3/4}\pi^{3/4}}\ ,\quad
   \eta(1+\I) =  \frac{e^{\I \pi/12} \Gamma(\tfrac14) }{2\pi^{3/4}}\ ,\quad
 \ee
Using these relations for $\lambda=-1$ we find
 \be
 \varpi_2=- \left(1+\frac{4\I \log2}{\pi} \right) \varpi_2^B + \frac{\I}{\pi} \varpi_3^B, \quad 
 \varpi_3=-\frac{\pi^2}{3} + (8\log 2-\I\pi)  \varpi_2^B -2  \varpi_3^B
 \ee
 Using \eqref{defTTDm}, we find, for $\lambda=-1$, 
 \be
 T=   \frac{1}{(2\pi\I)^2}  \left(
 (8 \log 2 - 2 \pi \I ) \varpi_2^B -2\varpi_3^B \right), \quad 
 T_D =  \frac38 + \frac{1}{(2\pi\I)^2} \left( \left(4 \log 2+\frac{\I\pi}{2} \right) \varpi_2^B - \varpi_3^B\right)
 \ee
 These relations are found to persist for any $m$ with any For $0<\Re m<1, \Im m>0$, with a judicious choice of constant term, invariant under $M_B(0)$,
 \be
 \begin{split}
   \label{TTDBp}
 T=&\tfrac12 + \frac{1}{(2\pi\I)^2} \left( 8 \log 2\,  \varpi_2^B -2 \varpi_3^B\right), \quad \\
 T_D=& \tfrac{1+m}{4} + \frac{1}{(2\pi\I)^2} \left( \left(4 \log 2+\frac{\I\pi}{2} \right) \varpi_2^B - \varpi_3^B\right)
 \end{split}
 \ee
 For $0<\Re m<1, \Im m<0$, 
  we find instead 
  \be
  \label{TTDBm}
  \begin{split}
 T=&\tfrac12-m + \frac{1}{(2\pi\I)^2} \left( (8 \log 2-2\pi\I)\,  \varpi_2^B -2 \varpi_3^B\right) \\
 T_D=& \tfrac{1-m}{4} + \frac{1}{(2\pi\I)^2} \left( \left(4 \log 2-\tfrac{\I\pi}{2} \right) \varpi_2^B - \varpi_3^B\right)
 \end{split}
 \ee
 where the constant term is invariant under $\widetilde{M}_B(0)$.
 These relations are consistent with the value of $ \tau_B= \frac{\de T_D}{\de T}$, except for 
 a shift by $\frac12$ and $\Gamma_0(4)$ action, 
 \be
 \begin{array}{ll}
 \Im m>0: & \tau_B= \frac{\de T_D}{\de T}\vert_{\tau=\tau_B} = \frac12 + \frac{\I K(1-\lambda)}{4K(\lambda)}
 \\
\Im m<0: & 
\tau_B=\frac{2K(\lambda)-\I \,K(1-\lambda)}{4 K(\lambda)-4 \I \,K(1-\lambda)}
\quad   
 \Leftrightarrow  \quad 
 \frac{3\tau_B-1}{4\tau_B-1}=  \frac12 + \frac{\I K(1-\lambda)}{4K(\lambda)} 
\end{array}
 \ee
Note that $\tau_B$ evaluates to $\frac{3+\I}{4}$ or $\frac{1+\I}{4}$ 
as $\lambda\to -1$ with $\Im m>0$ or $\Im m<0$, respectively.

\subsubsection{Convergence domains and numerical evaluation}

\begin{figure}[t]
    \centering
\includegraphics[height=7cm]{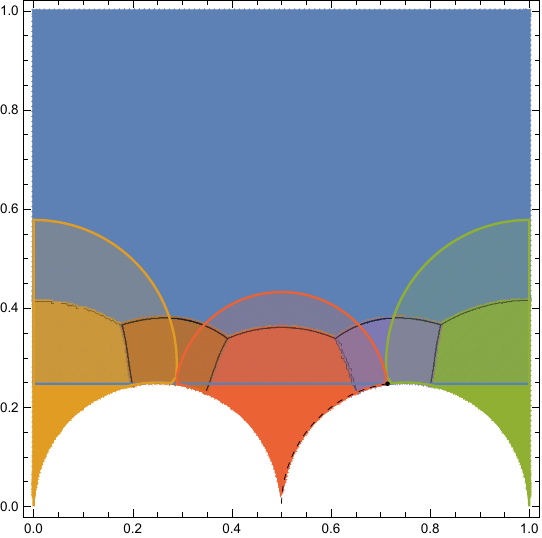} \includegraphics[height=7cm]{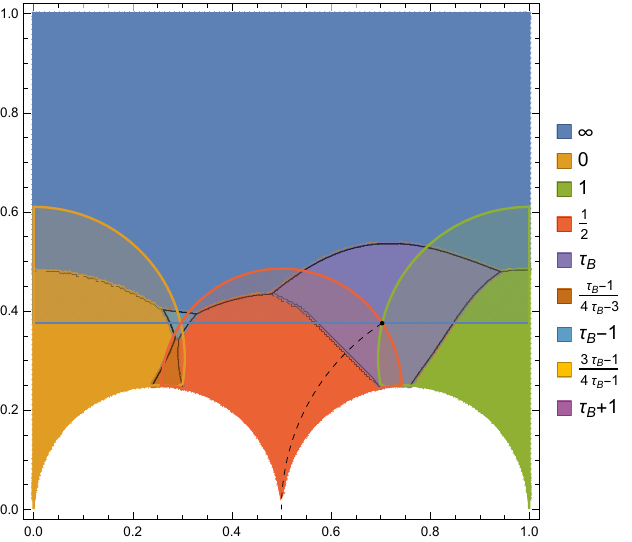}
       \caption{Convergence regions of the expansions around $\tau=\I\infty,0,1,\tfrac12$ (halfplane and disks intersected with the fundamental domain, shaded in blue, orange, green, red, respectively), and domains where each of these expansions are used, as well as expansion around images of the branch point $\tau_B$ (black dot, connected to $\tau=\frac12$ by the cut).
         Different values of $m$ require different subsets of the expansions.
         Left: $m=0.4$, involving the first six expansions.  Right: $m=0.4+0.3\I$, involving the first seven expansions (the last two would become relevant for other values of $m$).
     \label{ConvDomain}}
\end{figure}

The convergence domain of the expansions near $\tau=\I\infty,0,1/2$
 are determined by the nearest branch point, namely $\tau_B$ in the $\tau$ plane, $-1/(4\tau_B)$ or $-1/(4\tau_B-4)$ in
 the $\tau'$ plane and $\tau_B/(2\tau_B-1)$ in the $\tau''$ plane (see  Fig.~\ref{ConvDomain}).
Together with the expansion around $\tau=1$ obtained by translation, they cover the whole fundamental domain.  However, all three expansions break down both as $\tau$ approaches the branch point, and as $\tau$ approaches $1-\overline{\tau_B}$, which also lies at the boundary of the convergence disks.  Instead of completing the set of expansions using the large $z$ expansion \eqref{TTDBp}--\eqref{TTDBm}, we find it more efficient to solve numerically the Picard--Fuchs equation order by order in powers of $\tau''-\tau_B''$, for a fixed value of~$m$.  The convergence radius of the resulting series expansion is determined by the $\Gamma_1(4)$ image of $\tau_B''$ nearest to~$\tau_B''$.

We find numerical evidence that the large volume expansion agrees with the expansions around $0, \frac12,1$ in their common domain of convergence. In contrast, the expansions around $\tau=0$ and 
$\tau=\frac12$ agree for $\Im m>0$, but satisfy
\be
\begin{split}
T'(\tau,m) =& - T''(\tau,m) + 1 -2m, \quad \\  T_D'(\tau,m) =& - T_D''(\tau,m) + \frac{1-m}{2} 
\end{split}
\ee
when $\Im m<0$. This corresponds to the action of the monodromy matrix $\widetilde{M}_{B,0}$ in \eqref{MBk}, and and is consistent with the value of $(T,T_D)=(\frac12-m,\frac{1-m}{4})$ at the branch point, see \eqref{Eichlerlm}.   
Similarly, the expansions around $\tau=1$ and 
$\tau=\frac12$ agree for $\Im m<0$, but satisfy
\be
\begin{split}
T'(\tau-1,m)+1 =& - T''(\tau,m) + 1  \quad \\  T_D'(\tau-1,m)+T'(\tau-1,m)+\frac{1+m}{2} =& - T_D''(\tau,m) + \frac{1+m}{2} 
\end{split}
\ee
when $\Im m>0$. This corresponds to the action of the monodromy matrix $M_{B,0}$ in \eqref{MBk}, and is consistent with the value of $(T,T_D)=(\frac12,\frac{1+m}{4})$ at the branch point \eqref{Eichlerl}. 

\medskip

As a result, we arrive at the following prescription for computing $T(\tau,m), T_D(\tau,m)$ for any
$\tau$ in the Poincar\'e upper half-plane on the principal sheet (defined below \eqref{Eichlercst}). First, we bring $\tau$ in the fundamental domain $\cF$ by a suitable sequence of $\Gamma_0(4)$ 
transformations $\tau\mapsto \tau+1$ and $\tau\mapsto \frac{\tau-1}{4\tau-3}$ (as well as their inverses), keeping track of its action on the period vector $\Pi^t$ via the following lifts to $\Gamma$: 
\be
\begin{split}
\tau\mapsto \tau+1: & \quad 
\label{MLV}
M_{LV} = \begin{pmatrix} 1 & 0 & 0 & 0 \\
0 & 1 & 0 & 0 \\
1 & 0 & 1 & 0\\
\frac12 & \frac12 & 1 & 1 \end{pmatrix} 
\\
\tau\mapsto \frac{\tau-1}{4\tau-3}: & \quad M_S=  
\begin{pmatrix}
 1 & 0 & 0 & 0 \\
 0 & 1 & 0 & 0 \\
 1 & -2 & -3 & 4 \\
 \frac{1}{2} & -\frac{1}{2} & -1 & 1 \\
\end{pmatrix} = M_C^{-1} M_{LV}^{-1} 
\end{split}
\ee
Second, when $\tau$ belongs to $\cF$, for all $9$ expansions that are needed to cover the whole fundamental domain depending on the value of~$m$ (namely the expansions around $\tau=\infty,0,1/2,1,\tau_B,\allowbreak\frac{\tau_B-1}{4\tau_B-3},\tau_B-1,\frac{3\tau_B-1}{4\tau_B-1},\tau_B+1$), we compute the expansion parameters (such as $q$ or $\tau''-\tau_B''$), divided by the radius of convergence of the respective series.  The expansion for which this ratio is the smallest provides the best-converging series, which we use to perform the computation (see Fig.~\ref{ConvDomain}), taking into account the branch cut.  An important optimization is to compute the series expansion coefficients only once for each value of~$m$.
This algorithm is implemented as {\tt EichlerZ[\{r\_,d1\_,d2\_,ch2\_\},tau\_,m\_]}
in the Mathematica package {\tt F0Scattering.m} (see footnote \ref{foogit}).


\subsection{4D limit\label{sec_4d}}

While the scattering diagrams studied in this work determine the BPS spectrum in pure 5D super-Yang-Mills theory with $SU(2)$ gauge group (and vanishing theta angle) compactified on a circle, it is natural to ask if they reproduce the usual BPS spectrum of 4D super Yang-Mills theory \cite{Seiberg:1994rs} in the limit where the circle shrinks to zero size. Here we take some steps
in this direction, making contact with earlier results in the literature.

Starting from the representation \eqref{Qbf},  the authors of \cite{Banerjee:2020moh} show that 
the 4D limit is obtained by setting
\be
Q_b=-\Lambda^2 R^2, \quad Q_f=\frac12 + R^2 u_{\rm SW} + \dots 
\ee
and taking $R\to 0$ keeping $\Lambda, u_{\rm SW}$ fixed. In this limit, we have 
\be
\lambda\sim \frac{1}{\Lambda^4 R^4} \to \infty\ ,\quad 
u\sim -\frac{u_{\rm SW}}{R^2 \Lambda^4} \to\infty
\ee
Since the ratio $u/\sqrt{\lambda}$ goes to a constant, the modular parameter $\tau$ is fixed,
with $J_4'(\tau) = -8 u_{\rm SW} / \Lambda^2$. Up to a subleading term in $u$, this agrees with the geometric engineering limit
introduced in \cite{Katz:1996fh} (see \cite[(6.19)]{Huang:2013yta}), 
\be
z_1 = \frac14 \exp\left(-4 \eps^2 u_{\rm SW} \right)\ ,\quad z_2 = \eps^4 \Lambda^4, \quad \eps\to 0
\ee
keeping $u_{\rm SW} ,\Lambda$ fixed. 

We can also make contact with the periods of the SW curve as follows. In \cite[\S 4.2.4]{Alim:2023doi},
the authors provide a representation of the periods $Z_{\rm SW}(\gamma_1)$, $Z_{\rm SW}(\gamma_2)$ associated to the states that become massless at $u=1$ and $u=-1$ as quasi-modular forms 
\be
\label{ZSW}
\begin{split}
Z_{\rm SW} (\gamma_1) =& \frac{A^2-2B^2+E}{16 A} = q - 6 q^2 + 24 q^3 + \dots , \quad \\
Z_{\rm SW}(\gamma_2)  =& \frac{\pi \tau(A^2-2B^2+E)-2\I}{16 \pi A}  
\end{split}
\ee
where 
\be
A= \theta_3^2(2\tau)=\frac{\eta(2\tau)^{10}}{\eta(\tau)^4\eta(4\tau)^4}, \quad
B=\theta_4^2(2\tau) = \frac{\eta(\tau)^4}{\eta(2\tau)^2}, \quad
C=\theta_2^2(2\tau) = \frac{4\eta(4\tau)^4}{\eta(2\tau)^2}
\ee
\be
E= \partial_\tau \log B^2 C^2 = \frac13 E_2 + \frac23 B^2 + \frac13 C^2\ ,\quad A^2=B^2+C^2
\ee
The $u$ variable is
\be
u=-1+\frac{2B^2}{A^2} = \frac{24-J_4}{8+J_4}
\ee
such that $u=1,-1,\infty$ correspond to the cusps $\tau=0, \tau=\I\infty, \tau=\frac12$. Fricke transform
$\tau\mapsto -1/(4\tau)$ 
exchanges $B$ and $C$, $Z_{\rm SW} (\gamma_1)$ with $Z_{\rm SW} (\gamma_2)$  
and sends $u\to -u$.

The formulae 
\eqref{ZSW} are obtained by integrating the equations
\be
\partial_\tau 
\begin{pmatrix} 
Z_{\rm SW} (\gamma_1)\\
Z_{\rm SW} (\gamma_2)
\end{pmatrix} =C_{\rm SW}(\tau)  \begin{pmatrix} 
-1 \\ \tau
\end{pmatrix}
\ee
with 
\be
C_{\rm SW}(\tau) =  \frac{1}{16} \left(AB^2-\frac{B^4}{A} \right) = 
\frac{\eta(\tau)^{12} \eta(4\tau)^{12}}{\eta(2\tau)^{18}}
\ee
Using 
\be
J_4+8 = 16 \frac{A^2}{C^2} = 16 \frac{\eta(2\tau)^{24}}{\eta(\tau)^8\eta(4\tau)^{16}}
\ee
we see agreement with the limit $\lambda\to\infty$ of $C''$ in \eqref{Cppgen}, up to rescaling
by a factor of $\lambda^{1/4}$,
\be
C''(\tau,m) \sim  \frac{\eta^{8}(\tau)\eta^4(4\tau)}{\eta^6(2\tau)} \frac{8}{\lambda^{1/4} \sqrt{J_4+8}}
\ee

Of course, one can perform a Fricke-type transform $\tau\mapsto \tau/(2\tau+1)$ in the 
formulae from \cite{Alim:2023doi}, such that $u=1,-1,\infty$ get mapped to $\tau=0,1/2,\I\infty$.
Under this transformation, $B$ is invariant but $A$ and $C$ get exchanged, and 
$J_4 \mapsto \frac{256}{J_4+8}-8$.  As a result, the new identifications are 
\be
u = \frac18 J_4(\tau) \ , \quad C_{SW}(\tau) = \frac{\eta(2\tau)^{18}}{\eta(4\tau)^{12}}=
\frac{1}{q^{1/2}} -18 q^{3/2} + 147 q^{7/2} + \dots 
\ee
which is  recognized as  the 
$\lambda\to\infty$ limit of $C(\tau)$, up to a factor of $2\lambda^{1/4}$,
\be
C(\tau,m) \sim  \frac{\eta(\tau)^4 \eta(2\tau)^6}{\eta(4\tau)^4} 
\frac{\sqrt{J_4+8}}{2\lambda^{1/4}} = 
\frac{\eta(2\tau)^{18}}{2\lambda^{1/4} \eta(4\tau)^{12}}
\ee
In fact, we can rewrite the periods appropriate to the weak coupling regime (see e.g.~\cite{Lerche:1996xu})
\be
a= \frac{u^{1/2}}{\sqrt2}\, _2F_1\left( -\frac14,\frac14,1;1/u^2\right), \quad 
a_D = \frac{\I}{4} (u^2-1) \, _2F_1\left( \frac34,\frac34,2;1-u^2\right),
\ee
in terms of Eichler-type integrals of $C_{SW}=\sum_{n\geq 0} c_n q^{2n-\frac12}$:
\be
a= -\frac{1}{4\sqrt2} f_1, \quad a_D = \frac{1}{4\pi\I} \left( f_2 - f_1 \log q \right)
\ee
with
\be
\begin{split}
f_1 =& \sum_{n\geq 0} \frac{c_n}{2n-\frac12} q^{2n-\frac12} = 
-\frac{2}{q^{1/2}} - 12 q^{3/2} + 42 q^{7/2} + \dots = 2\frac{E_2 - C^2 + 5 A^2}{3C}
\\
f_2 =& \sum_{n\geq 0} \frac{c_n}{(2n-\frac12)^2} q^{2n-\frac12} =
\frac{4}{q^{1/2}} - 8 q^{3/2} + 12 q^{7/2} + \dots=
\frac{16}{C}
\end{split}
\ee
We expect (but cannot show yet) that in the limit $\lambda\to +\infty$, or $m\to -\I\infty$, the scattering diagram $\cD^{\Pi}_{m,\psi}$ should simplify drastically, leaving only the primary scatterings of the initial rays $\cO(k,k)$ and $\cO(k+1,k)$ in every strip $\Re\tau\in (k,k+1)$, reproducing the spectrum of monopoles and dyons in the weakly coupled 4D limit $\tau\to\I\infty$. Independently of this, 
it would be interesting to unfold the $u$-plane analysis of attractor flow trees for  4D $SU(2)$ 
super Yang-Mills theory in \cite{Alim:2023doi} into the modular upper half-plane, which would
avoid the need for branch cuts. It is also interesting to ask if a similar representation in terms of quasi-modular forms
exists for the Eichler integrals \eqref{f12} away from the 4D limit $\lambda=\infty$.


\begin{thebibliography}{10}

\bibitem{Bousseau:2022snm}
P.~Bousseau, P.~Descombes, B.~Le~Floch, and B.~Pioline, ``{BPS Dendroscopy on
  Local $\mathbb {P}^2$},'' {\em Commun. Math. Phys.} {\bf 405} (2024), no.~4,
  108, \href{http://www.arXiv.org/abs/2210.10712}{{\tt 2210.10712}}.

\bibitem{Bridgeland:2006bzr}
T.~Bridgeland, ``{Stability conditions on triangulated categories},'' {\em
  Annals Math.} {\bf 166} (2007), no.~2, 317--345,
  \href{http://www.arXiv.org/abs/math/0212237}{{\tt math/0212237}}.

\bibitem{Toda:2011aa}
Y.~Toda, ``{Bogomolov-Gieseker type inequality and counting invariants},''
  \href{http://www.arXiv.org/abs/1112.3411}{{\tt 1112.3411}}.

\bibitem{Feyzbakhsh:2020wvm}
S.~Feyzbakhsh and R.~P. Thomas, ``{Curve counting and S-duality},'' {\em
  Épijournal de Géométrie Algébrique} {\bf 7} (5, 2023).

\bibitem{Feyzbakhsh:2021rcv}
S.~Feyzbakhsh and R.~P. Thomas, ``{Rank r DT theory from rank 0},'' {\em Duke
  Math. J.} {\bf 173} (2024), no.~11, 2063--2116,
  \href{http://www.arXiv.org/abs/2103.02915}{{\tt 2103.02915}}.

\bibitem{Feyzbakhsh:2022ydn}
S.~Feyzbakhsh, ``{Explicit formulae for rank zero DT invariants and the OSV
  conjecture},'' \href{http://www.arXiv.org/abs/2203.10617}{{\tt 2203.10617}}.

\bibitem{Alexandrov:2023zjb}
S.~Alexandrov, S.~Feyzbakhsh, A.~Klemm, B.~Pioline, and T.~Schimannek,
  ``{Quantum geometry, stability and modularity},'' {\em Commun. Num. Theor.
  Phys.} {\bf 18} (2024), no.~1, 49--151,
  \href{http://www.arXiv.org/abs/2301.08066}{{\tt 2301.08066}}.

\bibitem{bridgeland2016scattering}
T.~Bridgeland, ``{Scattering diagrams, Hall algebras and stability
  conditions},'' {\em Alg. Geo.} {\bf 4} (2017) 523--561,
  \href{http://www.arXiv.org/abs/1603.00416}{{\tt 1603.00416}}.

\bibitem{Denef:2001xn}
F.~Denef, B.~R. Greene, and M.~Raugas, ``{Split attractor flows and the
  spectrum of BPS D-branes on the quintic},'' {\em JHEP} {\bf 05} (2001) 012,
\href{http://www.arXiv.org/abs/hep-th/0101135}{{\tt hep-th/0101135}}.

\bibitem{Denef:2007vg}
F.~Denef and G.~W. Moore, ``{Split states, entropy enigmas, holes and halos},''
  {\em JHEP} {\bf 1111} (2011) 129,
\href{http://www.arXiv.org/abs/hep-th/0702146}{{\tt hep-th/0702146}}.

\bibitem{Alexandrov:2018iao}
S.~Alexandrov and B.~Pioline, ``{Attractor flow trees, BPS indices and
  quivers},'' {\em Adv. Theor. Math. Phys.} {\bf 23} (2019), no.~3, 627--699,
\href{http://www.arXiv.org/abs/1804.06928}{{\tt 1804.06928}}.

\bibitem{Arguz:2021zpx}
H.~Arg\"uz and P.~Bousseau, ``{The flow tree formula for Donaldson-Thomas
  invariants of quivers with potentials},'' {\em Compositio Mathematica} {\bf
  158} (2022), no.~12, 2206--2249,
  \href{http://www.arXiv.org/abs/2102.11200}{{\tt 2102.11200}}.

\bibitem{Bousseau:2019ift}
P.~Bousseau, ``{Scattering diagrams, stability conditions, and coherent sheaves
  on $\mathbb{P}^2$},'' {\em J. Algebraic Geom.} {\bf 31} (2022) 593--686,
  \href{http://www.arXiv.org/abs/1909.02985}{{\tt 1909.02985}}.

\bibitem{Bridgeland:2005my}
T.~Bridgeland, ``{Stability conditions on a non-compact Calabi-Yau
  threefold},'' {\em Commun. Math. Phys.} {\bf 266} (2006) 715--733,
  \href{http://www.arXiv.org/abs/math/0509048}{{\tt math/0509048}}.

\bibitem{Bayer:2009brq}
A.~Bayer and E.~Macri, ``{The space of stability conditions on the local
  projective plane},'' {\em Duke Math. J.} {\bf 160} (2011) 263--322,
  \href{http://www.arXiv.org/abs/0912.0043}{{\tt 0912.0043}}.

\bibitem{graefnitz2020tropical}
T.~Graefnitz, ``{Tropical correspondence for smooth del Pezzo log Calabi-Yau
  pairs},'' {\em J. Alg. Geom.} {\bf 31} (2022), no.~4, 687--749,
  \href{http://www.arXiv.org/abs/2005.14018}{{\tt 2005.14018}}.

\bibitem{Aganagic:2002qg}
M.~Aganagic, M.~Marino, and C.~Vafa, ``{All loop topological string amplitudes
  from Chern-Simons theory},'' {\em Commun. Math. Phys.} {\bf 247} (2004)
  467--512, \href{http://www.arXiv.org/abs/hep-th/0206164}{{\tt
  hep-th/0206164}}.

\bibitem{Beaujard:2020sgs}
G.~Beaujard, J.~Manschot, and B.~Pioline, ``{Vafa\textendash{}Witten Invariants
  from Exceptional Collections},'' {\em Commun. Math. Phys.} {\bf 385} (2021),
  no.~1, 101--226, \href{http://www.arXiv.org/abs/2004.14466}{{\tt
  2004.14466}}.

\bibitem{Mozgovoy:2020has}
S.~Mozgovoy and B.~Pioline, ``{Attractor invariants, brane tilings and
  crystals},'' {\em Annales Inst. Fourier} {\bf 75} (12, 2025) 1331,
  \href{http://www.arXiv.org/abs/2012.14358}{{\tt 2012.14358}}.

\bibitem{Longhi:2021qvz}
P.~Longhi, ``{Instanton Particles and Monopole Strings in 5D SU(2)
  Supersymmetric Yang-Mills Theory},'' {\em Phys. Rev. Lett.} {\bf 126} (2021),
  no.~21, 211601, \href{http://www.arXiv.org/abs/2101.01681}{{\tt 2101.01681}}.

\bibitem{DelMonte:2021ytz}
F.~Del~Monte and P.~Longhi, ``{Quiver Symmetries and Wall-Crossing
  Invariance},'' {\em Commun. Math. Phys.} {\bf 398} (2023), no.~1, 89--132,
  \href{http://www.arXiv.org/abs/2107.14255}{{\tt 2107.14255}}.

\bibitem{DelMonte:2022kxh}
F.~Del~Monte and P.~Longhi, ``{The threefold way to quantum periods: WKB, TBA
  equations and q-Painlev\'e},'' {\em SciPost Phys.} {\bf 15} (2023), no.~3,
  112, \href{http://www.arXiv.org/abs/2207.07135}{{\tt 2207.07135}}.

\bibitem{Xiong:2025isclocal}
Y.~Xiong, ``{Invariant stability conditions on local $\mathbb{P}^1\times
  \mathbb{P}^1$ (after Del Monte-Longhi)},''
  \href{http://www.arXiv.org/abs/2404.05232}{{\tt 2404.05232}}.

\bibitem{Bridgeland:2024iscCY3}
T.~Bridgeland, F.~Del~Monte, and L.~Giovenzana, ``{Invariant Stability
  Conditions on Certain Calabi-Yau Threefolds},''
  \href{http://www.arXiv.org/abs/2412.08531}{{\tt 2412.08531}}.

\bibitem{Kim:2025pidkdi}
H.~Kim, J.~Manschot, G.~W. Moore, R.~Tao, and X.~Zhang, ``{Path Integral
  Derivations Of K-Theoretic Donaldson Invariants},''
  \href{http://www.arXiv.org/abs/2509.23042}{{\tt 2509.23042}}.

\bibitem{lpr-in-progress}
B.~Pioline and R.~Raj, 
\newblock In progress.

\bibitem{Eager:2016yxd}
R.~Eager, S.~A. Selmani, and J.~Walcher, ``{Exponential Networks and
  Representations of Quivers},'' {\em JHEP} {\bf 08} (2017) 063,
\href{http://www.arXiv.org/abs/1611.06177}{{\tt 1611.06177}}.

\bibitem{Banerjee:2020moh}
S.~Banerjee, P.~Longhi, and M.~Romo, ``{Exponential BPS graphs and D-brane
  counting on toric Calabi-Yau threefolds: Part II},''
  \href{http://www.arXiv.org/abs/2012.09769}{{\tt 2012.09769}}.

\bibitem{Gaiotto:2010be}
D.~Gaiotto, G.~W. Moore, and A.~Neitzke, ``{Framed BPS States},'' {\em Adv.
  Theor. Math. Phys.} {\bf 17} (2013), no.~2, 241--397,
\href{http://www.arXiv.org/abs/1006.0146}{{\tt 1006.0146}}.

\bibitem{Andriyash:2010qv}
E.~Andriyash, F.~Denef, D.~L. Jafferis, and G.~W. Moore, ``{Wall-crossing from
  supersymmetric galaxies},'' {\em JHEP} {\bf 1201} (2012) 115,
\href{http://www.arXiv.org/abs/1008.0030}{{\tt 1008.0030}}.

\bibitem{Okounkov:2003sp}
A.~Okounkov, N.~Reshetikhin, and C.~Vafa, ``{Quantum Calabi-Yau and classical
  crystals},'' {\em Prog. Math.} {\bf 244} (2006) 597,
  \href{http://www.arXiv.org/abs/hep-th/0309208}{{\tt hep-th/0309208}}.

\bibitem{Mozgovoy:2008fd}
S.~Mozgovoy and M.~Reineke, ``{On the noncommutative Donaldson-Thomas
  invariants arising from brane tilings},'' {\em Advances in mathematics} {\bf
  223} (9, 2010) 1521--1544, \href{http://www.arXiv.org/abs/0809.0117}{{\tt
  0809.0117}}.

\bibitem{Jafferis:2008uf}
D.~L. Jafferis and G.~W. Moore, ``{Wall crossing in local Calabi Yau
  manifolds},''
\href{http://www.arXiv.org/abs/0810.4909}{{\tt 0810.4909}}.

\bibitem{Manschot:2011ym}
J.~Manschot, ``{BPS invariants of semi-stable sheaves on rational surfaces},''
  {\em Lett. Math. Phys.} {\bf 103} (2013) 895--918,
  \href{http://www.arXiv.org/abs/1109.4861}{{\tt 1109.4861}}.

\bibitem{drezet1985fibres}
J.-M. Dr{\'e}zet and J.~Le~Potier, ``Fibr{\'e}s stables et fibr{\'e}s
  exceptionnels sur $\mathbb{P}_2$,'' in {\em Annales scientifiques de
  l'{\'E}cole Normale Sup{\'e}rieure}, vol.~18, pp.~193--243.
\newblock 1985.

\bibitem{coskun2021existence}
I.~Coskun and J.~Huizenga, ``{Existence of semistable sheaves on Hirzebruch
  surfaces},'' {\em Advances in Mathematics} {\bf 381} (2021) 107636.

\bibitem{Mozgovoy:2013zqx}
S.~Mozgovoy, ``{Invariants of moduli spaces of stable sheaves on ruled
  surfaces},'' \href{http://www.arXiv.org/abs/1302.4134}{{\tt 1302.4134}}.

\bibitem{Manschot:2016gsx}
J.~Manschot and S.~Mozgovoy, ``{Intersection cohomology of moduli spaces of
  sheaves on surfaces},'' {\em Selecta Mathematica} {\bf 24} (2018) 3889--3926, 
  \href{http://www.arXiv.org/abs/1612.07620}{{\tt
  1612.07620}}.

\bibitem{maulik2023cohomological}
D.~Maulik and J.~Shen, ``{Cohomological $\chi$--independence for moduli of
  one-dimensional sheaves and moduli of Higgs bundles},'' {\em Geometry \&
  Topology} {\bf 27} (2023), no.~4, 1539--1586.

\bibitem{Chiang:1999tz}
T.~M. Chiang, A.~Klemm, S.-T. Yau, and E.~Zaslow, ``{Local mirror symmetry:
  Calculations and interpretations},'' {\em Adv. Theor. Math. Phys.} {\bf 3}
  (1999) 495--565,
\href{http://www.arXiv.org/abs/hep-th/9903053}{{\tt hep-th/9903053}}.

\bibitem{Iqbal:2007ii}
A.~Iqbal, C.~Kozcaz, and C.~Vafa, ``{The Refined topological vertex},'' {\em
  JHEP} {\bf 10} (2009) 069,
\href{http://www.arXiv.org/abs/hep-th/0701156}{{\tt hep-th/0701156}}.

\bibitem{Huang:2013yta}
M.-X. Huang, A.~Klemm, and M.~Poretschkin, ``{Refined stable pair invariants
  for E-, M- and $[p, q]$-strings},'' {\em JHEP} {\bf 11} (2013) 112,
  \href{http://www.arXiv.org/abs/1308.0619}{{\tt 1308.0619}}.

\bibitem{macrì2019lecturesbridgelandstability}
E.~Macrì and B.~Schmidt, ``{Lectures on Bridgeland Stability},'' in {\em
  Moduli of Curves: CIMAT Guanajuato, Mexico 2016}, pp.~139--211.
\newblock Springer, 2019.
\newblock \href{http://www.arXiv.org/abs/1607.01262}{{\tt 1607.01262}}.

\bibitem{ks}
M.~Kontsevich and Y.~Soibelman, ``{Stability structures, motivic
  Donaldson-Thomas invariants and cluster transformations},''
  \href{http://www.arXiv.org/abs/0811.2435}{{\tt 0811.2435}}.

\bibitem{joyce2012theory}
D.~Joyce and Y.~Song, {\em A theory of generalized Donaldson--Thomas
  invariants}, vol.~217.
\newblock American Mathematical Society, 2012.

\bibitem{Manschot:2010qz}
J.~Manschot, B.~Pioline, and A.~Sen, ``{Wall Crossing from Boltzmann Black Hole
  Halos},'' {\em JHEP} {\bf 07} (2011) 059,
  \href{http://www.arXiv.org/abs/1011.1258}{{\tt 1011.1258}}.

\bibitem{Bondal:1990}
A.~I. Bondal, ``Representation of associative algebras and coherent sheaves,''
  {\em Mathematics of the USSR-Izvestiya} {\bf 34} (2, 1990) 23.

\bibitem{Aspinwall:2004bs}
P.~S. Aspinwall and S.~H. Katz, ``{Computation of superpotentials for
  D-branes},'' {\em Commun. Math. Phys.} {\bf 264} (2006) 227--253,
  \href{http://www.arXiv.org/abs/hep-th/0412209}{{\tt hep-th/0412209}}.

\bibitem{Mozgovoy:2021iwz}
S.~Mozgovoy, ``{Operadic approach to wall-crossing},'' {\em J. Algebra} {\bf
  596} (2022) 53--88, \href{http://www.arXiv.org/abs/2101.07636}{{\tt
  2101.07636}}.

\bibitem{Descombes:2021snc}
P.~Descombes, ``{Cohomological DT invariants from localization},'' {\em J.
  Lond. Math. Soc.} {\bf 106} (2022), no.~4, 2959--3007,
  \href{http://www.arXiv.org/abs/2106.02518}{{\tt 2106.02518}}.

\bibitem{perling2003some}
M.~Perling, ``{Some quivers describing the derived categories of the toric del
  Pezzos},'' 2003.
\newblock unpublished.

\bibitem{Feng:2000mi}
B.~Feng, A.~Hanany, and Y.-H. He, ``{D-brane gauge theories from toric
  singularities and toric duality},'' {\em Nucl. Phys.} {\bf B595} (2001)
  165--200,
\href{http://www.arXiv.org/abs/hep-th/0003085}{{\tt hep-th/0003085}}.

\bibitem{Feng:2001xr}
B.~Feng, A.~Hanany, and Y.-H. He, ``{Phase structure of D-brane gauge theories
  and toric duality},'' {\em JHEP} {\bf 08} (2001) 040,
\href{http://www.arXiv.org/abs/hep-th/0104259}{{\tt hep-th/0104259}}.

\bibitem{Closset:2019juk}
C.~Closset and M.~Del~Zotto, ``{On 5D SCFTs and their BPS quivers. Part I:
  B-branes and brane tilings},'' {\em Adv. Theor. Math. Phys.} {\bf 26} (2022),
  no.~1, 37--142, \href{http://www.arXiv.org/abs/1912.13502}{{\tt 1912.13502}}.

\bibitem{bridgeland2010helices}
T.~Bridgeland and D.~Stern, ``{Helices on del Pezzo surfaces and tilting
  Calabi--Yau algebras},'' {\em Advances in Mathematics} {\bf 224} (2010),
  no.~4, 1672--1716.

\bibitem{mou2019scattering}
L.~Mou, ``Scattering diagrams of quivers with potentials and mutations,''
  \href{http://www.arXiv.org/abs/1910.13714}{{\tt 1910.13714}}.

\bibitem{arcara2013bridgeland}
D.~Arcara, A.~Bertram, and M.~Lieblich, ``{Bridgeland-stable moduli spaces for
  K-trivial surfaces},'' {\em J. Eur. Math. Soc.(JEMS)} {\bf 15} (2013), no.~1,
  1--38.

\bibitem{maciocia2014computing}
A.~Maciocia, ``{Computing the walls associated to Bridgeland stability
  conditions on projective surfaces},'' {\em Asian Journal of Mathematics} {\bf
  18} (2014), no.~2, 263--280, \href{http://www.arXiv.org/abs/1202.4587}{{\tt
  1202.4587}}.

\bibitem{arcara2015bridgeland}
D.~Arcara and E.~Miles, ``{Bridgeland Stability of Line Bundles on Surfaces},''
  {\em Journal of Pure and Applied Algebra} {\bf 220} (2016), no.~4,
  1655--1677, \href{http://www.arXiv.org/abs/1401.6149}{{\tt 1401.6149}}.

\bibitem{Diaconescu:2007bf}
E.~Diaconescu and G.~W. Moore, ``{Crossing the wall: Branes versus bundles},''
  {\em Adv. Theor. Math. Phys.} {\bf 14} (2010), no.~6, 1621--1650,
  \href{http://www.arXiv.org/abs/0706.3193}{{\tt 0706.3193}}.

\bibitem{Katz:1996fh}
S.~H. Katz, A.~Klemm, and C.~Vafa, ``{Geometric engineering of quantum field
  theories},'' {\em Nucl. Phys. B} {\bf 497} (1997) 173--195,
  \href{http://www.arXiv.org/abs/hep-th/9609239}{{\tt hep-th/9609239}}.

\bibitem{Aganagic:2002wv}
M.~Aganagic, A.~Klemm, M.~Marino, and C.~Vafa, ``{Matrix model as a mirror of
  Chern-Simons theory},'' {\em Comm. Math. Phys.} {\bf 277} (2008) 771--819,
  \href{http://www.arXiv.org/abs/hep-th/0607100}{{\tt hep-th/0607100}}.
  
 \bibitem{Aganagic:2006wq}
M.~Aganagic, V.~Bouchard, and A.~Klemm, ``{Topological Strings and (Almost) Modular Forms},'' {\em JHEP} {\bf 02} (2004) 010,
  \href{http://www.arXiv.org/abs/hep-th/0211098}{{\tt hep-th/0211098}}.
 
  

\bibitem{Haghighat:2008gw}
B.~Haghighat, A.~Klemm, and M.~Rauch, ``{Integrability of the holomorphic
  anomaly equations},'' {\em JHEP} {\bf 10} (2008) 097,
  \href{http://www.arXiv.org/abs/0809.1674}{{\tt 0809.1674}}.

\bibitem{Huang:2010kf}
M.-x. Huang and A.~Klemm, ``{Direct integration for general $\Omega$
  backgrounds},'' {\em Adv. Theor. Math. Phys.} {\bf 16} (2012), no.~3,
  805--849, \href{http://www.arXiv.org/abs/1009.1126}{{\tt 1009.1126}}.

\bibitem{Huang:2011qx}
M.-x. Huang, A.-K. Kashani-Poor, and A.~Klemm, ``{The $\Omega$ deformed B-model
  for rigid $\mathcal{N}=2$ theories},'' {\em Annales Henri Poincare} {\bf 14}
  (2013) 425--497, \href{http://www.arXiv.org/abs/1109.5728}{{\tt 1109.5728}}.

\bibitem{Brini:2009nbd}
A.~Brini and A.~Tanzini, ``{Exact results for topological strings on resolved
  $Y^{p,q}$ singularities},'' {\em Commun. Math. Phys.} {\bf 289} (2009)
  205--252, \href{http://www.arXiv.org/abs/0804.2598}{{\tt 0804.2598}}.

\bibitem{Closset:2021lhd}
C.~Closset and H.~Magureanu, ``{The $U$-plane of rank-one 4d $\mathcal{N}=2$ KK
  theories},'' {\em SciPost Phys.} {\bf 12} (2022) 065,
  \href{http://www.arXiv.org/abs/2107.03509}{{\tt 2107.03509}}.

\bibitem{Aspman:2021vhs}
J.~Aspman, E.~Furrer, and J.~Manschot, ``{Cutting and gluing with running
  couplings in N=2 QCD},'' {\em Phys. Rev. D} {\bf 105} (2022), no.~2, 025021,
  \href{http://www.arXiv.org/abs/2107.04600}{{\tt 2107.04600}}.

\bibitem{Mohri:2000kf}
K.~Mohri, Y.~Onjo, and S.-K. Yang, ``{Closed submonodromy problems, local
  mirror symmetry and branes on orbifolds},'' {\em Rev. Math. Phys.} {\bf 13}
  (2001) 675--715, \href{http://www.arXiv.org/abs/hep-th/0009072}{{\tt
  hep-th/0009072}}.

\bibitem{Seiberg:1994rs}
N.~Seiberg and E.~Witten, ``{Electric - magnetic duality, monopole
  condensation, and confinement in N=2 supersymmetric Yang-Mills theory},''
  {\em Nucl. Phys. B} {\bf 426} (1994) 19--52,
  \href{http://www.arXiv.org/abs/hep-th/9407087}{{\tt hep-th/9407087}}.
  [Erratum: Nucl.Phys.B 430, 485--486 (1994)].

\bibitem{Alim:2023doi}
M.~Alim, F.~Beck, A.~Biggs, and D.~Bryan, ``{Special geometry, quasi-modularity
  and attractor flow for BPS structures},''
  \href{http://www.arXiv.org/abs/2308.16854}{{\tt 2308.16854}}.

\bibitem{Lerche:1996xu}
W.~Lerche, ``{Introduction to Seiberg-Witten theory and its stringy origin},''
  {\em Nucl. Phys. B Proc. Suppl.} {\bf 55} (1997) 83--117,
  \href{http://www.arXiv.org/abs/hep-th/9611190}{{\tt hep-th/9611190}}.

\end{thebibliography}

\providecommand{\href}[2]{#2}\begingroup\raggedright\endgroup

\end{document}